%% file: main.tex
\documentclass[12pt]{iopart}
\usepackage{graphicx}

\usepackage[colorlinks = true,
            linkcolor = blue,
            urlcolor  = blue,
            citecolor = blue,
            anchorcolor = blue]{hyperref}
\usepackage[utf8]{inputenc}
\usepackage{amssymb}
\expandafter\let\csname equation*\endcsname\relax
\expandafter\let\csname endequation*\endcsname\relax
\usepackage{amsmath}
\usepackage{pifont}
\usepackage{upgreek}
\usepackage{textgreek}
\usepackage{placeins}
\usepackage{color}
\usepackage{siunitx}
\usepackage{nth}
\usepackage{wrapfig}
\usepackage[labelfont=bf,textfont=md]{caption}
\usepackage{subcaption}
\DeclareSIUnit\year{yr}
\usepackage[normalem]{ulem}

\newcommand\snowmass{\begin{center}\rule[-0.2in]{\hsize}{0.01in}\\\rule{\hsize}{0.01in}\\
\vskip 0.1in Submitted to the  Proceedings of the US Community Study\\ 
on the Future of Particle Physics (Snowmass 2021)\\ 
\rule{\hsize}{0.01in}\\\rule[+0.2in]{\hsize}{0.01in} \end{center}}

\begin{document}

%\pubblock

\title[KATRIN: Status and Prospects]{KATRIN: Status and Prospects for the Neutrino Mass and Beyond}

\input{authorlist-IOP}

\begin{abstract}
The Karlsruhe Tritium Neutrino (KATRIN) experiment is designed to measure a high-precision integral spectrum of the endpoint region of T$_2$ $\upbeta$ decay, with the primary goal of probing the absolute mass scale of the neutrino. After a first tritium commissioning campaign in 2018, the experiment has been regularly running since 2019, and in its first two measurement campaigns has already achieved a sub-\si{\electronvolt} sensitivity. After \num{1000} days of data-taking, KATRIN's design sensitivity is \SI{0.2}{\electronvolt} at the \SI{90}{\percent} confidence level. In this white paper  we describe the current status of KATRIN; explore prospects for measuring the neutrino mass and other physics observables, including sterile neutrinos and other beyond-Standard-Model hypotheses; and discuss research-and-development projects that may further improve the KATRIN sensitivity.
\end{abstract}

\maketitle

\snowmass

\def\thefootnote{\fnsymbol{footnote}}
\setcounter{footnote}{0}
\setcounter{section}{-1}

\tableofcontents

\section{Executive Summary}
\label{sec:execsummary}

The Karlsruhe Tritium Neutrino (KATRIN) experiment, sited at the Karlsruhe Institute of Technology (KIT) in Germany, is designed to make a high-precision, high-statistics measurement of the endpoint region of the integral spectrum of T$_2$ $\upbeta$ decay. KATRIN's primary physics goal is the measurement of the absolute neutrino-mass scale via tritium-decay kinematics; recently, the experiment has set the world-leading limit of $m_\nu <$ \SI{0.8}{\electronvolt} (\SI{90}{\percent} C.L.), based on its first calendar year of data taking~\cite{knm2lett:2021}. $\upbeta$ decays are provided by a high-luminosity, windowless, gaseous, cryogenic, molecular tritium source, while the $\upbeta$ energies are analyzed by a MAC-E-filter spectrometer with \si{\electronvolt}-scale filter width. In this white paper, in addition to explaining the basics of the experiment, its analysis, and its recent results, we address KATRIN's status and prospects.

\begin{itemize}
    \item Two additional calendar years of regular neutrino-mass data have been acquired, and data-taking continues. Our sensitivity target on $m_\nu$ is \SI{0.2}{\electronvolt} at the \SI{90}{\percent} C.L.
    \item KATRIN's background is higher than anticipated. We enumerate the residual backgrounds affecting the measurement (Sec.~\ref{Subsec:Backgrounds}), and explore several complementary R\&D programs into further background mitigation  (Sec.~\ref{Subsec:BackgroundMitigation}). 
    \item The KATRIN collaboration is making continual progress on understanding and mitigating sources of systematic uncertainty, via theory, computational improvements, and dedicated measurements of systematics (Sec.~\ref{Subsec:SourcesOfSystematicUncertainty}).
    \item In addition to the neutrino-mass scale, KATRIN's exquisite spectral measurement enables a broad range of new-physics searches (Sec.~\ref{Sec:KatrinTestsOfNewPhysics}) ranging from sterile neutrinos to Lorentz-invariance violation to exotic weak currents. Additional physics observables, related to T$_2$ molecular physics and to the conversion-electron decays of our $^{83m}$Kr calibration standard, are accessible through KATRIN spectroscopy.
    \item By extending the KATRIN $\upbeta$-spectrum measurement to lower $\upbeta$ energies, we obtain sensitivity to sterile neutrinos at \si{\kilo\electronvolt} mass scales in a complementary measurement to astrophysical probes. We are actively developing new detector technology that will accommodate the higher rates at these lower energies (Sec.~\ref{Subsubsec:TristanDetector}).
\end{itemize}

\section{Introduction}
\label{Sec:Introduction}

The discovery of neutrino oscillations~\cite{Fukuda:SuperK1998, Ahmad:SNO2002} implies a non-vanishing rest mass of the neutrino, which is physics beyond the Standard Model (BSM). Unlike other BSM physics, such as dark matter or the baryon asymmetry of the Universe, neutrinos can be probed directly in the lab. 
In the standard picture, three massive Dirac (Majorana) neutrinos mix among each other, and seven (nine) physical parameters exist. Neutrino oscillations can measure six of these, but the neutrino mass-scale (and two Majorana CP phases) must be determined by other means. 
 Apart from being a fundamental parameter of nature, the neutrino-mass scale has deep implications. For instance, in many models for neutrino-mass generation the neutrino-mass scale is inversely proportional to the scale of its origin. Together with neutrinoless double-beta decay experiments~\cite{Dolinski:2019}, knowledge of (or constraints on) the neutrino-mass scale allows determination of the Majorana or Dirac character of neutrinos. The neutrino-mass scale is one of a very limited number of cosmological observables that is directly measurable in the laboratory. For instance, a signal in KATRIN accompanied by no signal in current and upcoming cosmology surveys~\cite{DES:CosmNuMass:2021} would imply that the $\Lambda$CDM model would need to be significantly modified, or that neutrinos possess exotic new features or interactions. Similarly, a signal in KATRIN but no signal in neutrinoless double-beta decay experiments would strongly imply that neutrinos are Dirac particles. No signal in KATRIN but a signal in neutrinoless double beta-decay experiments could mean that an alternative diagram leads to that decay, which would have interesting consequences for collider physics and for baryogenesis. As the only model-independent way to probe neutrino mass, direct searches such as KATRIN are crucial.

Building on some \num{70} years of kinematic searches~\cite{Formaggio:NuMassRev2021}, KATRIN probes the neutrino mass via a precise measurement of the electron energy spectrum resulting from the $\upbeta$ decay of molecular tritium:
\begin{equation}
    \mathrm{T}_2 \rightarrow {^3\mathrm{HeT}}^+ + {\mathrm{e}}^{-} + \overline{\nu}_{\mathrm{e}}.
\end{equation}
The nonzero rest mass of the neutrino results in a distortion of the $\upbeta$ spectrum that is statistically most significant at the endpoint of the spectrum. This shape distortion grants sensitivity to the square of the effective neutrino mass
\begin{equation}
    m_\nu^2 = \sum_i\lvert U_{ei} \rvert^2 m_i^2~ = m_\upbeta^2,
\end{equation}
an incoherent sum of the distinct neutrino-mass values $m_i$ weighted by their contributions to the electron-flavor state, given by the elements $U_{ei}$ of the Pontecorvo-Maki-Nakagawa-Sakata (PMNS) matrix. Here, we denote the observable $m_\nu$, but $m_\upbeta$ is often used in the literature for the same observable.  

Tritium is the isotope of choice for a number of reasons. First, it has the second lowest known $\upbeta$-decay endpoint energy, increasing the significance of the contribution of the neutrino mass. Second, its short half-life of \num{12.3} years enables high $\upbeta$-decay rates. Third, the relatively simple electron structure of both T and ${^3\mathrm{He}}^+$ facilitates calculation both for the atom and the molecule. Fourth, the low $Z$ of the nucleus reduces energy loss by inelastic scattering of the $\upbeta$ as it escapes the source. Finally, tritium $\upbeta$ decay is a super-allowed transition, and no shape correction to the spectrum is needed.

Only about one part in \num{e13} of the $\upbeta$ decays appear in the last \si{\electronvolt} of the spectrum; a statistically significant measurement at the endpoint of the spectrum therefore requires an extremely bright source and excellent resolution for measuring the energy of the decay electron. The Tritium Laboratory Karlsruhe (TLK) provides an intense \SI{e11}{Bq} gaseous tritium source while the electromagnetic spectrometer techniques pioneered by the earlier Mainz~\cite{Kraus2005} and Troitsk~\cite{Aseev2011} experiments provide sub-\si{\electronvolt} energy discrimination. The ability to precisely measure the $\upbeta$-decay spectrum grants sensitivity to additional physics topics such as sterile neutrinos, the number density of relic neutrinos and Lorentz-invariance violation. Such an experimental setup is not without its challenges, however. The intense gaseous source requires an understanding of the systematic uncertainties, e.g. plasma properties and instabilities. The low event rate at the endpoint of the spectrum demands identification and mitigation of sources of background. Significant progress has been made on these challenges as the KATRIN collaboration works to reach its \SI{0.2}{\electronvolt} design goal. %; beyond this it will be necessary to develop an atomic tritium source. 
A high-rate detector, denoted TRISTAN, is planned to expand the search for sterile neutrinos.

Section~\ref{Sec:TheKatrinApparatus} describes the KATRIN apparatus, including recent technological achievements (Sec.~\ref{Subsec:TechnicalDetails}), operational history (Sec.~\ref{Subsec:OperationalHistory}), and sources of background (Sec.~\ref{Subsec:Backgrounds}). General analysis strategies are summarized in Sec.~\ref{Sec:KatrinAnalysisToolsAndStrategies}, including a discussion of systematic uncertainties in Sec.~\ref{Subsec:SourcesOfSystematicUncertainty}. Section~\ref{Sec:MeasuringTheNeutrinoMassScaleWithKatrin} gives details about the inference of the neutrino mass, including KATRIN's recent results. Prospects for KATRIN tests of beyond-Standard-Model physics are described in Sec.~\ref{Sec:KatrinTestsOfNewPhysics}, with additional physics observables in Sec.~\ref{Sec:AdditionalPhysicsObservables}. Section~\ref{Sec:R&DForTheFutureOfKatrin} gives an overview of current research and development efforts, ranging from background mitigation (Sec.~\ref{Subsec:BackgroundMitigation}) to acceptance improvements (Sec.~\ref{Subsubsec:AcceptanceImprovements}) to plans for a future \si{\kilo\electronvolt}-scale sterile-neutrino search (Sec.~\ref{Sec:keVRandD}).  

% ############################################################
% The KATRIN apparatus
%# ##########################################################
\section{The KATRIN apparatus}
\label{Sec:TheKatrinApparatus}

\begin{figure}[tbp]
    \centering
    \includegraphics[width=\textwidth]{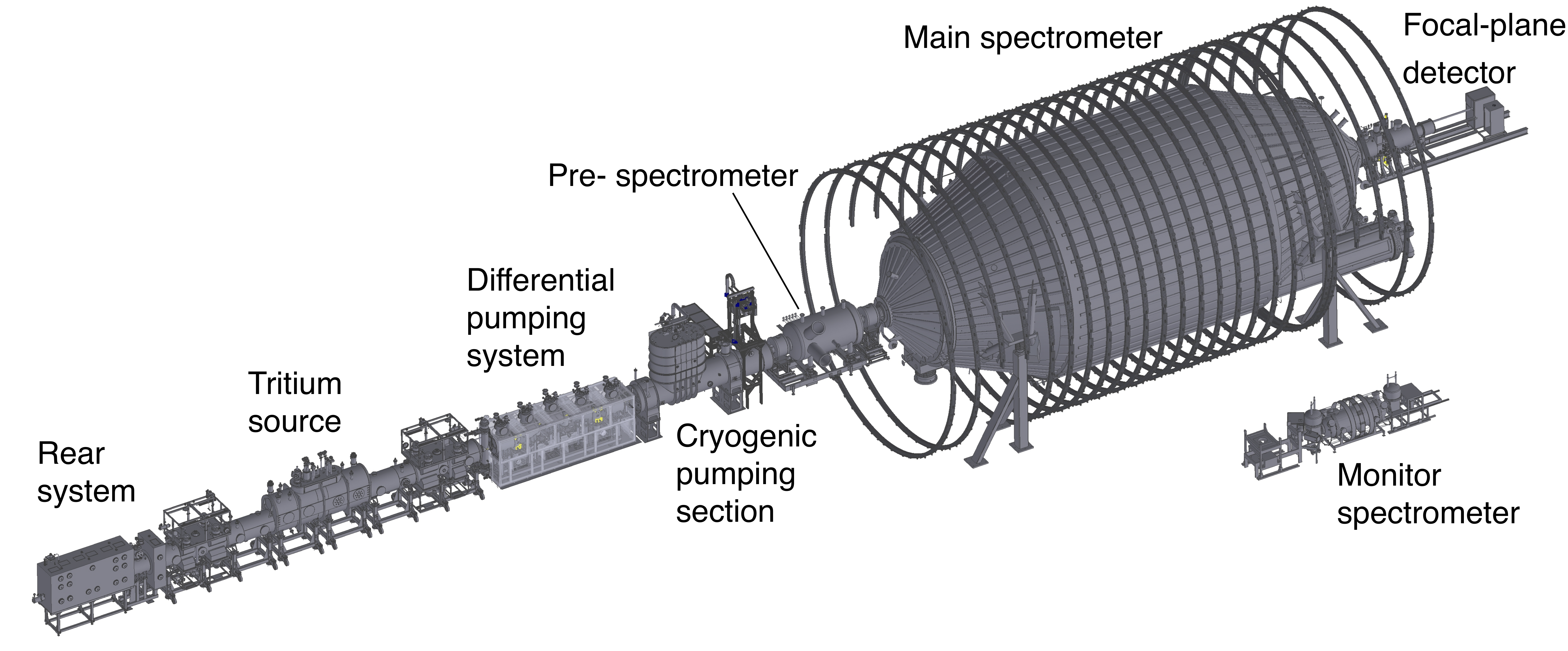}
    \caption{Engineering drawing of the KATRIN beamline.}
    \label{fig:beamline}
\end{figure}

The \SI{70}{\metre} KATRIN beamline is designed to perform high-precision energy analysis of $\upbeta$ electrons from a high-luminosity, gaseous T$_2$ source. Here, we give a high-level overview of the apparatus (Fig.~\ref{fig:beamline}); a complete technical description may be found in Ref.~\cite{Aker2021HW}, and details of selected systems, chosen based on recent technical progress, are given in Sec.~\ref{Subsec:TechnicalDetails}. Briefly, T$_2$ gas is purified in a tritium loop system (Sec.~\ref{Subsubsec:SourceGasCirculation}), which continuously delivers cold T$_2$ gas to the center of the source system. The gas diffuses to both ends of the source-system cryostat, where the first pumping stages are located. A fraction of the T$_2$ molecules experience $\upbeta$ decay during their flight within the source beam tube. Of the resulting $\upbeta$s, those that are emitted in the downstream direction are guided by magnetic field lines through the chicanes of two consecutive pumping stages that together reduce the flow of neutral tritium by some \num{14}~orders of magnitude. The first, differential pumping stage relies on turbomolecular pumps, while the second is a cryogenic pumping stage in which tritium is adsorbed onto an Ar frost layer that lines the beam tube.

Past the pumping systems, the $\upbeta$s reach a pair of tandem spectrometers designed according to the principle of magnetic adiabatic collimation with electrostatic filtering (MAC-E filters). This operating principle will be described in detail in Sec.~\ref{Subsubsec:MACEfilter}. Briefly, a MAC-E filter uses magnetic-field gradients to collimate the $\upbeta$-electron flux, allowing a longitudinal retarding potential $U$ to set a threshold on the total kinetic energy of the $\upbeta$s. A $\upbeta$ with energy $E > qU$, where $q$ is the electron charge, will pass through the spectrometer to the downstream exit; if $E < qU$, the electron cannot pass and is instead reflected in the upstream direction. To a good approximation, a MAC-E filter thus acts as an integrating high-pass filter, with a characteristic relative filter width set by the ratio of the minimum to the maximum magnetic fields. By scanning through a set of retarding energies $\{qU_i\}$, KATRIN is able to reconstruct an integral spectrum. 

In the original KATRIN design, the first MAC-E filter -- the pre-spectrometer -- was envisioned as a pre-filter, excluding the low-energy bulk of the spectrum to avoid backgrounds arising from collisions of $\upbeta$s with residual gas in the larger, high-resolution main spectrometer with a readily adjustable $qU$. However, following the discovery that tandem spectrometer operation led to an effectively energy-dependent background (Sec.~\ref{Subsec:Backgrounds}), KATRIN now runs without energizing the pre-spectrometer. $\upbeta$ electrons with sufficient energy pass through the main spectrometer and are counted in the focal-plane detector, a monolithic silicon p-i-n diode segmented into 148 equal-area pixels. A \SI{10}{\kilo\volt} post-acceleration electrode, immediately upstream of the detector, elevates the signal energies above local backgrounds.

Calibration and monitoring systems are located in several positions along the beamline. In the tritium circulation loops, a laser-Raman spectroscopy system measures the tritium purity of the gas being injected into the source. At its upstream end, the source section terminates in a gold-plated rear wall which can be held at a small voltage to better control the source plasma conditions. Behind the rear wall, silicon drift detectors monitor the source activity by detecting X-rays from $\upbeta$ interactions in the rear wall. An electron gun provides calibration electrons with controlled angle and energy; these electrons enter the source via a small aperture in the center of the rear wall, and travel the entire length of the beamline. Gaseous \textsuperscript{83m}Kr can be introduced into the source to circulate either by itself or along with other gases (Sec.~\ref{Subsubsec:Intense83mKrCalibrationSource}), providing monoenergetic electron lines through internal-conversion decays. 

Further downstream, in the cryogenic pumping section, the forward beam monitor uses silicon detectors to sample the $\upbeta$ rate at the edge of the flux tube. A condensed \textsuperscript{83m}Kr source is available, providing calibration electrons from a specific location and controlled starting potential. Parallel to the main spectrometer, a monitor spectrometer -- the refurbished MAC-E filter once used by the Mainz collaboration~\cite{Kraus2005} -- allows real-time comparison of the main-spectrometer retarding potential to a \textsuperscript{83m}Kr standard. The retarding potential is also monitored by high-precision voltage dividers. Finally, the detector section is fitted with electron and $\upgamma$ sources for energy calibration and efficiency measurements.

\subsection{Selected technical details}
\label{Subsec:TechnicalDetails}

Here, we highlight technical advances since the preparation of Ref.~\cite{Aker2021HW}. These advances include the system for circulating the source gas (Sec.~\ref{Subsubsec:SourceGasCirculation}), a new operational mode for the primary MAC-E filter (Sec.~\ref{Subsubsec:MACEfilter}), and a very intense, gaseous \textsuperscript{83m}Kr calibration source (Sec.~\ref{Subsubsec:Intense83mKrCalibrationSource}).

\subsubsection{Source-gas circulation}
\label{Subsubsec:SourceGasCirculation}

\begin{figure}[!ht]
    \centering
    \includegraphics[width=\textwidth]{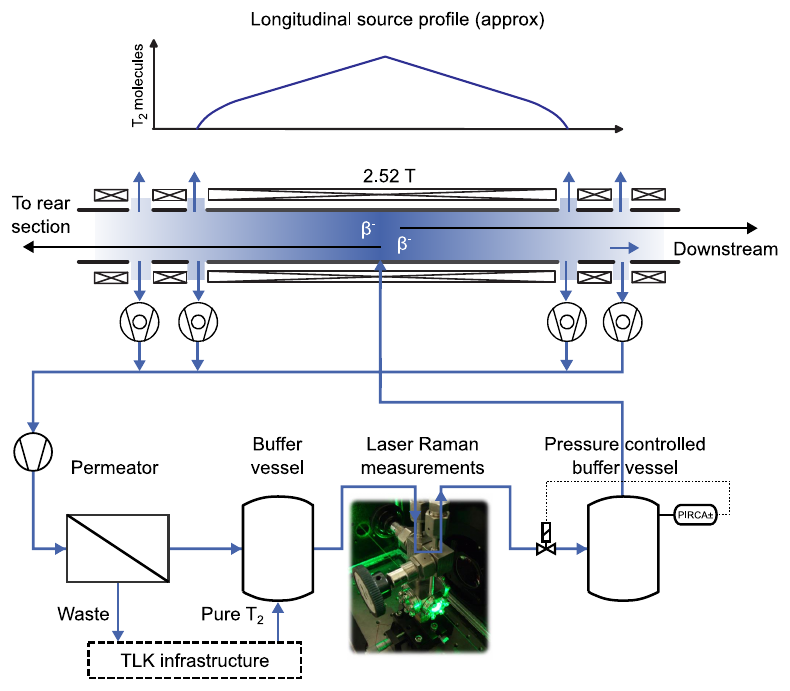}
    \caption{Simplified schematic of the windowless gaseous tritium source (WGTS) and tritium circulation loop. The temperature of the WGTS beam tube was \SI{30}{\kelvin} in the first two campaigns \cite{Aker2021Knm1, knm2lett:2021} and has been changed to \SI{80}{\kelvin} for subsequent campaigns.}
    \label{fig:wgts_source_density_profile_schematic}
\end{figure}

The Tritium Loop System supplies the windowless gaseous tritium source (WGTS) with a stable flow of tritium. Gas is fed into an injection chamber from two capillaries (the tritium and the tritium+krypton injection capillary) which are thermally coupled to the beam tube over a length of \SI{5}{\metre}. It is then injected into the central part of the WGTS beam tube via circumferential orifices~\cite{Kuckert2018}. Two differential pumping ducts are attached to both ends of the source tube, each equipped with two or four turbomolecular pumps (TMPs). A stationary gas-density profile inside the source tube is formed by continuous injection of gas in the middle of the beam tube and pumping at both ends~\cite{Kuckert2018}, as can be seen in Fig.~\ref{fig:wgts_source_density_profile_schematic}. The integral of this density profile along the length of the WGTS, the column density, is an experimentally accessible parameter for measurements and monitoring~\cite{Babutzka2012}.

The main requirements for the WGTS are 24/7 operation over long periods of time with tritium at a high purity of \SI{>95}{\percent}, and a stability of the column density better than \SI{<0.1}{\percent}~\cite{KATRIN2005,Aker2021HW}. A stable column density is achieved when the injection pressure and temperature of the beam tube are kept stable. Temperature stability is provided by the WGTS cryogenic system, while pressure stability and composition are the task of the Tritium Loop System. In order to achieve the necessary source activity, a stable cumulative throughput of up to \SI{\sim 40}{\gram~tritium\per day} is required. Such high flow rates of tritium can only be achieved in a closed-loop operation~\cite{Priester2015, Sturm:2021} as described below.

Tritium is injected from a pressure-controlled buffer vessel via the tritium-injection capillary and injection chamber into the WGTS beam tube. The majority of the gas (\SI{>99.9}{\percent}) is pumped out of the beam path with the first-line TMPs at both ends. The remaining gas is pumped off further along the beamline~\cite{Marsteller:2021}. The fore vacuum of the first-line TMPs, which directly attach to the pumping ducts, is provided by four second-stage TMPs, a scroll and a metal bellows pump in series. The tritium+krypton injection capillary connects the volume behind the second-stage TMPs with the injection chamber via a valve. The metal bellows pump pushes the gas through a palladium membrane filter (``permeator''). Only hydrogen isotopologs may pass through this permeator,
and all non-hydrogen impurities are thus removed~\cite{Bornschein:fst2005,Goto1970}. In order to avoid impurity accumulation and subsequent blocking of this filter, a small fraction (\SI{\sim 1}{\percent}) of the gas is continuously extracted in front of the permeator. The gas that passes through the permeator enters into an intermediate buffer vessel, supplemented with high-purity tritium to replace the gas extracted at the permeator. This high-purity tritium is drawn from a system of buffer vessels that serve as the interface to the TLK infrastructure. From there, the gas is led over a Laser Raman sampling cell~\cite{Zeller:sensors2020} and a regulation valve back into the pressure-controlled buffer vessel, completing the closed loop.

In addition to this standard configuration, the Tritium Loops system can be operated in further distinct modes~\cite{PhDMarsteller2020}, depending on the usage of the tritium- and the tritium+krypton injection capillary. When doing calibration measurements using $^{\mathrm{83m}}$Kr, the closed loop described above would filter out the krypton immediately, rendering the measurement impossible. If the tritium+krypton injection capillary is included in addition to the tritium-injection capillary, a fraction of the gas pumped out of the WGTS is directly re-injected. This so-called double-injection mode allows co-circulation of krypton through the source. However, a large part of the $^{\mathrm{83m}}$Kr is still filtered out of the gas stream. In order to achieve the maximum possible $^{\mathrm{83m}}$Kr activity in the source, the WGTS loop-only mode is used. By feeding all of the gas pumped out of the WGTS beam tube directly back into it, no $^{\mathrm{83m}}$Kr is lost. The small fraction of gas lost to pumps outside the WGTS is replenished using the tritium-injection capillary.

The design source temperature is \SI{30}{K}, achieved by a two-phase neon cooling at $10^{-3}$ stability \cite{Grohmann2013}. In order to perform $^{\mathrm{83m}}$Kr calibrations without condensation in the cold source, one needs to operate the source tube at a higher temperature (e.g. at \SI{120}{K} with two-phase argon cooling, or at \SI{80}{K} with two-phase nitrogen cooling).
The operation at higher temperature not only changes the Doppler broadening of the $\upbeta$ energy, but it also limits the maximum achievable column density for a given tritium throughput. Thus, krypton-calibration and neutrino-mass operation can only be performed under rather different states of the source with regard to column density and temperature, which both impact the secondary ionisation in the source plasma. Experimental studies indicated that the plasma effects could be more relevant than initial simulations indicated. It was therefore decided in 2020 (after the campaigns for which neutrino-mass results have been published~\cite{Aker2019, Aker2021Knm1, knm2lett:2021}), that neutrino-mass and calibration measurements should subsequently be performed at a common temperature of \SI{80}{K}. In this way, one can switch from one mode to the other to perform calibration measurements at any time, and the calibration is performed under more similar conditions to the primary measurement. The drawback of this choice, is however, that the maximum possible column density is reduced by about \SI{10}{\percent}. 

\subsubsection{MAC-E filter}
\label{Subsubsec:MACEfilter}

A sensitivity to the neutrino mass of less than \SI{1}{\electronvolt} requires energy discrimination at a similar level. Such resolution is obtained in KATRIN using the MAC-E Filter technique. 

\begin{figure}[tb]
    \centering
    \includegraphics[width=.9\textwidth]{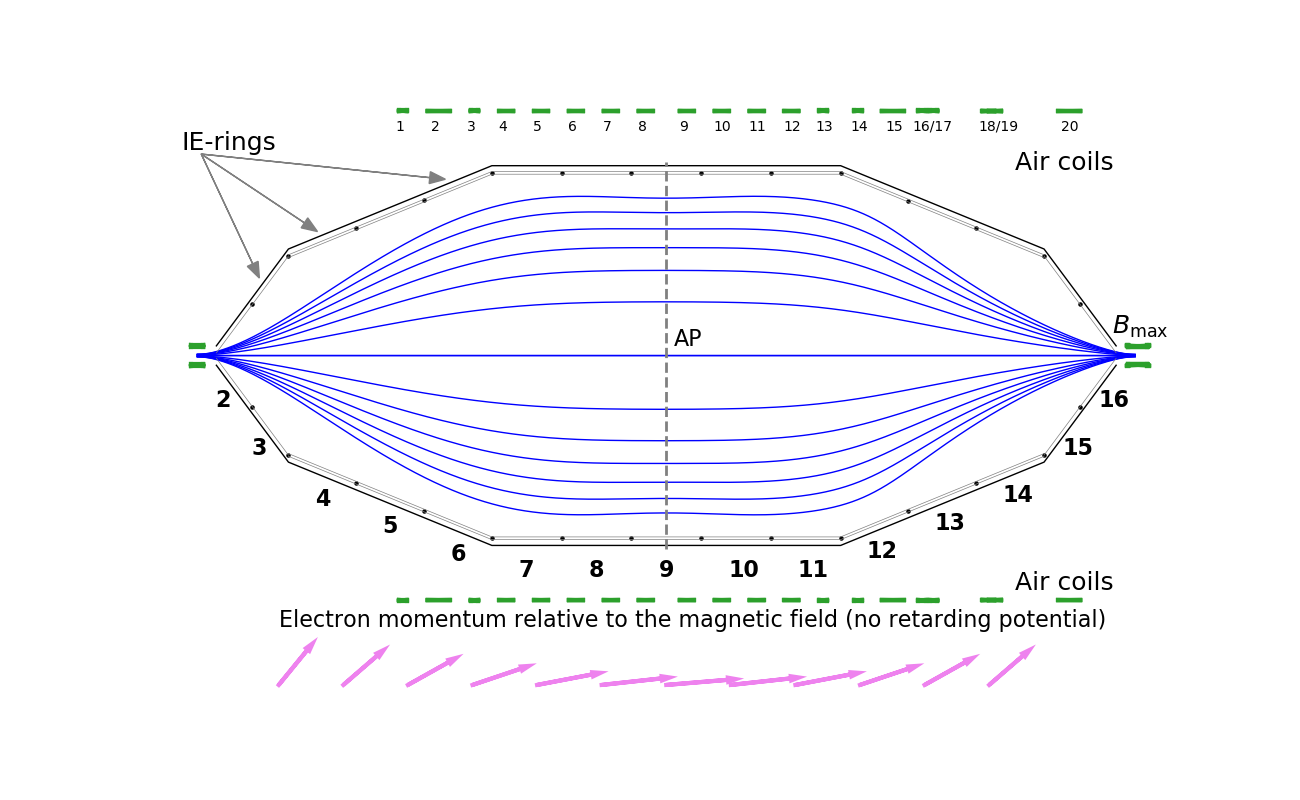}
    \caption{Schematic view of the KATRIN main spectrometer, a MAC-E Filter. Electrons from the tritium source are born in a high magnetic field and enter the spectrometer from the left, guided adiabatically along magnetic field lines (solid blue) to a region with low magnetic field. The lower part of the figure shows the evolution of the electron momentum due to the conservation of the orbital magnetic moment, in the absence of a retarding potential: the momentum is almost aligned with the magnetic field lines at the analyzing plane (dashed gray, AP). The magnetic field is defined by the stray fields of the superconducting solenoids at the entrance and exit of the spectrometer and by the system of \num{20} air coils (thick, green bars above and below the vessel). }
    \label{fig:MACEPrinciple}
    % Figure from the SAP idea paper, can be slightly modified or replaced.
\end{figure}

The basic working principle is shown in Fig.~\ref{fig:MACEPrinciple}. Electrons are born in a high magnetic field (\SI{2.5}{\tesla}) in the tritium source and are guided magnetically through the transport section towards the spectrometers, which analyze their kinetic energy. The electrons in the source have an isotropic starting-angle distribution. If an electron's motion is adiabatic and the magnetic-field strength $B$ is reduced by several orders of magnitude, the conserved orbital magnetic moment ($\mu = \frac{E_\perp}{B}$ in the non-relativistic limit) requires a corresponding reduction of the electron's kinetic energy due to motion perpendicular to the magnetic field lines, $E_\perp$. The total kinetic energy $E_\mathrm{tot} = E_\perp+E_\parallel$ is conserved and for large pitch angles (the angle $\theta$ between the magnetic field $\vec{B}$ and the electron momentum $\vec{p}$ at the electron's starting location) is mainly converted into $E_\parallel = E_\mathrm{tot}\cdot \cos{\theta}$. Without an electric field, $E_\perp$ reaches its minimum at the position where the magnetic field is minimal (typically, $\mathcal{O}$(\num{1}\si{\milli\tesla})). A high voltage $U$ (typically \SI{-18.6}{\kilo\volt}) applied to the spectrometer vessel results in an energy threshold $qU$ for an electron with charge $q$. The potential reaches its maximum in the central part of the spectrometer, the so-called analyzing plane (AP, see Fig.~\ref{fig:MACEPrinciple}), creating a barrier for electrons with insufficient kinetic energy $E_\parallel < qU$, while the other electrons are transmitted. The filter width of the spectrometer is given by $\Delta E = \frac{B_\mathrm{min}}{B_\mathrm{max}}\cdot E$ -- the maximal residual transverse energy $E_\perp$ for which an electron with the energy $E$ will be transmitted by the spectrometer.

The KATRIN main spectrometer has a design filter width of $\Delta E = E\cdot$\SI{0.3}{\milli\tesla}/\SI{6}{\tesla} $\approx$ \SI{1}{\electronvolt} for energies of \SI{18.6}{\kilo\electronvolt}. The $\upbeta$-spectrum is measured by scanning the retarding potential $U$ near the endpoint.

The magnetic flux of \SI{134}{\tesla\centi\metre\squared} is conserved along KATRIN's beamline. The high statistics required by the precise measurement of the shape of the tritium endpoint spectrum can be obtained by increasing the activity of the tritium source. The luminosity of the source, however, is limited in practice by the electron scattering in the source: a higher source length and column density would make it effectively opaque to electrons at relevant energies. Therefore the diameter of the source tube is the primary lever for increasing the luminosity. The KATRIN source has a diameter of \SI{9}{\centi\meter}. The reduction of the magnetic field in the main spectrometer increases the area of the flux, whose diameter reaches \SI{\sim9}{\metre} for the design magnetic-field configuration. The inner diameter of the main spectrometer is \SI{9.8}{\metre}.

Adiabatic motion is another key requirement for the MAC-E Filter; that is, the gradients of the magnetic fields must be small. The adiabatic transformation of transverse into longitudinal motion occurs over a distance of several meters and sets the \SI{23.23}{\metre} length scale of the KATRIN main spectrometer.

To avoid collisions of electrons with the residual gas in the main spectrometer and to reduce backgrounds, the spectrometer vessel volume of \SI{1240}{\metre\cubed} is kept at \SI{e-11}{\milli\bar}. Ref.~\cite{Arenz2016} describes the vacuum systems that maintain this pressure.

The magnetic field inside the main spectrometer is defined by the superconducting solenoids at the entrance and the exit of the main spectrometer. The solenoid at the exit of the spectrometer, the so-called pinch magnet, provides the highest magnetic field in the beam line. The magnetic field is fine-tuned by a system of air-cooled magnet coils (``air coils'') surrounding the main spectrometer (see Fig.~\ref{fig:MACEPrinciple}). A high-sensitivity magnetometer system monitors the fields at several external positions during measurements.

The retarding potential $U$ is generated by a high-precision, high-voltage supply with a distribution and monitoring system. The stability and precision of the voltage applied to the spectrometer vessel and inner electrode (IE) reach the ppm (\num{e-6}) level~\cite{Rodenbeck:2022iys}. The inner-electrode system also allows fine shaping of the retarding-potential profile along the symmetry axis of the spectrometer to ensure the required transmission properties.

In the design configuration of the main spectrometer, the maximum of the absolute retarding potential and the minimum of the magnetic field are placed in the center of the spectrometer. The cylindrical part of the vessel and IE (rings 7-11 in Fig.~\ref{fig:MACEPrinciple}) provides a homogeneous potential, and the magnetic field is kept minimal and constant along the symmetry axis of the vessel. With the analyzing plane in the center of the spectrometer, the variation of the electric potential and magnetic field across the analyzing plane is minimized. The symmetrical field configuration for the neutrino-mass measurements was defined using simulations~\cite{Glueck2013}.

The observation of an elevated background level (see Section \ref{Subsec:Backgrounds} for details) motivated a modification of the nominal magnetic flux in the spectrometer to improve the statistical sensitivity to the neutrino mass. The most significant background contributions scale with the imaged volume of the main spectrometer -- that is, the number of counted background electrons is proportional to the volume of the flux tube within the spectrometer vessel. The first modification was made during the first measurements of the tritium spectrum (Sec.~\ref{Subsec:OperationalHistory}). The magnetic field in the analyzing plane was increased by a factor of \num{3}, reducing the corresponding flux-tube volume at the cost of a broader filter width. 

The next background-mitigation measure was a new field configuration: the shifted analyzing plane (SAP)~\cite{lokhov2022background}. This operating mode further reduces the effective volume from which any background electrons can reach the detector, by shifting the maximum of the absolute retarding potential and the minimum of the magnetic field towards the detector. The low-energy background electrons cannot overcome the potential barrier and are reflected back to the source side. Only those that are emitted between the analyzing plane and the detector are counted. With this mode the background of the main spectrometer is reduced by a factor of two. The challenge of finding the optimal configuration is to minimize variations of the electromagnetic field in the analyzing plane while preserving the transmission conditions and narrow filter width of the spectrometer. For that purpose we use the air coils and IE system to place the magnetic field minima and the electric potential maxima at the same positions for all the radii in the spectrometer. Nonetheless, the increased field variation requires more fine-grained analysis of spectra recorded in different regions of the detector. After calibration measurements and assessment of the related systematic uncertainties, the SAP configuration was implemented as the default operating mode of the KATRIN main spectrometer. 

\subsubsection{Intense \textsuperscript{83m}Kr calibration source}
\label{Subsubsec:Intense83mKrCalibrationSource}

The monoenergetic conversion electrons from the radioactive decay of the \textsuperscript{83m}Kr isomeric state provide a unique calibration tool, since \textsuperscript{83m}Kr gas can be injected into the source where the conversion electrons are produced in a similar environment to the $\upbeta$ electrons. The continuous $\upbeta$ spectrum is distorted due to the energy losses of the $\upbeta$ electrons in the source gas as well as due to the time instabilities and spatial inhomogeneity of the source plasma. Measuring the spectra of the individual electron conversion lines provides information on the distortions. At the same time, due to its short half-life ($T_{1/2} =$ \SI{1.8620}{\hour}), \textsuperscript{83m}Kr does not remain in the source long-term.

At the Nuclear Physics Institute, Czech Academy of Sciences, {\v R}e{\v z}, KATRIN's \textsuperscript{83m}Kr source~\cite{Venos2014} was developed, based on the deposition of the parent radionuclide \textsuperscript{83}Rb into a few zeolite spherules (15-30 pieces). The krypton generator for injection of \textsuperscript{83m}Kr into the KATRIN source was then constructed using Swagelok components~\cite{Sentkerestiova2018}. The long-lived \textsuperscript{83}Rb ($T_{1/2} =$ \SI{86.2}{\day}) remains firmly attached to the aluminosilicate skeleton of zeolite while about \SI{80}{\percent} of the \textsuperscript{83m}Kr atoms born in the $^{83}$Rb decay emanate from zeolite and are thus available for injection. The parent $^{83}$Rb was produced in the proton-induced reactions $^{\rm{nat}}$Kr(p,xn)$^{83}$Rb at the U-120M and TR-24 cyclotrons  of the NPI within the Centre of Accelerators and Nuclear Analytical Methods (CANAM), where $^{\rm{nat}}$Kr refers to Kr gas with natural isotope abundances. Increasing activity demands resulted in gradual development of several generations of dedicated, pressurized, $^{\rm{nat}}$Kr gas targets. Continuous optimization of the target allowed us to enhance the $^{83}$Rb production rate from the initial value of \SI{14}{\mega\becquerel\per\hour} to the current value of \SI{132}{\mega\becquerel\per\hour}. Several \textsuperscript{83}Rb/\textsuperscript{83m}Kr emanation sources with a typical $^{83}$Rb activity of \SI{\sim 1.2}{\giga\becquerel} were provided for krypton measurements at KATRIN.

The first krypton measurement was accomplished in 2017, when \textsuperscript{83m}Kr was allowed to propagate freely from the krypton generator output into the source~\cite{Arenz2018,Altenmueller2020:Krspectr}. Then the Gaseous Krypton Source (GKrS) was assembled~\cite{Aker2021HW}, consisting of the krypton generator (in secondary containment to prevent potential tritium contamination), WGTS tube and necessary loops. The GKrS system allows three principal application modes: \textsuperscript{83m}Kr alone, T$_2$+\textsuperscript{83m}Kr and D$_2$+\textsuperscript{83m}Kr. The decision to neglect plasma effects when analyzing the first KATRIN neutrino-mass data, when only limited initial $\upbeta$-spectrum statistics were available, was based on the L$_3$-32 krypton line spectrum~\cite{Aker2019}. In 2020, extensive krypton measurements were performed to quantitatively study plasma effects (Sec.~\ref{Subsec:SourcesOfSystematicUncertainty}). To achieve the KATRIN design goal, an ultra-high-intensity \textsuperscript{83m}Kr source with \SI{\sim 10}{\giga\becquerel} of the parent  $^{83}$Rb was required, see also~\cite{knm2lett:2021}. The amount of $^{83}$Rb corresponds to the activity limit defined in the handling license of the Tritium Laboratory Karlsruhe for this radionuclide.

The fabrication of this source presented several technical challenges related to the large-scale production of $^{83}$Rb: long-lasting bombardment of the single target with a high-current proton beam, semi-automated and remote processing of the target, and deposition of the $^{83}$Rb into the zeolite. The new technology had to provide a high-quality \textsuperscript{83}Rb/\textsuperscript{83m}Kr emanation source, while minimizing the contamination risk and the radiation burden on personnel.

The final target system, in particular the beam entrance window, demonstrated high stability during the longest continuous bombardment performed on the TR-24 cyclotron: \SI{4.5}{days} with a \SI{24}{\mega\electronvolt}, \SI{50}{\micro\ampere} beam. Remote elution of $^{83}$Rb from the target chamber was performed in a dedicated hot cell equipped with telemanipulators. Roughly \SI{94}{\percent} of the $^{83}$Rb present in the target was recovered into the aqueous solution. Subsequent deposition of $^{83}$Rb into the zeolite spherules was a smooth process resulting in overall deposition efficiency of $\sim$\SI{92}{\percent}. In order to reduce unnecessary radiation exposure, some routine test measurements of the source were minimized, a measure made possible by the experience and data obtained during previous investigations of the sources and their behavior. The calibration source was successfully used at KATRIN in summer 2021.

\subsection{Operational history}
\label{Subsec:OperationalHistory}

The KATRIN collaboration was founded in 2001, and a comprehensive baseline design of the experiment was completed in 2004~\cite{KATRIN2005}. Initial measurements were conducted in 2012 to study transmission properties and background processes of the main spectrometer and focal-plane detector, which were the first major parts of the experimental apparatus to become operational~\cite{Fraenkle2017}. 

``First Light'' was achieved in 2016, when photoelectrons and inactive ions from an ultraviolet source were transported for the first time through the complete 70-m-long setup, including the source and transport systems inside the Tritium Laboratory Karlsruhe. The following year, 2017, saw a first campaign of precision spectroscopy, using monoenergetic conversion electrons from gaseous and condensed \textsuperscript{83m}Kr sources to demonstrate the high-resolution and high-stability performance of KATRIN's MAC-E type energy filter~\cite{Arenz2018,Altenmueller2020:Krspectr,Arenz2018b}.

In 2018 the KATRIN collaboration celebrated the official inauguration of the beamline with the first tritium operation of the source (albeit at low tritium content). Initial $\upbeta$-decay spectra showed an excellent match to the model description~\cite{Aker2020a}. Subsequently, the column density of the source was increased in successive steps; at about \SI{20}{\percent} of the nominal column density, the first neutrino-mass data were taken in Spring 2019~\cite{Aker2021Knm1,Aker2019}. After a final tritium burn-in phase of the source section in June 2019, stable conditions at high tritium throughput and isotopic purity were achieved in advance of the second neutrino-mass measurement campaign~\cite{knm2lett:2021}.

Since entering regular operations, the annual measurement program aims for at least \num{210} calendar days of scientific data taking, which has been continuously achieved from 2018 onwards. Typically, the beamline and tritium laboratory are shut down for maintenance work from December until February. The measurement program, starting in March, is split into three campaigns per year to allow a two- to three-week break in between each pair of campaigns to regenerate the tritium-loaded cryogenic pumping section. At least once per year, there is a measurement campaign with krypton (pure or mixed with tritium; see Secs.~\ref{Subsubsec:SourceGasCirculation} and~\ref{Subsubsec:Intense83mKrCalibrationSource}) for calibration and systematics studies.

\subsection{Backgrounds}
\label{Subsec:Backgrounds}

The sensitivity of the KATRIN experiment is currently limited by a higher background than anticipated, underlining the importance of understanding all background processes. During KATRIN's early operations, many different background sources and processes were investigated, understood, and mitigated. Detector background is comparatively small and can easily be measured by decoupling the detector from the rest of the system. Since it is nearly homogeneous and constant in the energy region of interest (ROI), consideration in the final analysis is not an issue; muon-induced events in the detector system are mitigated with a muon-veto system~\cite{Amsbaugh2015, PhDSchwarz2014}. The overall background in KATRIN is dominated by low-energy electrons that originate in the main spectrometer. This type of background is especially pernicious because it is not subject to energy discrimination: since signal $\upbeta$s are decelerated by the retarding potential until they are almost at rest in the central part of the main spectrometer (Sec.~\ref{Subsubsec:MACEfilter}), before being re-accelerated by the retarding potential as they travel toward the detector, low-energy secondary electrons created near the analyzing plane will arrive at the detector with close to the same energy that the surviving signal $\upbeta$s do.

External sources -- cosmic muons and gammas from environmental radioactivity -- can stimulate the emission of low-energy secondary electrons from the spectrometer walls. By design, these charged secondaries are excluded from the flux tube via electrostatic reflection (from the inner electrodes, which are held at a more negative potential than the walls) and via magnetic reflection (from the fields that contain the flux tube). Dedicated studies have confirmed that no significant spectrometer background is associated with tagged muons~\cite{Altenmueller2018} or with enhanced or suppressed external radioactivity~\cite{Altenmuller2019}.

Charged particles, traveling down the beamline, can create secondary electrons in the main spectrometer via ionization interactions with the residual gas. Tritium ions, created in the source via $\upbeta$ decay or scattering with $\upbeta$ electrons, are guided along the beamline by the same magnetic fields that guide $\upbeta$ electrons, and are not simply pumped away like neutral tritium gas. Dedicated electrodes in the beamline of the differential pumping system have been demonstrated to block and remove these ions before they can generate background in the spectrometers~\cite{PhdFriedel2020, PhDVizcaya2021}. Meanwhile, during the fourth neutrino-mass campaign it was realized that tandem spectrometer operation resulted in a background that increased with the amount of time spent at a given $qU$ setting, due to the trapping of $\upbeta$s in the potential minimum between the pre-spectrometer exit and the main-spectrometer entrance. This trap is cleared by a wire electron catcher inserted during each $qU$ setpoint change~\cite{Aker2020b}, so that a longer measurement time at a given setpoint results in a longer time between emptyings of the trap~\cite{knm2lett:2021}. To avoid the trap and remove this effectively energy-dependent background source, KATRIN now runs with the pre-spectrometer vessel grounded and its downstream inner electrode held at a moderate voltage of -100~V. 
 
The most significant background contributions are those induced by radioactivity inside the spectrometers, by two major generation mechanisms. First, we discuss background electrons due to magnetically trapped particles inside the sensitive flux-tube volume~\cite{Fraenkle2011, Mertens2013}. Depending on their initial energy and pitch angle, electrons can be trapped within the magnetic bottle of the spectrometer; since only scattering can reduce the electron energy, trapping is very efficient in the \SI{4e-11}{\milli\bar} ultra-high vacuum of the main-spectrometer vessel. Trapped electrons can scatter with residual gas and ionize atoms to generate low-energy secondary electrons, which may then propagate to the detector if they are created with low enough energy. A single initial trapped electron may produce a cluster of secondary electrons at the detector, so that the resulting background is non-Poissonian.

In KATRIN, this mechanism is fed by radon decays in the volume of the main spectrometer. There are three naturally occurring isotopes of radon, all of which emanate from the extensive non-evaporable getter material that (along with TMPs) achieves the spectrometer vacuum: $\mathrm{^{219}Rn}$, $\mathrm{^{220}Rn}$, and $\mathrm{^{222}Rn}$. With half-lives of \SI{56}{\second} and \SI{3.8}{\day}, respectively, the latter two are efficiently pumped out of the volume by TMPs before decaying. With a short half-life of about \SI{4}{\second}, $\mathrm{^{219}Rn}$ in the main-spectrometer volume will typically decay there, releasing shakeoff electrons with energies up to several \si{\kilo\electronvolt}. However, this background source can be efficiently mitigated with a baffle system in the pump ports, consisting of copper plates cooled with liquid nitrogen, on which $\mathrm{^{219}Rn}$ atoms adsorb and decay~\cite{Drexlin2017,Goerhardt2018}. The shifted-analyzing-plane configuration of the MAC-E filter (Sec.~\ref{Subsubsec:MACEfilter}) further reduces this background.

\begin{figure}[tb]
    \centering
    \includegraphics[width=.9\textwidth]{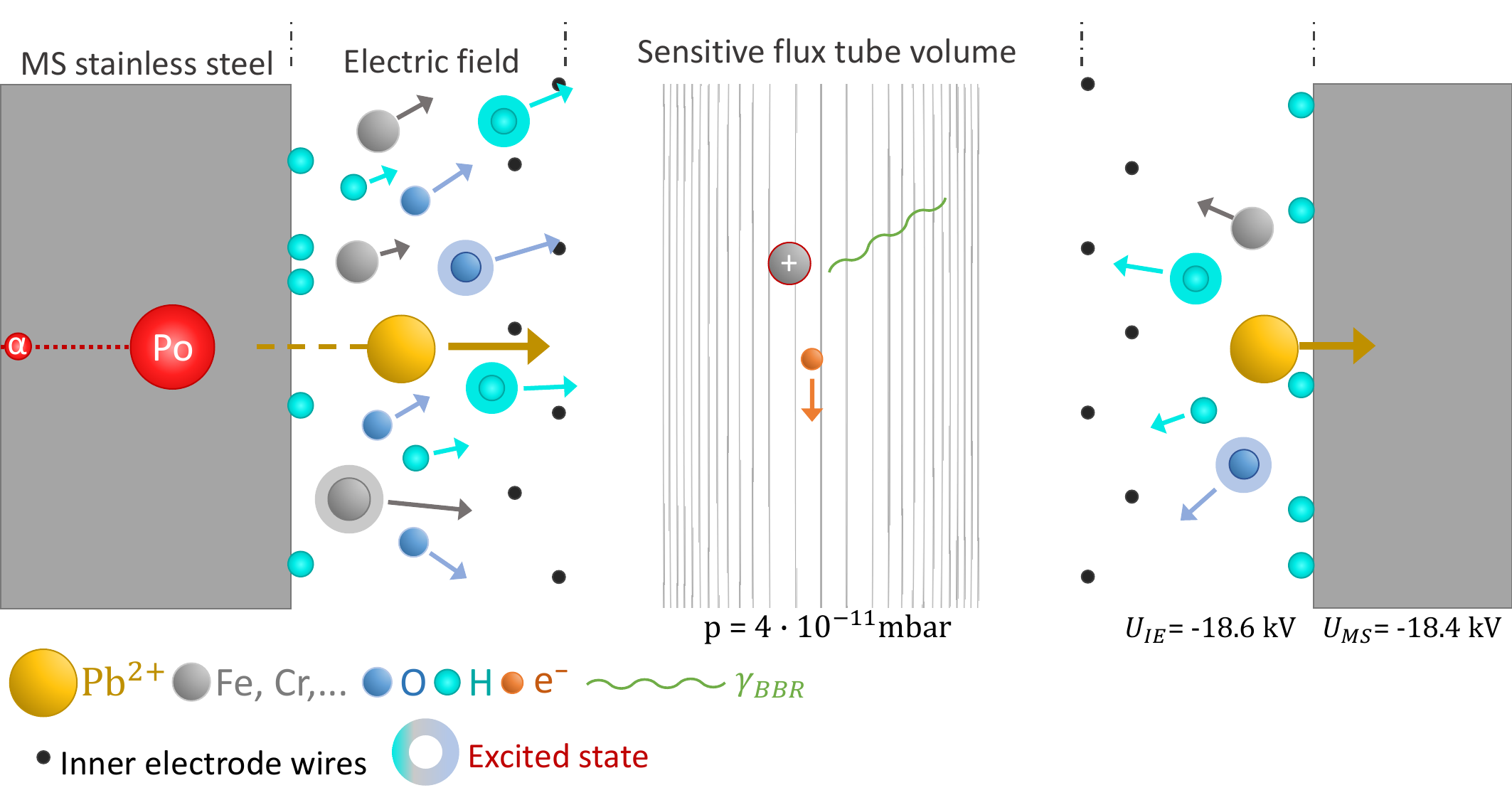}
    \caption{Illustration of Rydberg-atom generation (not to scale). Various atomic species are sputtered by the energetic recoil ion. Additional sputtering occurs if the ion leaves the vessel surface and strikes the opposite surface. Blackbody radiation (BBR) can ionize such excited atoms, releasing low-energy electrons that are either guided to the source or to the focal plane detector by the magnetic field lines, creating the sensitive flux tube volume.}
    \label{fig:RydbergProcess}
\end{figure}

The remaining background is generated by emission from the spectrometer walls of neutral particles, which can penetrate the electromagnetic shielding of the flux tube and be ionized within the main-spectrometer volume. These neutral messengers are highly excited Rydberg atoms, sputtered from the walls due to naturally implanted radioactivity within the steel; because of their high initial quantum numbers $n$, they can be ionized by blackbody radiation. 

During installation of the inner electrodes, ambient air circulated within the main spectrometer, and naturally occurring $\mathrm{^{222}Rn}$ adsorbed onto the surface, decayed and implanted radioactive impurities into the vessel walls. Eventually, the long-lived daughter $\mathrm{^{210}Pb}$ ($t_{1/2} =$~\SI{22}{\year}) accumulated, feeding a decay chain leading to the alpha emitter $\mathrm{^{210}Po}$ that decays to stable $\mathrm{^{206}Pb}$ with \SI{5.4}{\mega\electronvolt} excess energy. The recoil ion can sputter atoms from the inner vessel surface, as shown in Fig.~\ref{fig:RydbergProcess}.
These sputtered atoms could be iron, chromium, or oxygen from the steel, or hydrogen adsorbed on the surface; the total sputtering yield (released atoms per alpha decay) depends on the surface conditions. Depending on the excitation state of these Rydberg atoms, the ionization electrons will have energies in the \si{\milli\electronvolt} range. This hypothesis is supported by detailed investigations, including short-term contamination with shorter-lived decay chains from \textsuperscript{223}Ra and \textsuperscript{228}Th sources~\cite{Fraenkle2022}. Since the rate of ionization electrons detected depends on the volume imaged by the detector, this background can also be mitigated by the shifted analyzing plane (Sec.~\ref{Subsubsec:MACEfilter}).

At about \SI{150}{\milli cps}, the current background level is about 15 times larger than initially projected. An increased background level reduces the sensitivity to the spectral imprint of the neutrino mass. Further reducing the background level is of high importance. KATRIN's background-related research-and-development efforts are presented in Sec.~\ref{Subsec:BackgroundMitigation}.

% ############################################################
% KATRIN analysis tools and strategies
%# ##########################################################
\section{KATRIN analysis tools and strategies}
\label{Sec:KatrinAnalysisToolsAndStrategies}
% Responsible: Stephanie Hickford (KIT)

The squared neutrino mass, $m_{\nu}^{2}$, is obtained from a fit of the calculated analytical spectrum to the measured data. An overview of the KATRIN model is given in Sec.~\ref{Subsec:TheAnalyticModel}; the full details of the source and spectrum calculation may be found in Ref.~\cite{Kleesiek2018}. The spectrum fitting and limit setting procedures are briefly discussed in Sec.~\ref{Subsec:SpectralFittingAndLimitSetting}, and the data combination methods are described in Sec.~\ref{Subsec:DataCombination}. Section~\ref{Subsec:SourcesOfSystematicUncertainty} summarizes major sources of systematic uncertainty.

\subsection{The analytic model}
\label{Subsec:TheAnalyticModel}

The analytical spectrum is a convolution of the theoretical differential spectrum with the experimental response function. The response function describes the number of $\upbeta$ electrons that are transported through the flux tube and counted at the detector, including both the energy loss experienced as the $\upbeta$ electrons propagate through the source and the transmission properties of the MAC-E filter (Sec.~\ref{Subsubsec:MACEfilter}).

The differential decay rate of a tritium nucleus is described by Fermi's golden rule,
\begin{equation}
    \begin{split}
        \frac{d\Gamma}{dE} = \frac{G_{\text{F}}^{2}\cos^{2}\left(\theta_{\text{C}}\right)}{2\pi^{3}}\text{\textbar}M_{\mathrm{nuc}}\text{\textbar}^{2}F\left(Z,E\right)p\left(E+m_{e}\right)\\
        \cdot\sum_{f}P_{f}\epsilon_{f}\sqrt{\epsilon_{f}^{2}-m_{\nu}^{2}}\Theta\left(\epsilon_{f}-m_{\nu}\right),
\end{split}
    \label{Eq:DifferentialSpectrum}
\end{equation}
where $G_{\text{F}}$ is the Fermi constant, $\cos^{2}\left(\theta_{\mathrm{C}}\right)$ is the Cabibbo angle, $\text{\textbar}M_{\mathrm{nuc}}\text{\textbar}^{2}$ is the nuclear matrix element, and $F\left(Z,E\right)$ is the Fermi function with the atomic charge of the helium daughter nucleus $Z = 2$. $\epsilon_{f} = E_{0} - V_{f} - E$ is the neutrino energy with an energy correction from the molecular final state $f$ of T$_{2}$, where $E_{0}$ is the effective tritium endpoint and $V_{f}$ is the molecular excitation energy populated with probability $P_{f}$, as calculated in quantum chemical theory~\cite{Saenz2000, Doss2006, Aker2021Knm1}. The Heaviside function $\Theta\left(\epsilon_{f}-m_{\nu}\right)$ ensures conservation of energy. Doppler broadening and broadening arising from plasma effects within the source are emulated in the differential spectrum. The Fermi function is treated relativistically and radiative theoretical corrections on the atomic and nuclear level are included in the analytical model.

\begin{figure}[tb]
    \centering
    \includegraphics[width=.8\textwidth]{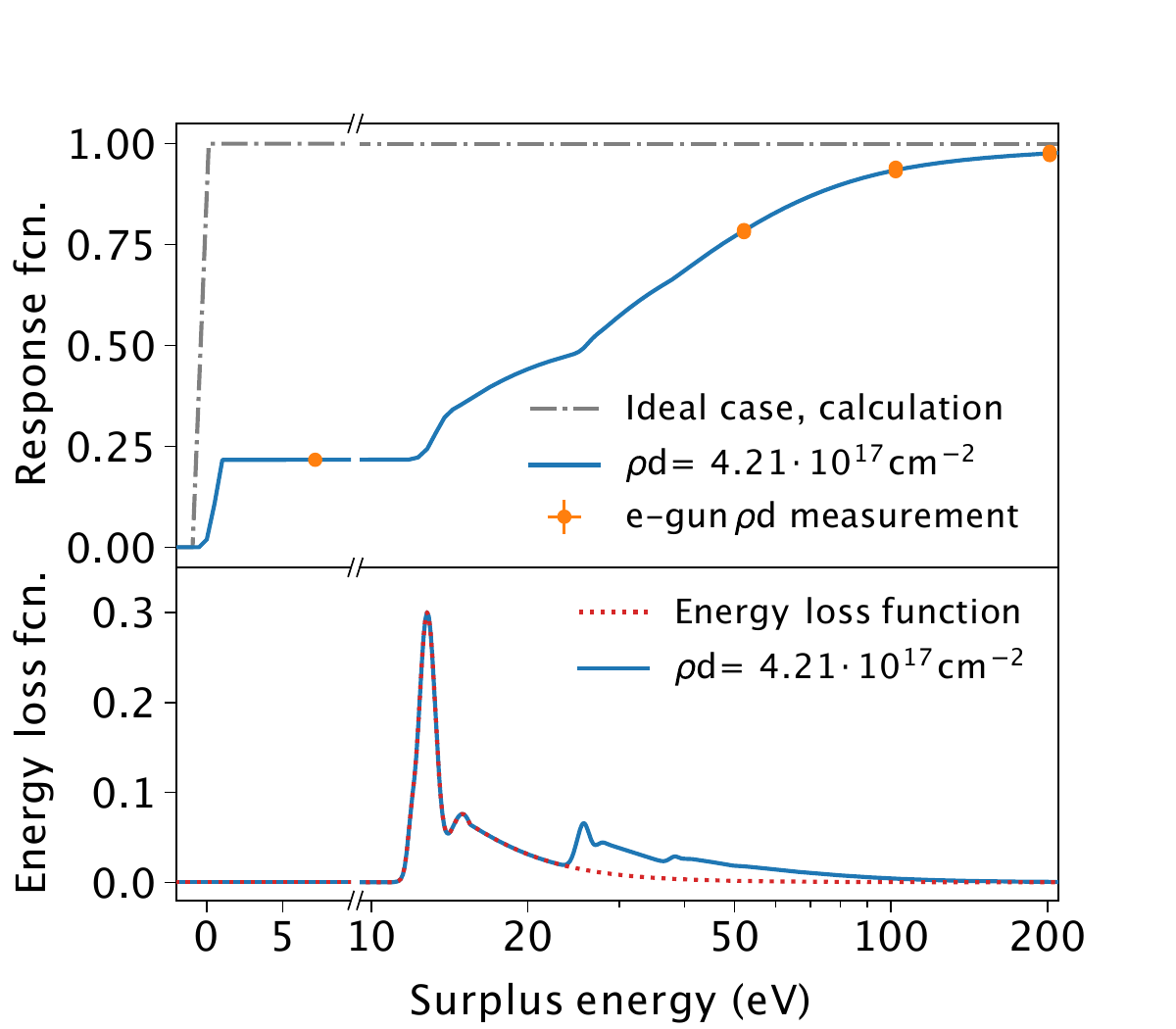}
    \caption{\textbf{Top:} Transmission function of the KATRIN spectrometer (gray dash-dotted line). When convolved with the energy-loss function for a given column density, the full response function for $\upbeta$-decay electrons is described (solid blue line). This agrees well with measurements from the electron gun with column density $\rho d = $ \SI{4.21e17}{\per\centi\metre\squared} (orange dots). \textbf{Bottom:} Energy-loss function measured in situ with the KATRIN electron gun (red dotted line). It is convolved with itself $s$ times to describe the energy loss of electrons passing through the source with $s$-fold scatterings (blue solid line). Electron-gun measurements are described in detail in Ref.~\cite{Aker2021Eloss}.}
    \label{Fig:TransmissionElossResponse}
    % Figure by Jan Behrens, analysis reviewed for KNM2 Nature paper.
\end{figure}

Electrons experience energy loss as they scatter from molecules within the gaseous tritium source. The energy-loss function, $f\left(\epsilon\right)$, describes the probability of a certain energy loss $\epsilon$ in a scattering process, illustrated by the red dotted line in the bottom panel of Fig.~\ref{Fig:TransmissionElossResponse}. The scattering probability functions, $P_{s}\left(\theta\right)$, describe the probability of an electron with pitch angle $\theta$ to scatter $s$ times before leaving the source. Electrons leaving the source without scattering experience no energy loss, $f_{0}\left(\epsilon\right) = \delta\left(\epsilon\right)$. For $s$-fold scattering the energy-loss function is convoluted $s$ times with itself leading to a total energy loss function of
\begin{equation}
    \sum_{s}P_{s}\left(\theta\right)f_{s}\left(\epsilon\right) = P_{0}\left(\theta\right)\delta\left(\epsilon\right) + P_{1}\left(\theta\right)f\left(\epsilon\right) + P_{2}\left(\theta\right)\left(f\otimes f\right)\left(\epsilon\right) + ....
    \label{Eq:EnergyLoss}
\end{equation}
illustrated by the blue solid line in the bottom panel of Fig.~\ref{Fig:TransmissionElossResponse}.

The transmission condition, $\mathcal{T}$, is either \num{1} or \num{0} (i.e. electrons are either transmitted or not through the MAC-E filter) depending on the starting energy $E$ and the starting pitch angle $\theta$ of the electron, and on the retarding energy $qU$. For an isotropic electron source the angular distribution of electrons is $\omega\left(\theta\right) d\theta = \sin\left(\theta\right)d\theta$ and $\mathcal{T}$ is integrated over the electron pitch angle $\theta$,
\begin{equation}
    \begin{split}
        T\left(E, qU\right) & = \int_{0}^{\theta_{\text{max}}}\mathcal{T}\left(E, \theta, qU\right)\cdot\sin\left(\theta\right)d\theta \\
        & =
        \begin{cases}
            0 & \quad \left(E - qU\right) < 0 \\
            1 - \frac{\sqrt{1 - \frac{\left(E - qU\right)}{E} \left(\frac{B_{\text{src}}}{B_{\text{ana}}}\right)}}{1 - \sqrt{1 - \left(\frac{B_{\text{src}}}{B_{\text{max}}}\right)}} & \quad 0 \le \left(E - qU\right) \le \Delta E, \\
            1 & \quad \left(E - qU\right) > \Delta E
        \end{cases}
    \end{split}
    \label{Eq:TransmissionFunction}
\end{equation}
where $\left(E-qU\right)$ is the surplus energy. The transmission function is therefore mainly determined by the magnetic fields of the source, $B_{\text{src}}$, the analyzing plane, $B_{\text{ana}}$, and the beamline maximum, $B_{\text{max}}$. The magnetic reflection imposes an upper limit on the pitch angle given by $\theta_{\text{max}} = \arcsin\sqrt{\left(B_{\text{src}}/B_{\text{max}}\right)}$ which reduces the effective width, or energy resolution, of the transmission function. These MAC-E-filter properties are described in Sec.~\ref{Subsubsec:MACEfilter} and illustrated by the grey dash-dotted line in the top panel of Fig.~\ref{Fig:TransmissionElossResponse}. Synchrotron energy losses due to the electron motion through magnetic fields is also included in the transmission-function calculation.

The final experimental response is the convolution of the energy-loss function with the transmission function, given by
\begin{equation}
    R\left(E, qU\right) = \int_{0}^{E}\int_{0}^{\theta_{\text{max}}} \mathcal{T}\left(E, \theta, qU\right) \cdot \sin\left(\theta\right) \cdot \sum_{s} P_{s}\left(\theta\right)f_{s}\left(\epsilon\right) d\theta d\epsilon
    \label{Eq:ResponseFunction}
\end{equation}
The response function is illustrated by the blue solid line in the top panel of Fig.~\ref{Fig:TransmissionElossResponse}. This experimental response function, together with the theoretical differential spectrum, constitutes the analytical model which is fit to the measured data. It can be computed for any region of the detector under analysis.

\subsection{Spectral fitting and limit setting}
\label{Subsec:SpectralFittingAndLimitSetting}

The number of observed counts measured at the detector is compared to the expected number of counts predicted from the analytical model (Sec.~\ref{Subsec:TheAnalyticModel}). The probability to have an observed outcome given the predicted number of counts, that is, the likelihood function, can be maximized to obtain the parameters of interest. In a ``statistics-only" fit there are four fit parameters:
\begin{itemize}
    \item Squared neutrino mass, $m_{\nu}^{2}$
    \item Effective tritium endpoint, $E_{0}$
    \item Signal amplitude, Sig
    \item Background rate, Bg
\end{itemize}
In practice the $\chi^{2}$, or negative log-likelihood, 
\begin{equation}
    \chi^{2} = -2\log\mathcal{L}\left(m_{\nu}^{2}, E_{0}, \text{Sig}, \text{Bg}\right)
    \label{Eq:Likelihood}
\end{equation}
is minimized to obtain the best-fit spectral parameters. If systematic uncertainties are treated via the ``pull term" method (Sec.~\ref{Subsec:FittingStrategies}) the number of fit parameters can be higher. 

If the sensitivity of a spectral measurement is not good enough to resolve the neutrino mass, then statistically we expect that half the time the measurement will show too many counts in the endpoint region, and half the time it will show too few. However, Eq.~\ref{Eq:DifferentialSpectrum} is valid only for $m_\nu^2 \geq 0$. In order to perform a well-behaved fit at this sensitivity boundary, the spectral function must be extended to negative $m_\nu^2$. The KATRIN analysis performs this extension by allowing $m_\nu^2$ to be negative in Eq.~\ref{Eq:DifferentialSpectrum}; the resulting $\chi^2$ function is asymmetric, but the analysis of early data sets is robust against a different choice of spectral extension~\cite{Formaggio:NuMassRev2021}.

When a fit of the analytical model to the data has been completed, an upper limit on the neutrino mass is set from the Neyman construction~\cite{Neyman} of the confidence belt. Statistically unfluctuated, ``Asimov'' data sets are generated for a set of true squared neutrino mass values, $m_{\nu,\text{true}}^{2}$, using the fit results as input parameters. Each Asimov data set is fit over a range of hypothetical measured squared neutrino masses, $m_{\nu,\text{meas}}^{2}$, and the confidence belt is then constructed by integrating under the resultant likelihoods to the desired confidence level. The ordering principle can be taken from the Feldman-Cousins method~\cite{Feldman} or the Lokhov-Tkachov method~\cite{Lokhov:2014zna}. These methods yield consistent results for positive best-fit neutrino mass values; when the best-fit neutrino-mass value is negative, the Lokhov-Tkachov method gives the sensitivity as the upper limit.

\subsection{Data combination}
\label{Subsec:DataCombination}

Neutrino-mass measurements are performed in ``scans" that step through the retarding-potential set points over a few hours under stable source conditions. The tritium $\upbeta$-decay electrons are counted at the focal-plane detector on \num{148} pixels (see Sec.~\ref{Sec:TheKatrinApparatus}) which provide spatial resolution over the cross-sectional area of the flux tube. In order to perform the neutrino mass analysis, the spectra from scans and pixels are combined.
\begin{itemize}
    \item \textbf{Scan combination:} Scans performed under the same operating conditions (during one measurement phase for example) are combined via ``stacking". This entails summing the counts from each detector pixel, and averaging parameters such as the retarding potential, tritium purity, and column density.
    \item \textbf{Pixel combination:} Pixels that have similar transmission properties are combined into ``patches". The simplest combination is all pixels to a uniform detector, but pixels can alternatively be combined into subsets taking into account the field configuration and the alignment of the beamline. Pixel combination entails summing the counts from each detector pixel within the defined patch, and averaging parameters such as the analyzing-plane magnetic field and maximum magnetic field.
    \item \textbf{Measurement-period combination:} Neutrino-mass scanning periods with the same measurement conditions are combined in the same ways as the individual scans. However, if the parameters of the experimental setup cannot be averaged, the periods can be treated via a simultaneous fit of several periods using the squared neutrino mass as a common parameter, while other fit parameters and systematic contributions can be defined independently for different periods. Section~\ref{sec:combinecampaigns} describes alternative schemes for combining data from different measurement periods.
\end{itemize}
When the data combination has been defined, the neutrino-mass analysis is performed with a fit of the analytic model to the measured data. There are several models with differing input parameters, corresponding to each of the spectral combinations. The minimization of the parameters of interest is therefore over several spectra. The common parameter for all spectra is the squared neutrino mass, and the log-likelihood profile is the summation of the log-likelihoods from each spectrum $i$. For the ``statistics-only" case, Eq.~\ref{Eq:Likelihood} is extended to
\begin{equation}
    -\log\mathcal{L} = \sum_{i} -\log\mathcal{L}_{i}\left(m_{\nu}^{2}, E_{0i}, \text{Sig}_{i}, \text{Bg}_{i}\right).
    \label{Eq:MultiLikelihood}
\end{equation}
The endpoint, $E_{0}$, is not a common parameter, allowing it to absorb systematic shifts arising from, e.g., plasma effects. In practice, additional parameters are included when systematic uncertainties are considered. These parameters can be either common or independent for each scan and pixel combination.

\subsection{Sources of systematic uncertainty}
\label{Subsec:SourcesOfSystematicUncertainty}

Inference of the squared neutrino mass, $m_\nu^2$, from the measured $\upbeta$-decay spectrum requires very accurate knowledge of the theoretical as well as experimental ingredients entering the analytic expression of the spectral shape. This analytic model, which incorporates both the theoretical $\upbeta$ spectrum and the experimental response, forms the fit function, as detailed in Eqs.~\ref{Eq:DifferentialSpectrum} and \ref{Eq:ResponseFunction} in Sec.~\ref{Subsec:TheAnalyticModel}. Uncertainties of the parameters used to calculate the spectrum model translate into uncertainties of the parameters obtained from the fit. For the two neutrino-mass results released by KATRIN thus far (Sec.~\ref{subsec:RecentResults}), the budget of systematic uncertainties is presented in Table~\ref{tab:systematics_breakdown-knm2}.

It is instructive to group the sources of systematic uncertainties into signal-related and background-related effects. In the following we will inspect these two classes separately.

\paragraph{Signal-related systematic uncertainties}
The environment of the magnetized tritium plasma inside the windowless, gaseous tritium source determines the electric and magnetic fields in which the $\upbeta$ electrons are produced. These boundary conditions impact the $\upbeta$-decay spectrum directly or indirectly -- for instance through the acceptance angle of signal electrons, which is determined by the ratio of source magnetic field to maximum magnetic field strength encountered along the trajectory, or through the energy scale determined by the difference in electric potential between the point of origin in the source and the analyzing point in the spectrometer. 

The strong magnetic fields of the source ($B_\mathrm{src} =$ \SI{2.52}{\tesla}) and of the pinch magnet ($B_\mathrm{max} =$ \SI{4.74}{\tesla}) were measured with precision probes before final system integration, when the sensitive field region was still accessible. Their stability is monitored during operations. We estimate the field uncertainties at $\sigma(B_\mathrm{src}) =$ \SI{1.7}{\percent} and $\sigma(B_\mathrm{max}) =$ \SI{0.1}{\percent}, respectively. The much weaker magnetic field inside the main spectrometer of $\mathcal{O}(\SI{e-4}{\tesla})$ is inaccessible to direct measurement during operations, but can instead be inferred from the data delivered by an extensive network of precision magnetic sensors outside of the vacuum vessel, combined with simulations using the Kassiopeia field-modeling and particle-tracking software~\cite{Furse2017}. The resulting uncertainty is estimated at the \SI{1}{\percent} level. 

The molecular tritium gas forms a self-ionizing plasma which is bounded on the sides by the stainless steel walls of the beam tube and at the upstream end by a gold-plated rear wall which can be supplied with a variable bias voltage. While a stable, non-zero absolute source potential is no concern for the neutrino-mass analysis since it would be absorbed in the effective, fitted endpoint of the spectrum, temporal drifts or spatial nonuniformities can cause a broadening of the response function. To assess such possible spatial variations, KATRIN makes use of monoenergetic conversion electrons from \textsuperscript{83m}Kr co-circulated with the tritium gas (Sec.~\ref{Subsubsec:Intense83mKrCalibrationSource}). Plasma parameters can be inferred from the narrow lines in the spectrum of conversion electrons from this nuclear standard. In order to ensure that calibration conditions are directly applicable to neutrino-mass measurements, KATRIN has run both at an elevated source temperature of \SI{80}{\kelvin} since~2019 (Sec.~\ref{Subsubsec:SourceGasCirculation}).

A key component of the model applied to fit the data of the integrated $\upbeta$-decay spectrum is the instrumental response function. The response is composed of two main ingredients (Sec.~\ref{Subsec:TheAnalyticModel}): the transmission function of the main spectrometer as a high-pass energy filter, and the energy-loss description of signal electrons traversing the gaseous tritium column in the source. The working point of the tritium source (temperature, gas density or flow rate) is optimized to yield a high luminosity of $\upbeta$ electrons at an acceptable opacity from scattering. Depending on their point of origin inside the extended source beam tube, $\upbeta$ electrons have a sizeable probability to undergo scattering and thus lose a certain amount of energy in inelastic encounters. This energy loss has to be taken into account with high accuracy in the description of the response. The two relevant factors are the precise value of the column density ($\rho d$) and the inelastic scattering cross section ($\sigma_\mathrm{inel}$). KATRIN employs a custom-made precision photoelectron source for monoenergetic electrons emitted at selected pitch angle in order to directly measure the product $\rho d \cdot \sigma_\mathrm{inel}$ as well as the full energy-loss function~\cite{Aker2021Eloss}. The monitoring of $\rho d \cdot \sigma_\mathrm{inel}$ is done periodically during a neutrino-mass measurement campaign and is accompanied by a combined online monitoring of the source activity (through beta-induced X-ray spectroscopy at the upstream end of the beamline and through a forward beam monitor at the exit of the source and transport section) and of the tritium purity (through Raman spectroscopy), whereas the energy-dependent loss function was determined in dedicated measurements (Fig.~\ref{Fig:TransmissionElossResponse}). As a result of these efforts, the energy-loss function is now well characterized and understood based on a physics model, and the uncertainty on $m_\nu^2$ due to $\rho d \cdot \sigma_\mathrm{inel}$ is now at the level of \SI{0.01}{\electronvolt\squared}. 

In addition to the systematics related to operating conditions or experimental effects, another uncertainty is introduced at the level of the differential $\upbeta$-decay spectrum (analytic model, Sec.~\ref{Subsec:TheAnalyticModel}) by the energy distribution allocated to electronic and rotational-vibrational excited states of the (\textsuperscript{3}HeT)\textsuperscript{+} molecule in the final state of the decay.
This probability distribution over the various molecular states of the daughter $^3$HeT$^+$ ion has been computed in a series of very precise calculations adopting sophisticated molecular-structure theory~\cite{Saenz2000,Doss2006,Aker2021Knm1}. 
The resulting uncertainty on the neutrino mass has recently been re-assessed and found to be sufficiently small that it does not limit the neutrino-mass determination by the KATRIN experiment.

\paragraph{Background-related systematic uncertainties}
Understanding the background characteristics, such as a possible spatial dependence and variation with time or energy (Sec.~\ref{Subsec:Backgrounds}), is key to a thorough assessment of background-related uncertainties. These effects can best be mitigated by further reducing the overall background level; efforts in this direction are described in Sec.~\ref{Subsec:BackgroundMitigation}. 

Apart from the increase of the statistical uncertainty due to a high background level,  several effects related to the background could influence the shape and uncertainties of the measured spectrum. The first effect is related to the decay of radon in the volume of the main spectrometer; as described in Sec.~\ref{Subsec:Backgrounds}, each radon decay can result in several secondary electrons that can eventually reach the detector. Since they originate from the same primary electron, these counts are no longer independent and do not follow the Poisson distribution, but rather a wider distribution, approximated by a Gaussian. This over-dispersion effectively increases the spread of the measured count rates and therefore the statistical uncertainty of the spectrum. This was a dominant contribution to the total neutrino-mass uncertainty in the first two measurement campaigns (Table~\ref{tab:systematics_breakdown-knm2}).

The next effect is related to the possible dependence of the background rate on the absolute retarding potential applied to the spectrometer. The dependence is described by a linear function (slope) in the range of retarding energies near the tritium endpoint. An estimate of the slope is obtained from dedicated measurements of the background rate over a wider range of retarding potentials, with the valve to the tritium source closed. The typical size of the effect is $\mathcal{O}(\SI{1}{\milli cps/\kilo\electronvolt})$ and the contribution to the neutrino-mass uncertainty is shown in Table~\ref{tab:systematics_breakdown-knm2}.

Finally, the Penning trap in the region between the pre- and main spectrometer causes an increase of the background rate during each scan step (Sec.~\ref{Subsec:Backgrounds}), so that the measured rate depends on the measurement time spent at each setting. The magnitude of the effect is estimated by fitting a linear increase of the rate within each set point. The rate increase is of the order of $\mathcal{O}(\SI{1}{\micro cps/\second})$, which leads to an $\mathcal{O}(\SI{0.1}{\milli cps})$ rate increase  for a \SI{100}{\second} measurement step. The time spent at each scan step varies from \SIrange{20}{700}{\second} and therefore the additional rate is different for individual retarding-potential set points, leading to a shape distortion of the measured spectrum. To mitigate this significant source of  systematic uncertainty, the pre-spectrometer vessel was grounded beginning in 2021, thereby effectively lifting the trapping conditions (Sec.~\ref{Subsec:Backgrounds}).

Here and in the results reported in Sec.~\ref{subsec:RecentResults}, we take into account the systematics as assessed for KATRIN's initial two neutrino-mass phases. Significant progress has since been made in controlling both the signal- and background-related uncertainties, and will be reflected in future neutrino-mass results. 

% ############################################################
% Measuring the neutrino-mass scale with KATRIN
%# ##########################################################
\section{Measuring the neutrino-mass scale with KATRIN}
\label{Sec:MeasuringTheNeutrinoMassScaleWithKatrin}

As described in Sec.~\ref{Subsec:SpectralFittingAndLimitSetting}, the neutrino-mass analysis requires fitting model spectra -- incorporating the square of the neutrino mass among other fit parameters -- to the measured, integral spectrum. The KATRIN analysis strategy includes several complementary methods for handling systematic uncertainties, as detailed below. The fitting frameworks, developed by independent teams, rely on a shared implementation of the underlying spectral model and experimental response function. Additional physics searches, as described in Secs.~\ref{Sec:KatrinTestsOfNewPhysics} and~\ref{Sec:AdditionalPhysicsObservables}, can typically be performed by incorporating new terms into the spectral model. 

The KATRIN analysis is model-blinded: a designated collaborator generates an arbitrary, secret Gaussian broadening of the electronic ground state of the daughter molecule in T$_2$ $\upbeta$ decay, which alters the spectral model in a manner correlated much more strongly to the squared neutrino mass than to the other fit parameters. This broadening is implemented independently for each measurement campaign, in a software module that is automatically synchronized with the analysis packages. The analysis procedures are first developed and tested on Monte-Carlo data generated using slow-control data recorded during the measurement. Then, the model-blinded fits are performed on the actual data. The broadening is revealed, and the analysis performed with the correct model, only after all input parameters and systematic uncertainties have been finalized.

\subsection{Parameter inference}
\label{Subsec:FittingStrategies}

KATRIN's free parameters $\vec{\Theta}$ are inferred by minimizing the $\chi^2$ function, 
\begin{equation}
    \chi^2 = \left( \vec{R}_{\mathrm{data}}(q\vec{U}, \vec{r}) - \vec{R}(q\vec{U}, \vec{r}\, \text{\textbar} \vec{\Theta},\vec{\eta}) \right)^{T} \cdot C^{-1} \left( \vec{R}_\mathrm{data}(q\vec{U}, \vec{r})-\vec{R}(q\vec{U}, \vec{r}\, \text{\textbar} \vec{\Theta},\vec{\eta}) \right), \label{eq:cov}
\end{equation}
where $\vec{R}_\mathrm{data}(q\vec{U}, \vec{r})$ gives the measured count rates at a retarding energy $qU_i$ for the detector region $r_j$, $\vec{R}(q\vec{U}, \vec{r})$ gives the predictions of these rates, and $C$ is the covariance matrix that includes the statistical uncertainties and, in some cases, systematic uncertainties; systematic parameters are collected in $\vec{\eta}$. As described in Sec.~\ref{Subsec:DataCombination}, by combining data from multiple scans and multiple pixels, we ensure that the counts can be described by a Gaussian distribution, so that this approach is valid.

To propagate uncertainties on the systematic parameters $\vec{\eta}$, KATRIN pursues four parallel strategies as summarized below. Additional details can be found in Refs.~\cite{Aker2021Knm1, knm2lett:2021}. This analysis is very computationally intensive, but the recent development of a neural-network method for computing the KATRIN physics model promises to reduce the computational load of KATRIN analysis by up to three orders of magnitude~\cite{Netrium:2022}.

\paragraph{Covariance matrix}
This approach explicitly incorporates the systematic uncertainties in the covariance matrix $C$ that appears in Eq.~\ref{eq:cov}. The diagonal entries describe uncorrelated uncertainties for each $R(qU_i, r_j)$, while the off-diagonal terms describe the correlated uncertainties between different data points $R(qU_i, r_j)$. The covariance matrix is pre-computed from $\mathcal{O}(\num{e4})$ decay spectra that are simulated while varying the systematic parameters $\eta_i$ within their probability density functions.

\paragraph{Pull terms}
In this method, a systematic parameter $\eta_i$ with an estimated value $\hat{\eta_i}$ and an uncertainty $\sigma_{\eta_i}$, as determined from separate measurements, is treated as a free parameter in the fit. A corresponding pull term $\eta_i$ is added to the 
$\chi^2$ function from Eq.~\ref{eq:cov},
\begin{equation}
    \chi_{\mathrm{tot}}^2(\vec{\Theta},\vec{\eta}) = \chi^2(\vec{\Theta},\vec{\eta}) + \sum_i\left(\frac{\hat{\eta_i}-\eta_i}{\sigma_{\eta_i}}\right)^2 ~. \label{eq:pull}
\end{equation}
Each $\eta_i$ is therefore free to vary, but an increasing $\chi^2$ penalty is imposed the further it moves from its best estimation, depending on our uncertainty on this parameter.

\paragraph{Monte Carlo propagation}

This technique involves propagation of uncertainties via $\mathcal{O}$(\num{e5}) fits to simulated spectra. To assess the statistical uncertainty, we generate \num{e5} statistically randomized Monte Carlo spectra, which are fit with a constant model. To assess the systematic uncertainties, a statistically unfluctuated spectrum is simulated and fit \num{e5} times with the systematic parameters $\vec{\eta}$ varied each time according to their probability density functions. To obtain the total uncertainty, these steps are combined, producing a histogram of best-fit parameters. Each entry in this histogram is weighted by the likelihood of its corresponding fit, and the best-fit value and uncertainty are taken from the mode and width of the resulting distribution.

\paragraph{Bayesian}
In contrast to the frequentist methods described above, in Bayesian inference the posterior probabilities of $\vec{\Theta}$ are computed from a prior probability and the data, according to Bayes' theorem. The KATRIN analysis uses a flat, positive prior on $m_\nu^2$, restricting the posterior distribution to physically allowed values. In principle, the systematic effects $\vec{\eta}$ are included as free parameters in the Bayesian analysis, with each $\eta_i$ constrained by prior knowledge. Due to computational constraints, this is presently done only for the uncertainty on a $qU$-dependent background, with all other effects implemented either with a covariance matrix or with model variation in the inputs of a large number of Markov Chain Monte Carlo simulations.

\subsection{Combining data from different measurement campaigns}
\label{sec:combinecampaigns}

KATRIN measurement campaigns are separated by planned breaks during which maintenance activities are conducted -- such as pumping away the T$_2$ gas, cleaning the rear wall, regenerating the Ar frost in the cryogenic pumping system, baking out the main spectrometer, or replacing the wafer of the focal-plane detector. Often, commissioning or calibration measurements are conducted under conditions that differ significantly from neutrino-mass running. Different measurement campaigns may see different background levels, different energy scales (and hence effective endpoints), and different column densities (and hence energy-loss profiles), among other significant differences; see Table~\ref{tab:knm1-2-comparison}, further below, for a comparison of the first two neutrino-mass measurement campaigns. In general, it is not possible to describe and fit multiple campaigns with a single model spectrum.

As summarized in Ref.~\cite{knm2lett:2021}, the KATRIN collaboration has explored several approaches for combining multiple data sets, building on the analysis approaches described above in Sec.~\ref{Subsec:FittingStrategies}. For example, when the data sets are statistically dominated, a straightforward Frequentist approach is to combine the results of multiple measurement campaigns by summing their respective $\chi^2$ curves. Alternatively, one may perform a joint Frequentist fit of the two data sets, in which $m_\nu^2$ is a shared fit parameter for both campaigns, whereas the parameters $A_\mathrm{s}$, $R_\mathrm{bg}$, and $E_0$ are allowed to differ between the two campaigns (Sec.~\ref{Subsec:DataCombination}). Each systematic uncertainty may be treated jointly between the data sets, or separately for each data set, as appropriate. In a Bayesian framework, the posterior distribution from analyzing one data set can simply be used as prior information for the analysis of the next, neglecting correlations between the two spectra. As additional data sets are added and systematic uncertainties become more important, these first approaches will be further refined.

\subsection{Recent results}
\label{subsec:RecentResults}

The KATRIN experiment has been operated with tritium since 2018~\cite{Aker2020a}, and has been performing neutrino-mass measurement campaigns from 2019 on. The multi-month campaigns usually take place twice a year and are designated KNM, for ``KATRIN Neutrino Mass campaign'', followed by a sequential number. In spring 2022 KATRIN conducted its KNM7 campaign. At present, results from the KNM1 and KNM2 campaigns (both from 2019) have been published. We review these results here.
Table \ref{tab:knm1-2-comparison} summarizes the operational parameters and performance of these campaigns.

\begin{table}[tb]
    \centering
    \caption{Key operational parameters for the first (KNM1~\cite{Aker2019}) and second (KNM2~\cite{knm2lett:2021}) KATRIN neutrino-mass campaigns. The total $\upbeta$-electron count is given for the analysis interval extending from  $E_0-$\SI{40}{\electronvolt} to $E_0$. The $\upbeta$-electron-to-background ratio is given by the ratio of this number to the extrapolated background counts in the same energy range. Table reproduced from Ref.~\cite{knm2lett:2021}.}
    \begin{tabular}{llll}
        \hline
        & KNM1 & KNM2 \\
        \hline
        Number of scans                        & \num{274}               & \num{361} \\
        Total scan time                        & \SI{521.7}{\hour}       & \SI{743.7}{\hour} \\ 
        Background rate                        & \SI{290}{mcps}          & \SI{220}{mcps} \\
        T$_{2}$ column density                 & \SI{1.11e17}{\per\centi\metre\squared} &  \SI{4.23e17}{\per\centi\metre\squared} \\
        Source activity                        & \SI{2.5e10}{\becquerel} & \SI{9.5e10}{\becquerel} \\
        Total number of $\upbeta$-electrons    & \num{1.48e6}            & \num{3.68e6} \\
        $\upbeta$-electron-to-background ratio & \num{3.7}               & \num{9.9} \\
        \hline
    \end{tabular}
    \label{tab:knm1-2-comparison}
\end{table}

\paragraph{KNM1 - Spring 2019}

In the first neutrino-mass campaign, the KATRIN source was operated at a reduced column density of $\SI{1.11e17}{molecules/cm\squared}$, roughly one fifth of the design value. The lower tritium throughput of about \SI{4.9}{\gram\per\day} was necessary because materials previously unexposed to tritium have a high rate of impurity formation from radiochemical reactions~\cite{Sturm:2021}. These impurities can then accumulate and freeze in the tritium-injection capillary, gradually reducing the throughput. The reduced column-density setting was chosen to stabilize the tritium column density.  With this source setting, the KNM1 measurement campaign had a total duration of four weeks, yielding \num{272} spectral scans that passed data-quality selection. The 148 detector pixels were also subjected to quality cuts, which led to an exclusion of \num{31} pixels mainly in the outer detector rings. The spectra recorded from these selected pixels were combined into a single spectrum (Sec.~\ref{Subsec:DataCombination}) on which the final analysis was performed.  

Three different methods, two Frequentist and one Bayesian, were applied for the parameter inference (see Sec.~\ref{Subsec:FittingStrategies}), as described in detail in Ref.~\cite{Aker2021Knm1}. The choice of experimental input parameters for the KATRIN model directly affects the neutrino-mass observable, so a model-blind analysis was performed (see beginning of Sec.~\ref{Sec:MeasuringTheNeutrinoMassScaleWithKatrin}).  The best fit from this analysis was $m_\nu^2 = -1.0^{+0.9}_{-1.1}$\si{\electronvolt\squared}~\cite{Aker2019}. This result is consistent with a vanishing neutrino mass. The Lokhov-Tkachov prescription~\cite{Lokhov:2014zna} was applied to derive the resulting limit of $m_\nu <$ \SI{1.1}{\electronvolt} (\SI{90}{\percent} C.L.). Table~\ref{tab:systematics_breakdown-knm2} gives the uncertainty budget. Compared to the statistical uncertainty (\SI{0.97}{\electronvolt\squared}), the total systematic uncertainty (\SI{0.32}{\electronvolt\squared}) was of secondary importance. The largest systematic effect was attributed to the non-Poissonian nature of the radon-induced background (Sec.~\ref{Subsec:Backgrounds}).
 
The best-fit value for the effective endpoint, $E_0 =$ \SI{18573.7\pm0.1}{\electronvolt}, allows comparison of the KATRIN result to external measurements of the $^3$He--T atomic-mass difference~\cite{Myers2015}. Combined with our knowledge of the KATRIN energy scale, these values are in agreement~\cite{Aker2021Knm1}.

\paragraph{KNM2 - Fall 2019}
The issue of radiochemical impurity generation was resolved for the second campaign starting in September of the same year. After extensive tritium exposure in the loop system during and following KNM1, impurity formation ceased due to depletion of available surface reactants. For example, carbon present on the stainless-steel wall surface had been mostly converted to tritiated hydrocarbons, which were then removed by the PdAg-membrane filter in the tritium loop~\cite{Sturm:2021}. This allowed an increase of source activity by almost a factor of four, to \SI{9.5e10}{\becquerel}. Furthermore, KNM2 saw a \SI{30}{\percent} increase in total scan time recorded, in comparison to KNM1. All of these factors led to significantly higher statistics. At the same time, the background conditions were improved substantially: Since the copper baffles of the main spectrometer were regenerated during a spectrometer bakeout prior to the KNM2 campaign, the radon-induced non-Poissonian background contribution was reduced significantly, as reflected by the three-times-smaller corresponding systematic uncertainty (Table~\ref{tab:systematics_breakdown-knm2}). The overall background rate was approximately \SI{25}{\percent} lower in KNM2 compared to KNM1. Scan-to-scan fluctuations were also reduced, due to enhanced high-voltage reproducibility. 

\begin{figure}[tb]
    \centering
    \includegraphics[width=0.7\textwidth]{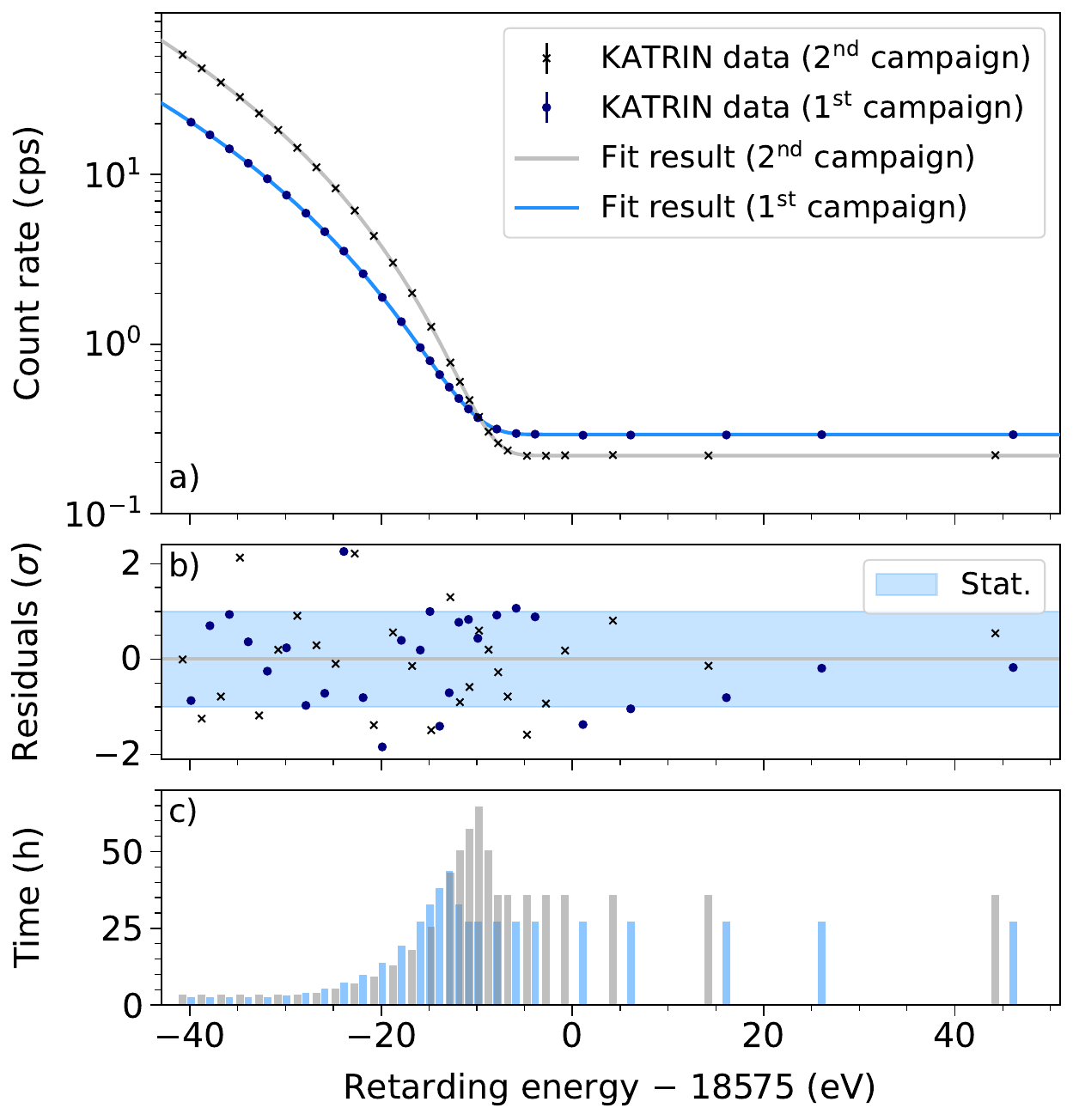}
    \caption{Measured tritium $\upbeta$ spectra and fit model for KATRIN's KNM1 (blue; \cite{Aker2019}) and KNM2 (gray; \cite{knm2lett:2021}) campaigns. (a) Simultaneous, uniform fit to both data sets. (b) Residuals for the uniform fit. (c) Distribution of measurement time near the endpoint, covering the signal and background regions. The share of measurement time allocated to individual scan steps is different for the two campaigns because it is optimized based on signal-to-background ratio, which was larger for the KNM2 data. Adapted from Ref.~\cite{knm2lett:2021}.}
    \label{fig:spectra_knm1_2}
\end{figure}

The analysis again uses 117 detector pixels after quality selection. By virtue of the improved statistics for KNM2, these pixels can now be grouped into \num{12} concentric rings or into four annuli, apart from being combined into a uniform detector. An analysis according to rings or annuli can probe for radial patterns in the resulting fit parameters, notably the effective endpoint $E_0$, which might be a sign of plasma effects impacting the energy scale at different regions of the beam cross section. Eventually, no such radial dependence was found at a significant level in the analysis of KNM2 data; Fig.~\ref{fig:spectra_knm1_2} therefore displays the result of a uniform fit to the data. A detailed account of the different analysis methods applied to the data is presented in Ref.~\cite{knm2lett:2021}. 

The best-fit values of squared neutrino mass and effective endpoint are $m_\nu^2 = 0.26^{+0.34}_{-0.34}\,\si{\electronvolt\squared}$ and $E_0 =$ \SI{18573.69\pm0.03}{\electronvolt}, respectively. Again, the results obtained with the different analysis methods are in agreement. Table~\ref{tab:systematics_breakdown-knm2} contains the breakdown of systematic uncertainties, highlighting numerous improvements made through dedicated calibration measurements, more precise determination of various model inputs, and background reduction leading up to KNM2. The total uncertainty on the squared neutrino mass is cut by a factor of three compared to KNM1. 

From the best-fit value of $m_\nu^2$, which is compatible with zero within uncertainties, we deduce an upper limit of $m_\nu <$ \SI{0.9}{\electronvolt} (\SI{90}{\percent} C.L.). Given that the central value is slightly positive, both the Feldman-Cousins and Lokhov-Tkachov methods result in the same numerical value of the Frequentist limit. The Bayesian approach yields an upper bound of $m_\nu <$ \SI{0.85}{\electronvolt} (\SI{90}{\percent} credible interval).

\begin{table}[tb]
    \centering
    \caption{Breakdown of the most significant uncertainties for the first (KNM1~\cite{Aker2021Knm1}) and second (KNM2~\cite{knm2lett:2021}) KATRIN neutrino-mass campaigns. Comprehensive tabulations are found in the respective publications. The statistical uncertainty, even though much smaller in KNM2 compared to KNM1, was the dominant uncertainty on the neutrino-mass square derived in both campaigns. Several systematics, such as the ones related to the background characteristics, are reduced substantially from KNM3 onwards.}
    \label{tab:systematics_breakdown-knm2}
    \begin{tabular}{lcc}
        \hline
        Effect ($1\sigma$ uncertainty on $m_\nu^2$)  & KNM1 (\si{\electronvolt\squared}) & KNM2 (\si{\electronvolt\squared}) \\
        Statistical	                                 & \num{0.97} & \num{0.29} \\
        \hline
        {Non-Poissonian background}                  & \num{0.30} & \num{0.11}  \\
        \hline
		Source-potential variations (plasma effects) & Neglected  & \num{0.09} \\
		Scan-step-duration-dependent {background}    & Neglected  & \num{0.07} \\
		$qU$-dependent {background}                  & \num{0.07} & \num{0.06} \\
        Magnetic fields                              & \num{0.05} & \num{0.04} \\
		Molecular final-state distribution           & \num{0.02} & \num{0.02} \\ 
		Column density $\times$ {inelastic scat.} cross-section $(\rho d \sigma)$ & \num{0.05} & \num{0.01}  \\
		\hline
		\textbf{Total uncertainty}                   & \num{1.02} & \num{0.34} \\
        \hline
    \end{tabular}
\end{table}

\paragraph{Combined result of KNM1 and KNM2 campaigns}

Several approaches are pursued to combine results from the two presently unblinded data sets, as detailed in Ref.~\cite{knm2lett:2021} and summarized in Sec.~\ref{sec:combinecampaigns}. One such method is a simultaneous fit of both data sets, as presented in Fig.~\ref{fig:spectra_knm1_2}. Here, the data from all pixels are merged into a uniform spectrum for each campaign. The joint fit uses $m_\nu^2$ as a shared parameter, while  $A_\mathrm{s}$, $R_\mathrm{bg}$, and $E_0$ are implemented as individual parameters for KNM1 and KNM2. In this analysis, systematic uncertainties are treated via pull terms; systematic inputs that differ from one measurement campaign to the next are allowed to vary for KNM1 and KNM2, whereas other pull terms are fit as common parameters. The result of the combined fit yields $m_\nu^2 =$ \SI{0.07\pm0.32}{\electronvolt\squared}, which translates to an upper limit of $m_\nu <$ \SI{0.75}{\electronvolt} (\SI{90}{\percent} C.L.). Alternatively, adding the $\chi^2$ curves from the fits to the two individual data sets gives a consistent upper limit of $m_\nu <$ \SI{0.81}{\electronvolt} (\SI{90}{\percent} C.L.). Our reported upper limit, $m_\nu <$ \SI{0.8}{\electronvolt} (\SI{90}{\percent} C.L.), is rounded to the digit where both approaches agree.

A Bayesian approach, using the posterior distribution from the analysis of the first data set as an informative prior for the analysis of the second one, finds $m_\nu <$ \SI{0.73}{\electronvolt} (\SI{90}{\percent} credible interval).

\begin{figure}[tb]
    \centering
    \includegraphics[width=\textwidth]{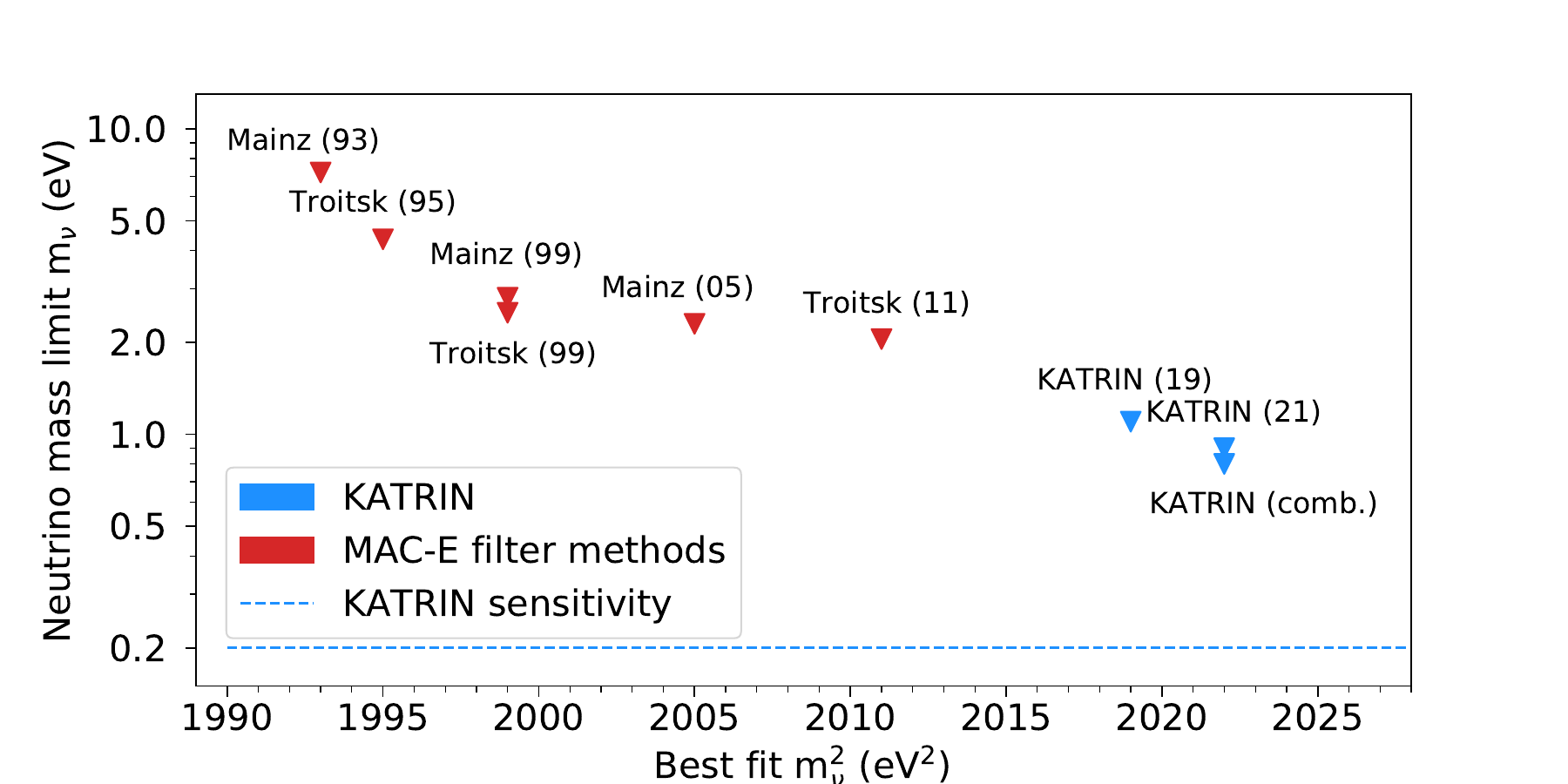}
    \caption{Overview of direct neutrino-mass measurements from the past three decades. Previous upper limits, all obtained using the MAC-E-filter technique, are marked in red; those from KATRIN are highlighted in blue. The projected sensitivity of KATRIN after \num{1000} full measurement days is indicated by the dashed line. References for data points: Mainz (93)	\cite{Weinheimer:1993pd}, Troitsk (95) \cite{Belesev:1995sb}, Mainz (99)	\cite{Weinheimer:1999tn}, Troitsk (99) \cite{LOBASHEV1999327}, Mainz (05) \cite{Kraus2005}, Troitsk (11) \cite{Aseev2011}, KATRIN (19) \cite{Aker2019,Aker2021Knm1}, KATRIN (comb.), KATRIN (21) \cite{knm2lett:2021}.}   	
    \label{fig:neutrinomasshistory}
\end{figure}

With the release of its second data set, KATRIN became the first kinematics-based neutrino-mass experiment to reach a sub-\si{\electronvolt} sensitivity. The relevance of this achievement is illustrated in Fig.~\ref{fig:neutrinomasshistory}, which places these initial results obtained by KATRIN in the historical context of tritium-based neutrino-mass experiments from the past three decades.

% ############################################################
% KATRIN tests of new physics
%# ##########################################################
\section{KATRIN tests of new physics}
\label{Sec:KatrinTestsOfNewPhysics}

%\section{General remarks on new physics in beta decay}
%\label{Sec:GeneralRemarksOnNewPhysicsInBetaDecay}

Testing physics beyond the light-neutrino standard picture adds exciting and potentially groundbreaking physics potential to the experiment. In the upcoming sections we will discuss the sensitivity of KATRIN to a variety of possible scenarios. The motivation of many of these scenarios originates in the facts that neutrino mass is generated in a different way from the other masses of Standard-Model (SM) particles, and that neutrinos are in general perfect candidates to be connected to BSM physics. Frequent features of such frameworks are additional sterile-neutrino states, new neutrino interactions, new particles coupling to neutrinos, etc. The scale at which new physics appears is in general not clear, in particular for sterile neutrinos. KATRIN is sensitive to two interesting mass scales, the \si{\electronvolt}-\ scale, which matches the design analysis interval of the experiment, and the \si{\kilo\electronvolt} scale, which falls within the $Q$-value of tritium decay. There are long-standing experimental hints for \si{\electronvolt}-scale sterile neutrinos~\cite{Abazajian:2012ys}, and \si{\kilo\electronvolt}-scale neutrinos are excellent candidates to be warm dark matter~\cite{Drewes:2016upu}. 

We separate our discussion into \si{\electronvolt}-scale sterile neutrinos (Sec.\ \ref{sec:eV}), \si{\kilo\electronvolt}-scale sterile neutrinos (Sec.\ \ref{sec:keV}), general neutrino interactions and charged currents beyond the SM (Sec.\ \ref{sec:CCBSM}),  new light bosons (Sec.\ \ref{Subsec:NewLightBosons}),  Lorentz-invariance violation (Sec.\ \ref{Subsec:LorentzInvarianceViolation}), and other new physics  (Sec.\ \ref{Subsec:OtherNewPhysics}). Moreover, we discuss the sensitivity to relic neutrinos (Sec.\ \ref{Subsec:RelicNeutrinos}), i.e.\ the ubiquitous background of very low-energy  neutrinos that froze out in the early Universe at temperatures of \si{\mega\electronvolt}, equivalent to an age of \SI{1}{\second} after the Big Bang. While this is standard physics, it is clear that if observed, there must be a mechanism that leads to large overdensities in our local neighborhood. 

The separation between topics is unavoidably blurry. For instance, \si{\electronvolt}- or \si{\kilo\electronvolt}-scale sterile neutrinos can have new interactions, or Lorentz-invariance violation may be present only for sterile neutrinos. Some of the scenarios we discuss here are better motivated than others, but more often than not neutrino physics has provided unexpected results, and one should remain open to all possibilities. 

We stress that the characteristics of the new physics that may show up in $\upbeta$ decay can be distinguished from the now-standard physics of active neutrino mass. The new physics can manifest in the measured spectrum in diverse ways, e.g.\,a kink-like spectral feature (heavy sterile neutrinos), shape distortion (exotic weak interactions, light sterile neutrinos, or new light bosons), appearance of a line feature (relic-neutrino capture) or a sidereal modulation of fit parameters (Lorentz-invariance violation). 

\subsection{Sterile neutrinos at the eV scale}
\label{sec:eV}

Right-handed neutrinos are a well-motivated SM extension. The introduction of a right-handed partner to the left-handed neutrinos provides a natural way to create neutrino masses~\cite{Abazajian:2012ys}.  Right-handed neutrinos, as opposed to the known left-handed neutrinos, would not take part in the weak interaction and would therefore be sterile. No gauge symmetry of the SM forbids the introduction of a Majorana mass term of arbitrary scale for the right-handed neutrino. As a consequence, new neutrino-mass eigenstates arise that can have an admixture with the active SM neutrinos~\cite{Volkas:2001zb}. In the following, these new mass eigenstates are referred to as ``sterile" neutrinos.

Light sterile neutrinos on the \si{\electronvolt}-mass range are motivated by long-standing and accumulating anomalies in short-baseline neutrino oscillation experiments, such as the reactor antineutrino anomaly~\cite{Mention:2011rk} and the recently reaffirmed gallium anomaly~\cite{Hampel:1997fc, Abdurashitov:2009tn, barinov2021results}. The postulated sterile-neutrino solutions of both anomalies cover a wide range of fourth mass eigenvalues. While the small-mass solutions, $\Delta m_{41}^2 <$ \SI{10}{\electronvolt}, are largely excluded by short-baseline reactor experiments~\cite{PROSPECT:2020sxr, Danilov:2019aef, MINOS:2020iqj, DoubleChooz:2020pnv, STEREO:2019ztb}, KATRIN is the only running laboratory experiment that is sensitive to the remaining large $\Delta m_{41}^2$ solutions.  

The sterile-neutrino search in KATRIN considers a minimal extension of the standard model by one additional sterile neutrino ($3\nu+1$), associated with a mass eigenvalue $m_4$ or $m_s$. Here, the standard $\upbeta$-decay spectrum $R_\upbeta(E,m_\nu^2)$ is complemented by a sterile branch $R_\upbeta(E,m_4^2)$:
\begin{equation}\label{eq:3nup1}
    R_\upbeta(E,m_\nu^2,m_4^2,\text{\textbar}U_{e4}\text{\textbar}^2) = (1-\text{\textbar}U_{e4}\text{\textbar}^2) \cdot R_\upbeta(E,m_\nu^2) +  \text{\textbar}U_{e4}\text{\textbar}^2 \cdot R_\upbeta(E,m_4^2)
\end{equation}
with $U$ being the $4\times4$ unitary PMNS matrix.
The sterile branch is weighted with the mixing of the electron flavor with the fourth mass eigenstate $\lvert U_{e4} \rvert^2$, simply referred to below as active-to-sterile mixing. The signature of a sterile neutrino in KATRIN is a kink-like distortion of the $\upbeta$ spectrum, most prominent at $E \approx E_0 - m_4$. Figure~\ref{fig:BetaSpectrum} is an illustration of the shape distortion for a \si{\kilo\electronvolt}-scale $m_4$.

The sterile-neutrino analyses of the first two measurement campaigns have been published in Refs.~\cite{Aker2021Sterile, Aker2022Sterile}. The analyses are sensitive to $m_4^2\lesssim\SI{1600}{\electronvolt\squared}$ and $\lvert U_{e4} \rvert^2 \gtrsim$ \num{6e-3}. As no sterile neutrino signal was found at \SI{95}{\percent} C.L., exclusion bounds were set. 
%The result of the second science run and the combination of first and second data set is presented in~\cite{Aker2022Sterile}. The analysis is sensitive to the sterile-mass range $m_4\lesssim\SI{1600}{eV^2}$ and an active-to-sterile mixing $\lvert U_{e4} \rvert^2 \gtrsim \num{6E-3}$. An improved exclusion limit has been reported, with respect to the first sterile neutrino analysis. 

\begin{figure}[tb]
    \centering
    \includegraphics[width=1.0\textwidth]{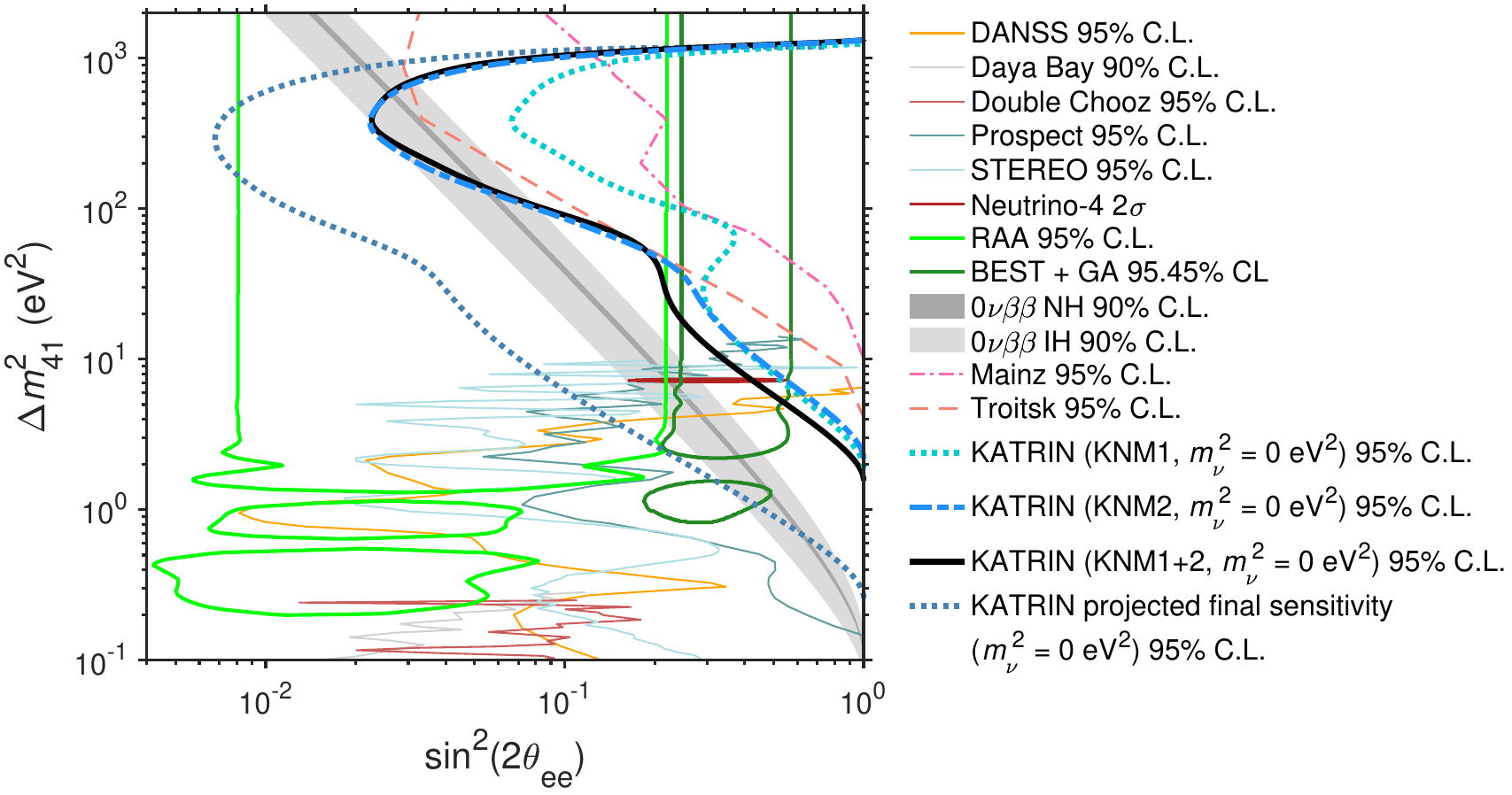}
    \caption{\SI{95}{\percent} C.L.\ exclusion contours of the first and second KATRIN science run with $m_\nu^2=\SI{0}{eV^2}$, as well as the combined exclusion from both campaigns. The final sensitivity is given by the projection described in Ref.~\cite{Aker2021Sterile}. 
    The second measurement is able to exclude a wider range than both the Mainz~\cite{Kraus:2012he} and Troitsk~\cite{Belesev:2012hx} experiments for $m_4^2\lesssim\SI{300}{\electronvolt\squared}$. Large $\Delta m_{41}^2$ can be excluded to a great extent for the solutions of the reactor antineutrino anomaly (RAA) and BEST+gallium (GA) anomalies~\cite{Mention:2011rk, barinov2021results}. The combined analysis is in tension with the positive results claimed by Neutrino-4~\cite{NEUTRINO-4:2018huq} for $\sin^2(2\theta_{ee}) \gtrsim 0.4$.
    KATRIN data improve the exclusion bounds set by short-baseline oscillation experiments for $\Delta m_{41}^2 \gtrsim$ \SI{10}{\electronvolt\squared}~\cite{PROSPECT:2020sxr, Danilov:2019aef, MINOS:2020iqj, DoubleChooz:2020pnv, STEREO:2019ztb}. Constraints from $0\nu\upbeta\upbeta$ with $m_{\upbeta\upbeta}<$\SI{0.16}{\electronvolt} are shown as gray bands~\cite{ParticleDataGroup:2020ssz,KamLANDZen:2016,PhysRevLett.125.252502}. Adapted from  Ref.~\cite{Aker2022Sterile}.}
    \label{fig:SterileExclusion}
\end{figure}

Figure~\ref{fig:SterileExclusion} displays the latest KATRIN results at \SI{95}{\percent} C.L.\ and compares them to other experiments. These exclusion contours assume $m_\nu^2=\SI{0}{eV^2}$, resulting in $\Delta m_{41}^2 \approx m_4^2$; Ref.~\cite{Aker2022Sterile} includes an extensive discussion of the relationship between active and sterile neutrino masses in such a search. Analysis of the KNM2 data improves on constraints set by the Mainz~\cite{Kraus:2012he} and Troitsk~\cite{Belesev:2012hx} experiments for the mass range $m_4\lesssim \SI{300}{\electronvolt\squared}$. As short-baseline experiments are sensitive to  different observables from kinematic searches, namely $\Delta m_{41}^2$ and $\sin^2\left(2 \theta_{ee}\right) = 4\lvert U_{e4}\rvert^2 \, (1-\lvert U_{e4}\rvert^2)$, the KATRIN, Mainz and Troitsk exclusion contours have been transformed accordingly. %, with $\Delta m_{41}^2 =  m_4^2 -m_\nu^2 $ and $\sin^2\left(2 \theta_{ee}\right) = 4 \lvert U_{e4} \rvert^2 \left(1-\lvert U_{e4} \rvert^2\right)$.
A considerable fraction, $\SI{20}{eV^2}\lesssim \Delta m_{41}^2 \lesssim$ \SI{1000}{\electronvolt\squared}, of the gallium-anomaly space can be excluded by KATRIN. Moreover, the reactor antineutrino anomaly is excluded for $\SI{50}{\electronvolt\squared} \lesssim \Delta m_{41}^2 \lesssim \SI{1000}{\electronvolt\squared}$. The hint of a signal from the Neutrino-4 collaboration is disfavored for $\sin^2(2\theta_{ee}) \gtrsim$ \num{0.4} at \SI{95}{\percent} C.L.

The present exclusion limits are dominated by statistical uncertainties. Therefore, additional measurement time and statistics from future KATRIN campaigns will improve its sensitivity considerably. The projected final sensitivity demonstrates that KATRIN will be able to examine the entire large-$\Delta m_{41}^2$ parameter space of the gallium anomaly and a significant fraction of the reactor antineutrino anomaly. %provide competitive results to short-baseline experiments on the eV-mass scale.

\subsection{Sterile neutrinos at the keV scale}
\label{sec:keV}

Sterile neutrinos with a mass in the kilo-electronvolt (\si{\kilo\electronvolt}) regime are promising dark-matter candidates~\cite{PhysRevLett.72.17, PhysRevLett.110.061801, PhysRevLett.82.2832}. Due to their mass scale and production mechanism, they can act as warm dark matter, which could mitigate tensions between cosmological observations and predictions of purely cold-dark-matter scenarios~\cite{PhysRevLett.72.17}.

\begin{figure}[tb]
    \centering
    \includegraphics[width=\textwidth]{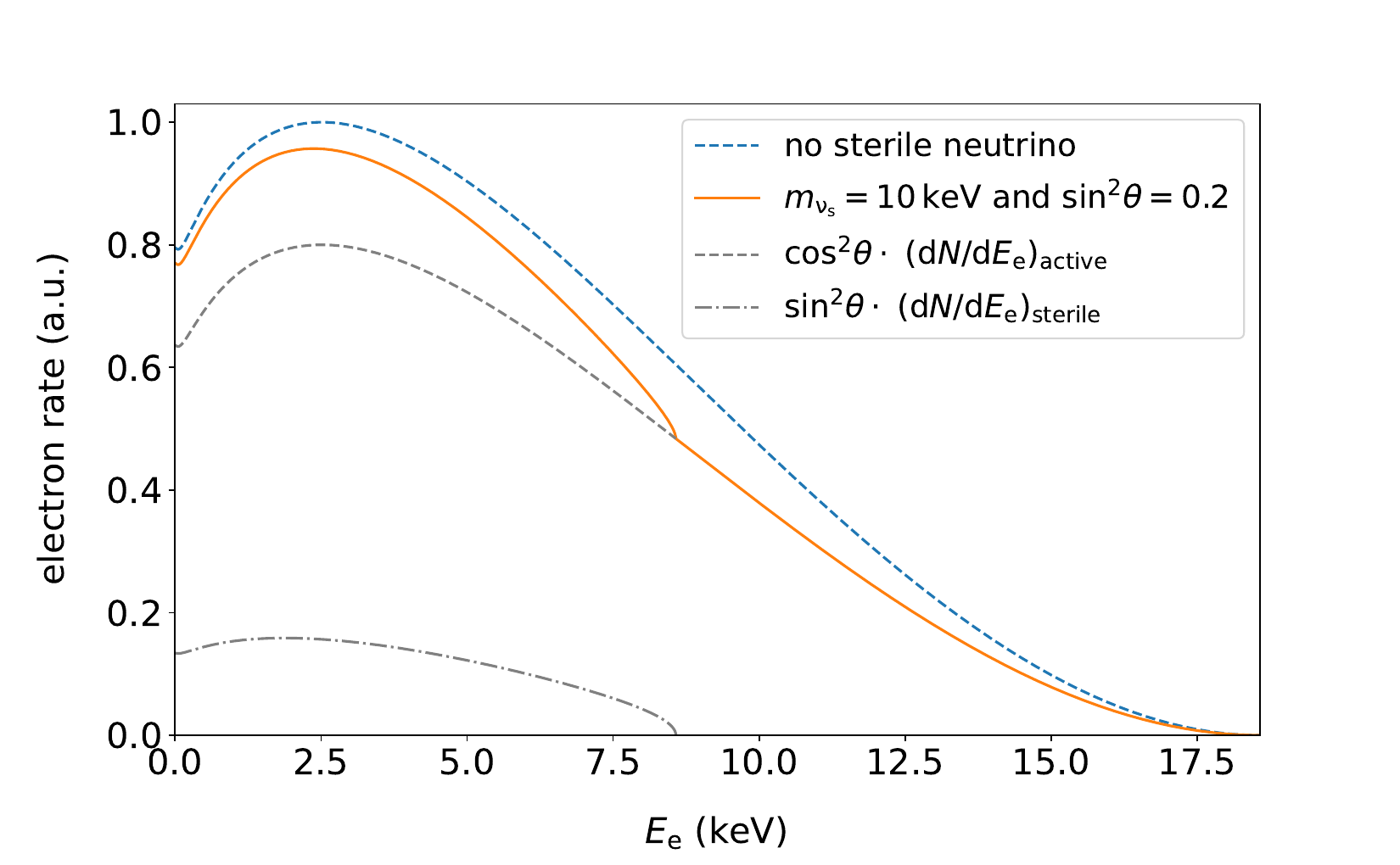}
    \caption{Signature of a 10-keV sterile neutrino in a differential tritium $\upbeta$-decay spectrum: The blue dashed line shows the spectrum without sterile neutrinos. The  solid orange line shows the spectrum with a sterile neutrino with $m_{\mathrm{s}} =$ \SI{10}{\kilo\electronvolt} and an exaggerated mixing amplitude of $\sin^2 \theta =$ \num{0.2}. The latter is composed of the active branch (gray dashed line) and the sterile branch (gray dot-dashed line).}
    \label{fig:BetaSpectrum}
\end{figure}

As for eV-scale sterile neutrinos (Sec.~\ref{sec:eV}), the presence of a sterile neutrino in the ~\si{\kilo\electronvolt} mass range would result in a characteristic kink-like signature (Fig.~\ref{fig:BetaSpectrum}) at energy $E_0-m_{\mathrm{s}}$, with a magnitude governed by the mixing amplitude~$\sin^2 \theta$ of the sterile neutrino with the active Standard Model electron neutrino~\cite{PhysRevLett.45.963, Mertens:2014nha}. 

The KATRIN experiment has several advantageous features for a \si{\kilo\electronvolt}-scale sterile neutrino search. With an endpoint of $E_0 =$\SI{18.6}{\kilo\electronvolt}, tritium $\upbeta$-decay permits a search for sterile neutrinos on a mass range of multiple \si{\kilo\electronvolt}. Due to the short-half life of \num{12.3} years, high decay rates can be achieved with reasonable amounts of tritium which in turn leads to rapid statistical significance on the sterile-to-active mixing amplitude.

The use of the KATRIN experiment for a \si{\kilo\electronvolt}-scale sterile neutrino search has been investigated in several studies (for example in~\cite{Mertens:2014nha,PhDKorzeczek2020,PhDHuber2020}). It was shown that the experiment can be used with only minor modifications for a first, low-statistics measurement. As a proof of principle, KATRIN data taken in 2018 has been used to search for \si{\kilo\electronvolt}-scale sterile neutrinos in a mass range of up to \SI{1.6}{\kilo\electronvolt}; a publication is in preparation.

For higher-statistics measurements, a new detector system for KATRIN is under development. For a technical description of the TRISTAN detector project, an overview of performance tests, and the current project status, see Sec.~\ref{Subsubsec:TristanDetector}.

\begin{figure}[!ht]
	\centering
    \includegraphics[width=\columnwidth]{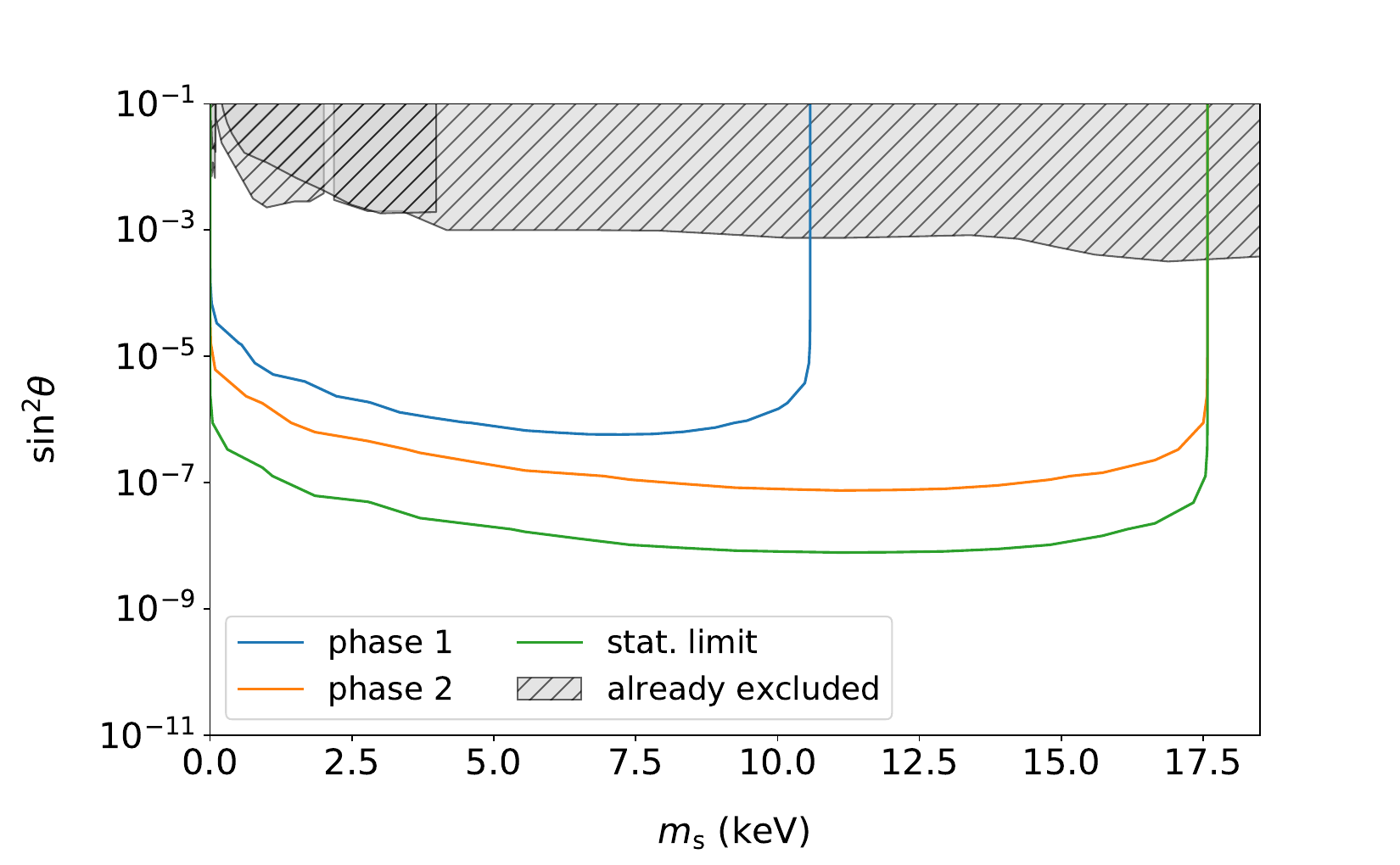}
	\caption{KATRIN sensitivity to \si{\kilo\electronvolt}-scale sterile neutrinos in different scenarios. The statistical limit assumes \num{e18} electrons over the full energy range, corresponding to a measurement time of one year at the full KATRIN source strength. Phase 1 denotes the first stage of TRISTAN operation with a reduced amount of TRISTAN modules (approx. \num{1000} pixels) and a lowered KATRIN source strength (\SI{0.3}{\percent} of the nominal column density). The spectrum is scanned on an energy interval of \SIrange{8}{18}{\kilo\electronvolt} with a total measurement time of $t_\mathrm{meas} =$ \SI{0.5}{yrs}. In the follow up stage (phase 2), the TRISTAN detector is operated with all \num{3500} pixels. The tritium spectrum is scanned on the full energy range at a reduced source activity of \SI{2}{\percent} of the nominal column density for $t_\mathrm{meas} =$ \SI{1.0}{yrs}. The already excluded parameter space is based on Refs.~ \cite{Kraus:2012he,Hiddemann,Holzschuh,Abdurashitov2017Sterile}.}
	\label{fig:tristan:sensitivity}
\end{figure}

With 1 year of KATRIN operation at full source strength, a statistical sensitivity of the order of $\sin^2 \theta =$ \num{e-8} can be reached. It is, however, extremely challenging to control systematic uncertainties at this level. Therefore, the targeted design sensitivity is  $\sin^2 \theta =$ \num{e-6}, which would surpass the sensitivity of previous laboratory-based searches and reach a region of cosmological interest. Note that while astrophysical constraints on the mixing are quite strong, there are ways to evade them and to bring the mixing up to scales reachable by the TRISTAN detector upgrade~\cite{Bezrukov:2017ike,Benso:2019jog}. Figure~\ref{fig:tristan:sensitivity} shows the sensitivity for different measurement scenarios, using the KATRIN experiment and a TRISTAN detector to search for sterile neutrinos.

\subsection{Generalized neutrino interactions and exotic charged currents}
\label{sec:CCBSM}

Generalized neutrino interactions (GNI) can be derived by embedding a basis of gauge-invariant dimension-six operators into the SM effective field theory, respecting the full $\mathrm{SU}(3)_{\mathrm{C}} \times \mathrm{SU}(2)_{\mathrm{L}} \times \mathrm{U}(1)_{\mathrm{Y}}$ gauge symmetry. This generalized theory includes a variety of interactions such as scalar, pseudoscalar, vector, axial vector, and tensor interactions with SM fermions~\cite{BISCHER2019114746}. The wide range of interactions could arise from heavy new physics and would cause energy-dependent distortions of the measured $\upbeta$-decay spectrum. As this generalized parameter space is rather complex, it is sensible to look at different scenarios separately. 

Precision kinematics of weak decays in particular allow searches for exotic forms of interactions, such as additional non-standard right-handed currents departing from the established $V-A$ structure of SM weak interactions as a special case of GNI. Left-right symmetric models are hypothetical extensions of the SM in which right-handed weak currents occur. The imprint of right-handed currents on the tritium $\upbeta$-decay spectrum with active neutrinos has been the subject of earlier studies (e.g., Refs.~\cite{Bonn:2007su,Stephenson:2000mw}). Dedicated experiments
%Experiments dedicated to test standard electroweak interactions in $\upbeta$-decay 
(most notably, neutron decay, reviewed in Ref.~\cite{Severijns:2006dr}) have yielded very stringent limits. The sensitivity of the standard KATRIN analysis is limited by the fact that the endpoint $E_0$ is one of the free fit parameters (Sec.~\ref{Subsec:SpectralFittingAndLimitSetting}). Sensitivity to right-handed currents with active neutrinos can potentially be improved by fixing the endpoint in the fit, which would require a very accurate measurement of the Q-value (through the \textsuperscript{3}H-\textsuperscript{3}He mass difference) and knowledge of the absolute KATRIN energy scale at the level of \SIrange[range-phrase=--]{30}{100}{meV} or better.  

Recently, the interplay of right-handed currents with potential sterile neutrinos has become a focus of interest (see, e.g., Refs.~\cite{Barry:2014ika,Ludl:2016ane,Steinbrink:2017uhw}). It has been shown in these works that KATRIN will be able to set bounds on the strength of the left-right interference terms as a function of $m_s$, a low-energy search complementary to LHC-based efforts. New limits could be obtained for a Fierz-like interference term that does not originate in the left-right symmetric model~\cite{Steinbrink:2017uhw}.

\subsection{New light bosons}
\label{Subsec:NewLightBosons}

New particles coupling to known particles are a generic prediction of BSM theories. As no heavy new particles have been found at collider experiments, focus  on light new physics has increased in recent years. 
Light scalar or vector bosons coupling to electrons, neutrinos, $u$ or $d$ quarks may be emitted in $\upbeta$ decay, if their mass is below the $Q$ value of tritium. KATRIN is particularly sensitive to this interesting low-mass regime. Ref.~\cite{Arcadi:2018xdd} has analyzed this in detail for emission from the leptons. 

\begin{figure}[tb]
    \centering
    \includegraphics[width=0.49\textwidth]{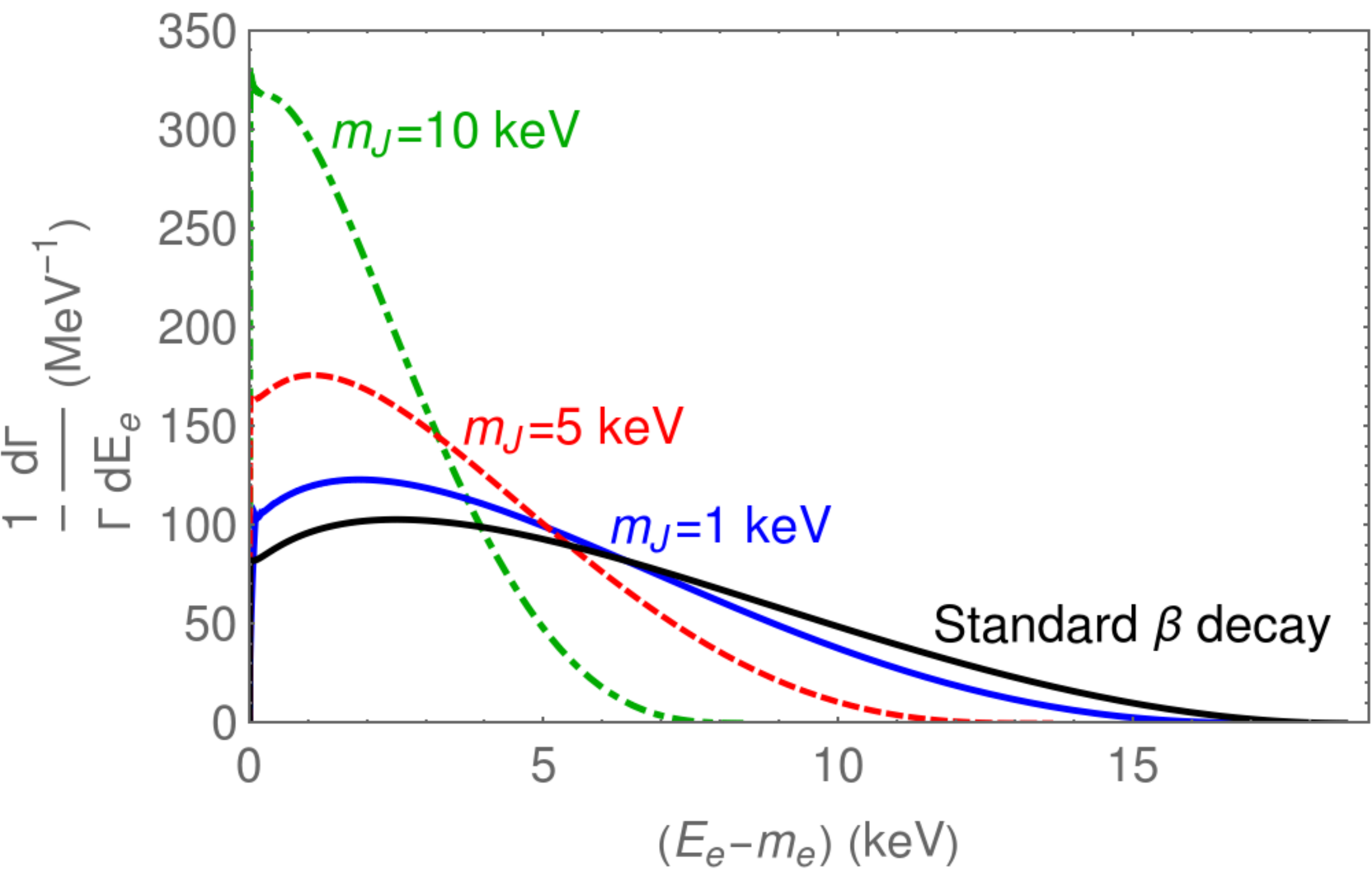}
    \hfill
    \includegraphics[width=0.49\textwidth]{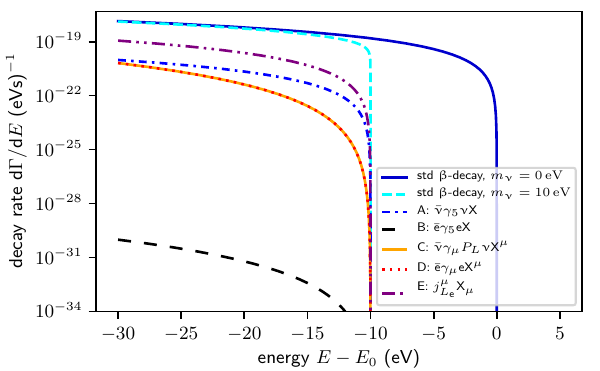}
    \caption{\textbf{Left:} Normalized electron-energy spectrum of the decay $^3\text{H} \to {^3\text{He}}^{+} + e^-+ \bar \nu_e + J$ (with the pseudoscalar $J$ emitted from the neutrino) and the standard $\upbeta$-decay spectrum. The coupling $g_{\nu J}$ has been set to \num{1}; the spectrum scales with $g_{ \nu J}^2$. \textbf{Right:} Differential spectrum for standard $\upbeta$ decay without (solid blue) and with \SI{10}{\electronvolt} (dashed cyan) neutrino mass, as well as five spectra expected for the additional emission of a light boson $X$ with $g_X = 1$, $m_X =$ \SI{10}{\electronvolt} and various coupling types. Figures taken from Ref.~\cite{Arcadi:2018xdd}.}
    \label{fig:pseudo}
\end{figure}

It is common to find massive CP-even (scalar) and CP-odd (pseudoscalar) fields in new physics theories. For instance, axions and axion-like particles  fall into this category. In some cases, these pseudoscalar particles play the role of mediators between SM particles and the dark-matter sector. In some other cases, they are connected to neutrino masses and lepton-number violation, most notably in so-called Majoron models. Agnostic to its possible origin, we can assume  that the pseudoscalar $J$ is coupled to neutrinos via $ i g_\nu  \bar\nu \gamma_5 \nu J $. A scalar coupling and a lepton-number-violating coupling have the same effect on $\upbeta$ decay. Figure~\ref{fig:pseudo} (left) shows the effect on the electron spectrum. 

Vector bosons $Z'$ are a common feature in many BSM frameworks with additional gauge symmetries. Their masses are typically determined by a (combination of) gauge coupling(s) and by the energy scale at which the additional gauge symmetry is spontaneously broken. There is a subtlety depending on whether the $Z'$ couples only to electrons or neutrinos, or to both. In the former case the current is not conserved and the theory is non-renormalizable, leading to a $1/m_Z'^2$ enhancement of the emission process. In contrast, if the $Z'$ couples to both the neutrino and electron, the theory is free of conceptual issues. Focusing on leptonic couplings, we thus have five relevant cases: a pseudoscalar coupling to neutrino, to electron, and a $Z'$ coupling to neutrino, electron and both. Fig.~\ref{fig:pseudo} (right) shows the differential spectrum for these cases. 

\begin{figure}[tb]
    \centering
     \includegraphics[width=0.49\textwidth]{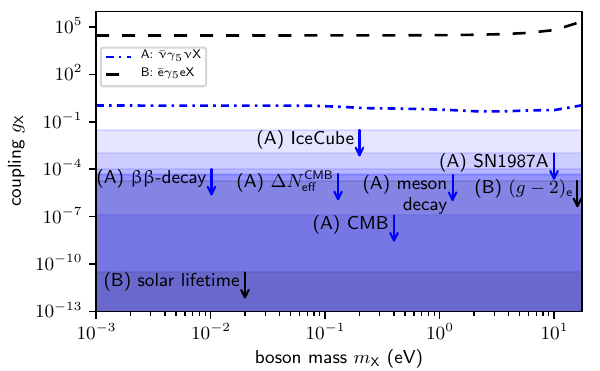}
     \hfill
    \includegraphics[width=0.49\textwidth]{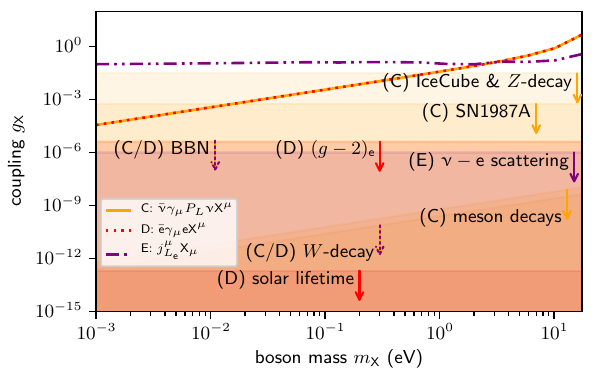}
    \caption{The lines show the statistical sensitivity (\SI{90}{\percent} C.L.) of KATRIN~for the detection of \si{\electronvolt}-scale light pseudoscalars emitted from neutrinos (A) or from electrons (B) (left) and vector bosons emitted from neutrinos (C), from electrons (D), or from both (E) (right). The shaded areas mark the parameter regions allowed from other constraints. Figures taken from Ref.~\cite{Arcadi:2018xdd}.}
    \label{fig:limits_bos}
\end{figure}

The light-boson spectrum superimposed on the standard $\upbeta$-decay spectrum has six fit parameters:
\begin{enumerate}
	\item neutrino mass squared $m_\nu^2$
	\item endpoint $E_0$
    \item the amplitude  ensuring correct normalization of the superposition of light-boson and standard $\upbeta$-decay spectra
	\item background rate %$Bg$
	\item light-boson mass $m_X$
	\item light-boson coupling $g_X$
\end{enumerate}
Using the standard settings defined in the design report~\cite{KATRIN2005}, including an analysis interval $[E_0 -\SI{30}{\electronvolt}, E_0 +$ \SI{5}{\electronvolt}], a true neutrino mass of zero, and a measuring time of three full years, gives expected limits shown in Fig.~\ref{fig:limits_bos}.

As Fig.~\ref{fig:limits_bos} shows, KATRIN's sensitivity region is apparently already excluded by previous laboratory, astrophysical and cosmological searches. However, the laboratory constraints are extrapolated from searches for heavier states, conducted at a far larger energy scale; meanwhile, the astrophysical and cosmological limits are highly sensitive to specific flux and source hypotheses. KATRIN's data, acquired directly at the relevant energy scale, will thus offer an important complement to these earlier searches. 
Furthermore, KATRIN is an excellent probe for  exotic scenarios with strongly energy-dependent couplings~\cite{Dvali:2016uhn}. 

\subsection{Lorentz-invariance violation }
\label{Subsec:LorentzInvarianceViolation}

A variety of BSM theories predict deviations from Lorentz invariance, the possibilities for which are parameterized in the Standard Model Extension~\cite{PhysRevD.55.6760, PhysRevD.58.116002, PhysRevD.69.105009, PhysRevD.85.096005}. In the neutrino sector, many of the possible Lorentz-invariance-violating parameters are strongly constrained by time-of-flight or neutrino-oscillation experiments. However, some ``countershaded'' parameters, so named because they are virtually invisible to a wide range of experiments, are still constrained only weakly, if at all~\cite{PhysRevD.85.096005, Diaz2013, Lehnert:2009qv}. 

The complex oscillation-free parameter $\left( a_{\text {of}}^{(3)} \right)_{jm}$, where $\{j,m\}$ are angular-momentum quantum numbers with $j=\{0,1\}$, would introduce a preferred direction in space. In a precise measurement of the $\upbeta$-decay spectrum, this could manifest as a variation of the spectral endpoint depending on the latitude and orientation of the $\upbeta$-electron beam, or as a sidereal oscillation of the spectral endpoint $E_0$~\cite{PhysRevD.85.096005,Lehnert:2021tbv}. Because KATRIN analyzes only those $\upbeta$-electrons that are emitted within \SI{50.4}{\degree} relative to the direction of the beam axis, it is in principle sensitive to this parameter. 

KATRIN has performed a search for Lorentz-invariance violation based on its first science run. This campaign featured about 300 $\upbeta$-scans over a time period of two weeks. For the best sensitivity, we defined an effective time point for each scan, based on the ordering of scan steps and the region of the scanned spectrum that most dominates determination of the spectral endpoint. Based on these timepoints and the best-fit endpoint of each scan, KATRIN searched for an oscillatory signature with the required period of \SI{23.7}{\hour}. This analysis finds no sign of Lorentz-invariance violation, thus setting the first limit on $a_{\text {of}}$~\cite{lorentz-inprep}. With additional statistics acquired over more sidereal cycles, this limit will be improved.

\subsection{Relic neutrinos}
\label{Subsec:RelicNeutrinos}

Modern cosmology predicts the existence of a cosmic (or relic) neutrino background (C$\nu$B) that fills the entire universe. Cosmic neutrinos decoupled from the rest of the Universe about one second after the Big Bang. Nowadays relic neutrinos cluster around galaxies, and the local relic-neutrino overdensity $\eta$ is predicted to be \num{1.2} to \num{20} for neutrino masses below \SI{0.6}{\electronvolt}~\cite{Ringwald_2005,de_Salas_2017}. The direct detection of these ancient neutrinos would provide direct insight into the early history of the universe. However, this measurement remains one of the greatest challenges in neutrino physics. 

\begin{figure}[tb]
\begin{centering}
	\includegraphics[width=0.75\textwidth]{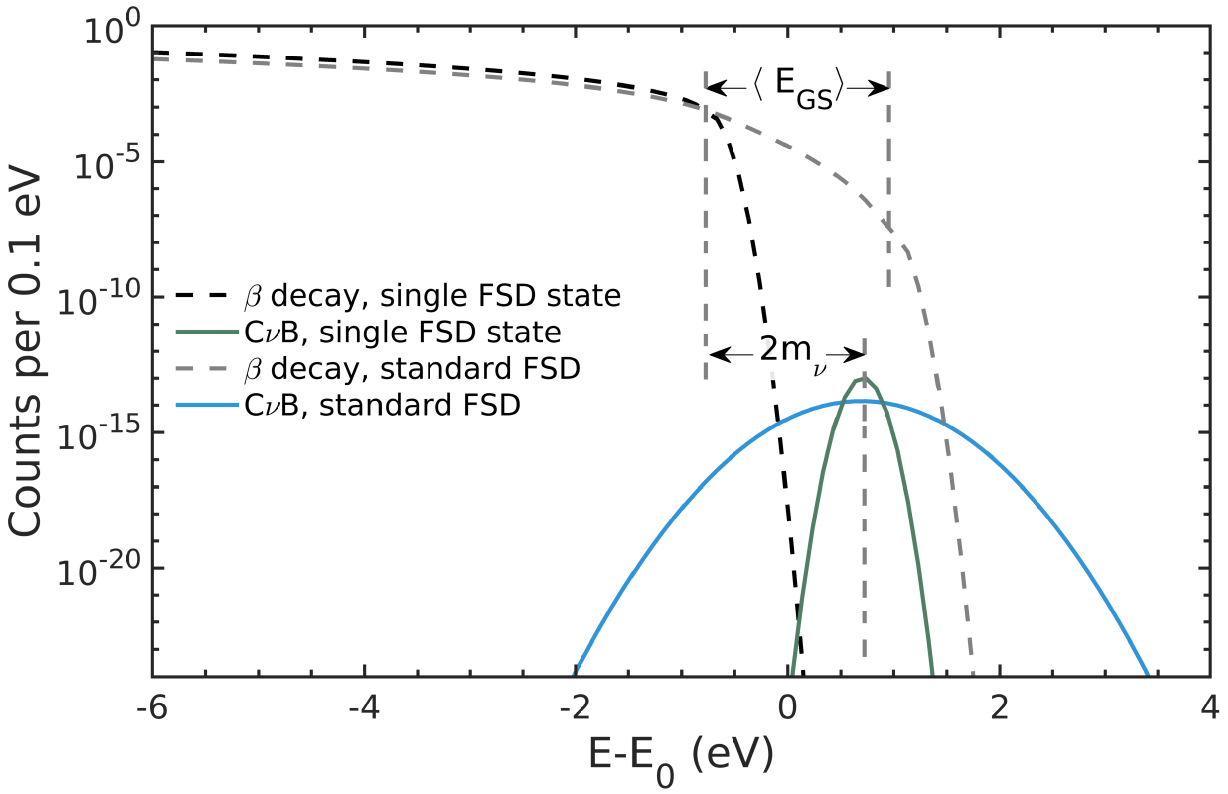}
	\caption{Simulation of the effects of the molecular final-state distribution (FSD) on the differential spectra from $\upbeta$ decay and from relic-neutrino capture, assuming $m_\nu = 0.7$~eV and $\eta = 1$; details are in Ref.~\cite{KATRIN:2022kkv}, from which this figure is adapted. Doppler broadening is included for all spectra. $\langle E_{\mathrm{GS}} \rangle \approx$~\SI{1.7}{\electronvolt} is the mean energy of the electronic ground state. \label{fig:simrelicspectrum}}
\end{centering}
\end{figure}

Relic neutrinos can interact with the tritium located in the KATRIN source via the neutrino capture reaction $\nu_e+{}^3\mathrm{H}\rightarrow {}^3\mathrm{He}^+ +e^-$~\cite{PhysRev.128.1457,Faessler_2017}, leading to a monoenergetic electron signal just beyond the endpoint. Fig.~\ref{fig:simrelicspectrum} shows the resulting spectral contributions as well as the broadening imposed by the molecular final-state distribution (FSD) of T$_2$ decay. KATRIN's high-precision spectral measurement can thus provide new constraints on the local overdensity of relic neutrinos. 

\begin{figure}[tb]
	\includegraphics[width=0.55\textwidth]{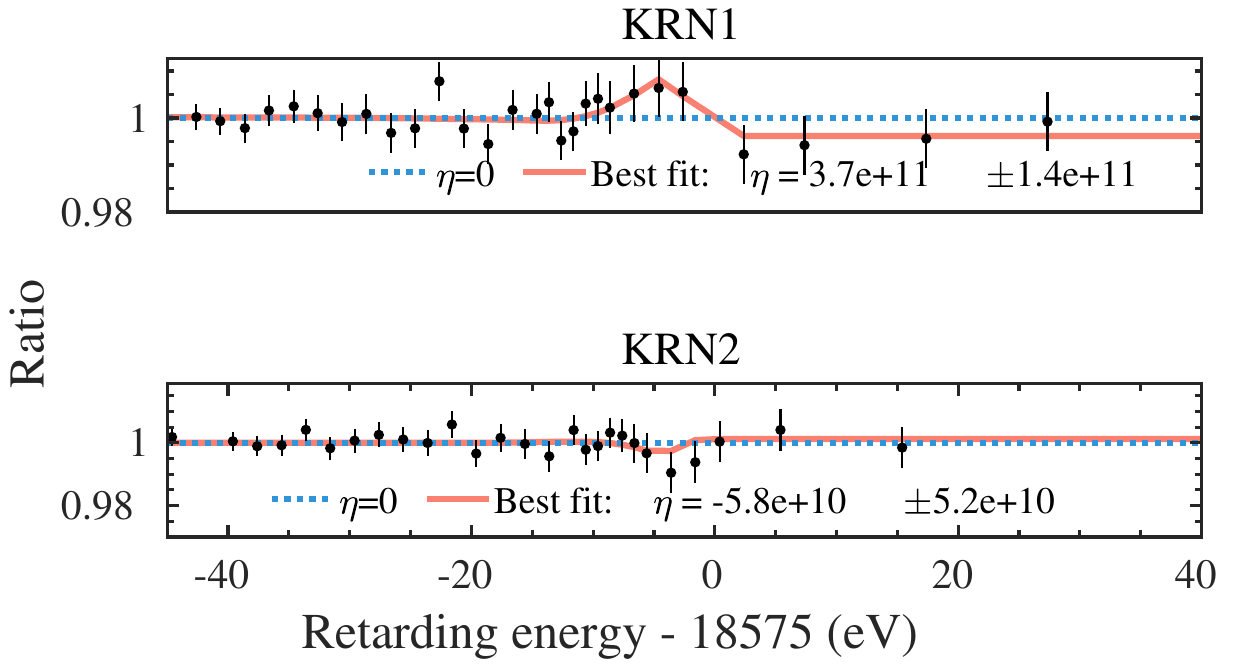}
	\includegraphics[width=0.45\textwidth]{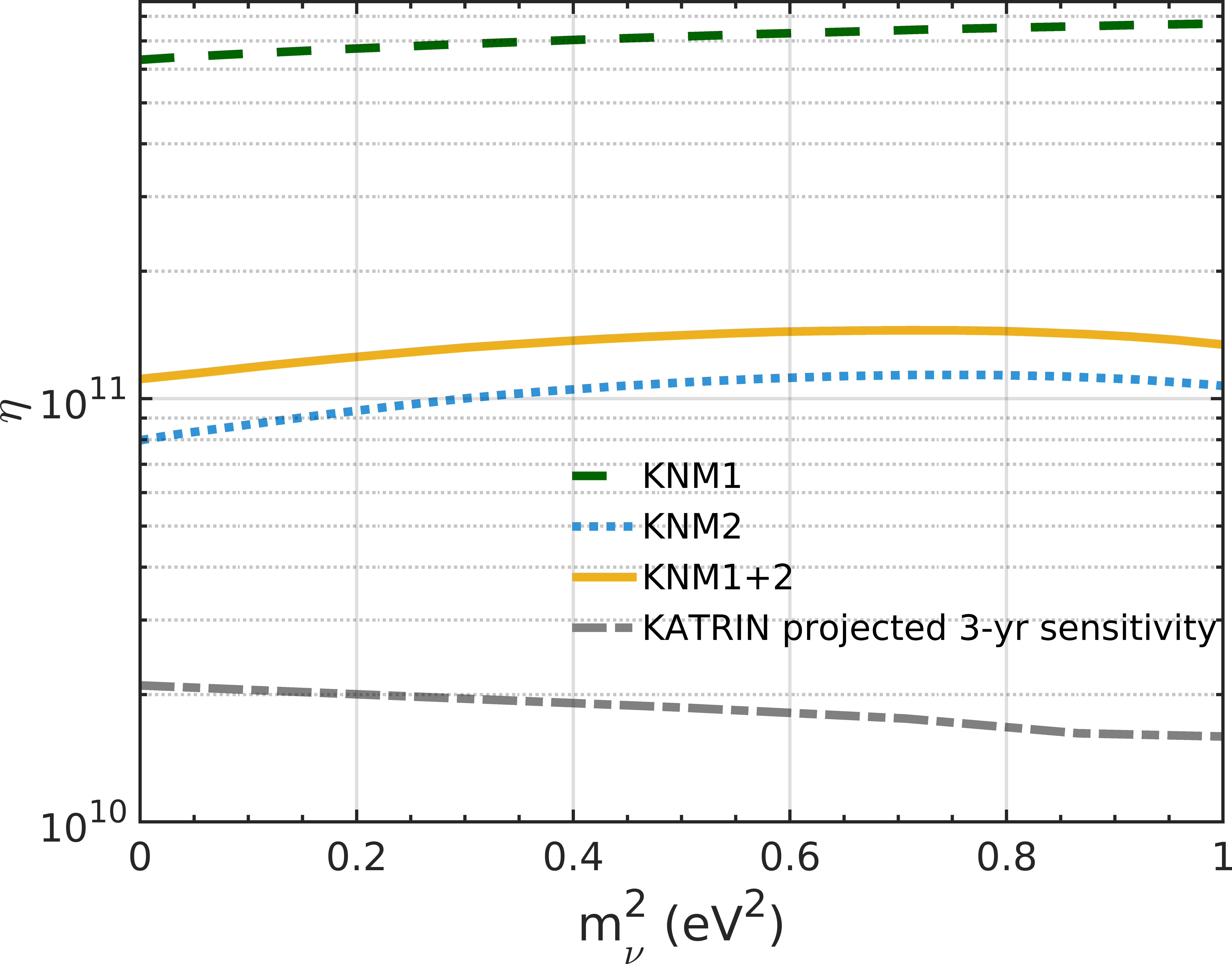}
	\caption{Left: \label{fig:krn12} Spectral ratio of the best fits for relic neutrinos with respect to the null hypothesis $\eta =$ \num{0} for the first two measurement campaigns. The non-zero values of the best fit of $\eta$ are both consistent with background fluctuations. Right: Exclusion contours for the relic-neutrino overdensity $\eta$ at \SI{99}{\percent} C.L. for both measurement campaigns and their combination, and the projected final KATRIN sensitivity. Adapted from Ref.~\cite{KATRIN:2022kkv}.}
\end{figure}

Using the data from the first two science runs conducted in 2019~\cite{Aker2019, Aker2021Knm1, knm2lett:2021}, KATRIN has established new constraints on the local overdensity of relic neutrinos~\cite{KATRIN:2022kkv}. The data set includes \num{5.16e6} $\upbeta$ decay electrons and \num{0.72e6} background events below the endpoint. No significant relic-neutrino signal is observed, and the $\eta$ parameter was constrained to be smaller than \num{9.7e10} (\num{1.1e11}, \num{1.3e11}) at \SI{90}{\percent} (\SI{95}{\percent}, \SI{99}{\percent}) C.L. for neutrino masses less than \SI{1}{\electronvolt}. This new result, shown in Fig.~\ref{fig:krn12}, improves on previous upper limits set by direct kinematic experiments at Los Alamos~\cite{Robertson:1991} and Troitsk~\cite{LOBASHEV1999327}.

KATRIN is continuing to operate toward the goal of collecting 1000 days of data by 2024. The current measurements at increased background triggered a reassessment of the final sensitivity for the relic neutrino overdensity, now at $ \eta< $\num{1.0e10} (\num{1.4e10}, \num{1.8e10}) at \SI{90}{\percent} (\SI{95}{\percent}, \SI{99}{\percent}) for a background rate of \SI{130}{mcps} summed over all detector pixels. We note that the irreducible background from $\upbeta$ decay, as broadened by the FSD, makes it infeasible to distinguish a relic-neutrino signal for $m_\nu <$~\SI{0.85}{\electronvolt} unless $\eta >$~\num{e6}; simply increasing the tritium mass does not improve sensitivity, since the $\upbeta$-decay rate scales in the same way as the neutrino-capture rate~\cite{KATRIN:2022kkv}. 

\subsection{Other new physics}
\label{Subsec:OtherNewPhysics}

In this section we briefly describe additional new physics scenarios that could be probed by KATRIN, but that the collaboration has not yet studied in detail. 

\begin{figure}[t!]
    \centering
    \includegraphics[width=0.45\textwidth]{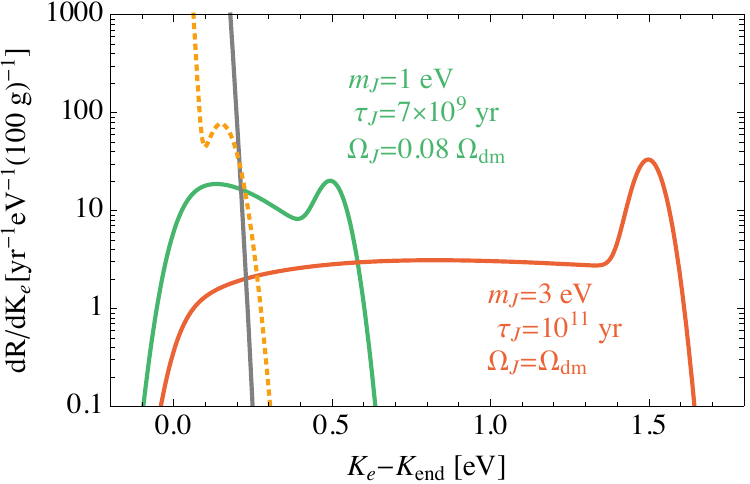}
    \includegraphics[width=0.45\textwidth]{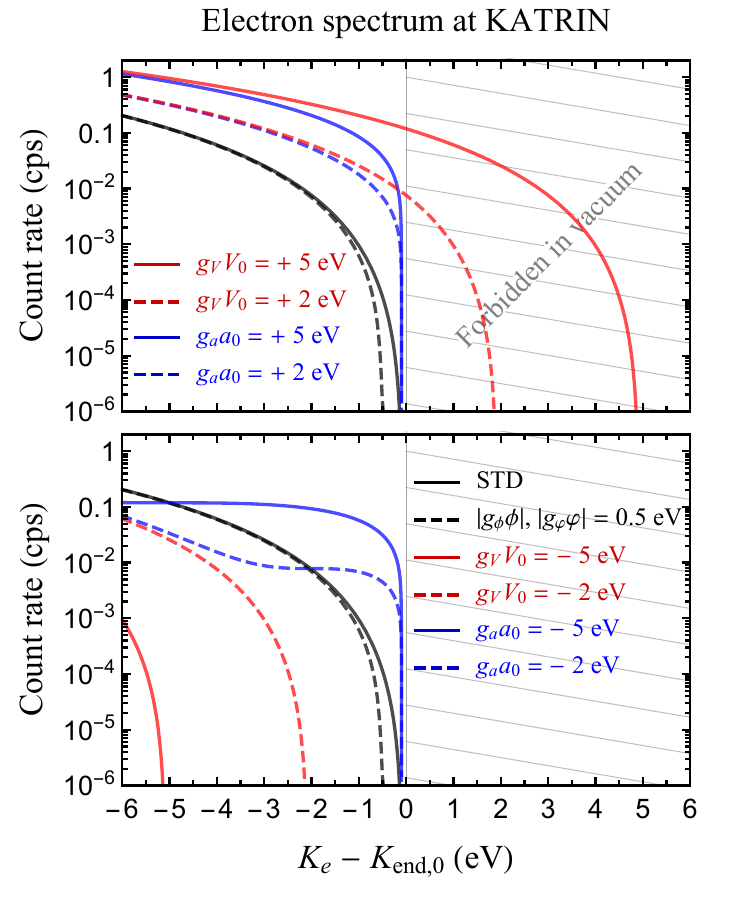}
    \caption{\textbf{Left:} Event rates for capture on tritium for neutrinos from Majoron decay $J\to\nu\nu$ for $m_J =$ \SI{1}{\electronvolt}, $\tau_J =$ \SI{7e9}{yr},  $\Omega_J/\Omega_{\rm dm} =$ \num{0.08} (green, solid) and $m_J =$ \SI{3}{\electronvolt}, $\tau_J =$ \SI{e11}{yr}, $\Omega_J/\Omega_{\rm dm} =$ \num{1} (red, solid) and natural $\upbeta$ decay (gray, solid) with an electron energy resolution of \SI{0.1}{\electronvolt}. For comparison, we plot the total rate for signal plus background for a Majorana relic-neutrino signal with $m_\nu =$ \SI{0.15}{\electronvolt} (yellow, dashed). Taken from~\cite{McKeen:2018xyz}. \textbf{Right:} The integrated $\upbeta$-decay spectra at KATRIN{} in various types of media: scalar, pseudoscalar and two different vector and axial-vector potentials, with neutrino mass fixed to  \SI{0.1}{\electronvolt}. Taken from~\cite{Huang:2021zzz}.}
    \label{fig:otherNP}
\end{figure}

If a dark-matter particle is light, and decays into neutrinos, tritium may capture them and generate via inverse $\upbeta$ decay an electron with energy beyond the endpoint. This interesting application of a neutrino-mass experiment as an indirect dark-matter experiment has been proposed in Refs.~\cite{McKeen:2018xyz,Chacko:2018uke}, and applied to the proposed PTOLEMY experiment, see Fig.~\ref{fig:otherNP} (left). Of course, KATRIN could already set limits on such scenarios. 

Another example is $\upbeta$-decay in a medium, i.e.\ a  background field, which could be formed by dark matter, dark energy or a fifth-force potential. Neutrinos couple to this background field, and this modifies their dispersion relation, which has an impact on the measurable electron spectrum (Fig.~\ref{fig:otherNP}, right). The interaction can be of scalar, pseudoscalar, vector or axial-vector type. Scalar and pseudoscalar interactions mimic the effect of neutrino mass, and therefore could explain a disagreement between cosmological and KATRIN results on neutrino mass. Even more interesting are the axial-vector and vector forms. The former can lead to a characteristic spectrum distortion; the latter can feature the spectacular effect of the  $\upbeta$ spectrum  extending beyond the normal kinematic limit, because processes which are not kinematically allowed in vacuum can take place if the medium modifies the dispersion relations. Taking publicly available information, Ref.~\cite{Huang:2021zzz} set limits on the interaction potentials that are expected to dominate, depending on the explicit model realization.

% ############################################################
% Additional physics observables
%# ##########################################################
\section{Additional physics observables}
\label{Sec:AdditionalPhysicsObservables}

\subsection{Molecular final states from T$_{2}$ decay}
\label{Subsec:MolecularFinalStatesFromT2Decay}

With a molecular tritium (T$_2$) source, KATRIN measures a spectrum that is modified by the molecular final-state distribution: the spectrum of rotational, vibrational and electronic modes excited in the daughter $^3$HeT$^+$ molecule by the beta decay of one tritium nucleus in the parent molecule. As a result, the neutrino-mass analysis requires knowledge of the energy levels of these distributions, as well as the probability that each is excited in the beta decay. KATRIN's spectral analysis relies on precise calculations of this distribution, computed using the geminal-basis method over the last 20-odd years~\cite{Saenz2000, Doss2006, Aker2021Knm1}. It is impractical to measure the entire final-state distribution, but observables predicted by this theoretical framework have good agreement with a wide range of experiments, summarized in Ref.~\cite{Bodine2015}. Recently, the TRIMS experiment followed up on the one area of discrepancy with existing experiments, the branching ratio to the bound molecular ion $^3$HeT$^+$. The TRIMS results agree closely with the theoretical calculations for both HT and TT parent molecules~\cite{trims:2020}.

The KATRIN spectrum provides a unique sensitivity to an observable that has never yet been measured: the onset to the electronic excited states. This observable is of interest beyond the neutrino-mass community. Given the prevailing abundance of H and He in the universe, $^4$HeH$^+$ -- which is described using the same theoretical framework as its isotopolog $^3$HeT$^+$ -- played an important role in the chemistry of the early universe, and has recently been detected for the first time in an astrophysical source, in the planetary nebula NGC 7027 viewed by SOFIA~\cite{Gusten:2019njc}. Although ample spectroscopic data are available for the electronic ground states of this molecule (reviewed in Refs.~\cite{Jonsell1999, Bodine2015}), the electronic excited states have a purely repulsive character in the Franck-Condon window and are very difficult to probe through traditional spectroscopic methods. The available experimental data on these states are very limited~\cite{PhysRevLett.98.223202,PhysRevA.82.023415,PhysRevLett.121.073203}.

KATRIN's sensitivity to the electronic excited states arises from the theoretical prediction of a separation of about \SI{20}{\electronvolt} between the highest-lying states of the electronic ground-state manifold, and the first electronic excited state. This relatively sudden onset should cause a distortion in the spectral shape, through which KATRIN can measure the first electronic excited state's location and amplitude. 

\subsection{Precision spectroscopy of \textsuperscript{83m}Kr conversion-electron lines}
\label{Subsec:83mKrSpectroscopy}

KATRIN is not limited to tritium spectroscopy, but can in principle be used to measure other electron spectra with comparable energies. In fact, the first measurement of a gaseous source at KATRIN was in 2017 with the \textsuperscript{83m}Kr calibration source described in Sec.~\ref{Subsubsec:Intense83mKrCalibrationSource}. These repeated measurements are performed for their crucial contributions to the understanding of important systematic effects, but they also allow spectroscopy of the source; some results of the 2017 measurement campaign have already been published~\cite{Altenmueller2020:Krspectr,PhDSeitz2019}. This section discusses the possibility of further spectroscopic measurements, based on the experience of several \textsuperscript{83m}Kr campaigns since 2017; however, some findings should be transferable to the spectroscopy of other nuclides.

Isomeric \textsuperscript{83m}Kr decays via two consecutive $\upgamma$ transitions with energies $E_\upgamma$ of roughly \num{32.2} and \SI{9.4}{\kilo\electronvolt}. The direct transition with an energy of \SI{41.6}{\kilo\electronvolt} is suppressed but possible, and KATRIN has observed clear indications of this transition. In addition to $\upgamma$ emission the transition can also occur by internal conversion, where the transition energy is transferred to an electron in the atomic shell $s$, which is then emitted. The ratio of emitted electrons to $\upgamma$ photons is called the conversion coefficient $\alpha^{(s)}$. To calculate the energy $E_\mathrm{ce}^{(s)}$ of the emitted electron, the binding energy $E_\mathrm{ce}^{(s)}(\mathrm{bind})$ of the shell and the recoil energy of the nucleus ($E_\gamma(\mathrm{recoil})$ and $E_\mathrm{ce}(\mathrm{recoil)}$) need to be subtracted from the transition energy, giving~\cite{Venos2018}
\begin{equation}
    E_\mathrm{ce}^{(s)}=E_\gamma+E_\gamma(\mathrm{recoil})-E_\mathrm{ce}^{(s)}(\mathrm{bind})-E_\mathrm{ce}(\mathrm{recoil)}~.
    \label{Eq:KrConversionElectronEnergy}
\end{equation}
In X-ray notation the electron shell is given as $s\in\{\mathrm{K},\mathrm{L}_{l_1},\mathrm{M}_{l_2},\mathrm{N}_{l_3},~\dots\}$, where the capital letters $\{\mathrm{K},\mathrm{L},\mathrm{M},\mathrm{N},~\dots\}$ denote the principal quantum number and the index $l_i$ orbitals of different angular momentum.

The binding energies vary strongly between the shells, ranging from around \SI{14}{\kilo\electronvolt} for the innermost K-shell electrons to \SI{14}{\electronvolt} for the outermost N-shell electrons. The recoil energies are very small in comparison. Consequently, the electron spectrum is divided into several line groups. An overview of the \num{32.2}\,keV \textsuperscript{83m}Kr conversion spectrum is shown in Table~\ref{tab:KryptonSpectrum}.

All electron conversion lines of the \SI{9.4}{\kilo\electronvolt} transition, and the K conversion line of the \SI{32}{\kilo\electronvolt} transition, are below the tritium $\upbeta$-spectrum endpoint of \SI{18.6}{\kilo\electronvolt} and are therefore hidden in the $\upbeta$ spectrum, when the source operates at the nominal tritium activity. This puts a first constraint on the spectroscopy of those lines, which would require a tritium-free source, possibly including a cleaning of residual tritium from the rear-wall surface. Consequently, only the conversion electrons of the \SI{32.2}{\kilo\electronvolt} transition have been used in most of the \textsuperscript{83m}Kr campaigns since the first exposure with tritium.

\begin{table}
    \centering
    \caption{Mean line position $E_\mathrm{ce}$, width $\mathrm{\Gamma}$ and conversion coefficient $\alpha$ of the 32.2\,keV transition of the \textsuperscript{83m}Kr conversion spectrum (values from Ref.~\cite{Venos2018}).}
    \label{tab:KryptonSpectrum}
\begin{tabular}{lccc}
\hline
Line		&	$E_\mathrm{ce} (\mathrm{eV})$	&	$\mathrm{\Gamma (eV)}$	&	$\alpha$	\\
\hline
K				&	17824.2(5)			&	2.71(2)			&	478.0(50)	\\
$\mathrm{L}_1$	&	30226.8(9)			&	3.75(93)		&	31.7(3)		\\
$\mathrm{L}_2$	&	30419.5(5)			&	1.25(25)		&	492.0(50)	\\
$\mathrm{L}_3$	&	30472.2(5)			&	1.19(24)		&	766.0(77)	\\
$\mathrm{M}_1$	&	31858.7(6)			&	3.5(4)			& 	5.19(5)		\\
$\mathrm{M}_2$	&	31929.3(5)			&	1.6(2)			&	83.7(8)		\\
$\mathrm{M}_3$	&	31936.9(5)			&	1.1(1)			&	130.0(13)	\\
$\mathrm{M}_4$	&	32056.4(5)			&	0.07(2)			&	1.31(1)		\\
$\mathrm{M}_5$	&	32057.6(5)			&	0.07(2)			&	1.84(2)		\\
$\mathrm{N}_1$	&	32123.9(5)			&	0.40(4)			&	0.643(6)	\\
$\mathrm{N}_2$	&	32136.7(5)			&	0.03			&	7.54(8)		\\
$\mathrm{N}_3$	&	32137.4(5)			&	0.03			&	11.5(1)		\\
\hline
\end{tabular}
\label{tab:KrypSpec}
\end{table}

The analysis of the krypton spectrum needs no less attention than that of the tritium spectrum, especially since all measurements of the gaseous source are exposed to the same source systematic effects. This is exactly what one makes use of in calibration measurements: If it is given that the experimental conditions are the same and that one knows the krypton spectrum very well, one can use the krypton data to determine systematic effects and take them into account in the modeling of the tritium measurement. One of the systematic effects which the krypton measurement aims to quantify is the electric starting potential of the source, whose mean value shifts the position of the selected krypton line and whose inhomogeneity broadens the line. Conversely, one can only determine the absolute positions and widths of the krypton lines well if one knows the source potential well.

Thus, for spectroscopic measurements the electric potential of the source is a systematic effect. When deciding which lines to use for calibration or for purely spectroscopic measurements, two additional systematic effects must be considered in addition to typical indicators such as the signal-to-background ratio:
\begin{itemize}
    \item \textbf{Inelastic scattering:} It is usually advisable to use a carrier gas in addition to the krypton to achieve high rates and rate stability. Moreover, the usage of tritium also leads to a smaller inhomogeneity of the electric potential of the source, since the $\upbeta$ decay creates a plasma which shields inhomogeneities of the source-tube work function. On the other hand, inelastic scattering of the krypton conversion electrons on the carrier gas produces a shape distortion of the krypton spectrum starting typically around \SI{10}{\electronvolt} below the line, such that only the electron conversion lines of outer shells can be measured without precise modeling of the scattering.
    \item \textbf{Background slope:} Due to the integral measurement principle of KATRIN, the conversion spectra of higher-shell electrons contribute to the background in any spectrum measurement of the inner-shell conversion electrons of \textsuperscript{83m}Kr. Since the transmission conditions are non-adiabatic for high surplus energies, leading to a loss of rate, this background can have an energy slope, which needs to be modeled precisely.
\end{itemize}
For the regular calibration measurements the N$_{2,3}$-32 doublet is typically chosen for several reasons: 
since the N$_{2,3}$ electrons belong to the outermost atomic shell of krypton, there is no background contribution from other conversion electrons from the \SI{32.2}{\kilo\electronvolt} transition.
Also, since there is no filling up of the N$_{2,3}$ vacancies after emission of the conversion electrons from outer shells, they have a long lifetime and the lines therefore have a quasi-vanishing width. The ratio of the line width and the standard deviation of the source potential dictates which uncertainties dominate; for the N$_{2,3}$-32 doublet the $\mathcal{O}$(\SI{10}{\milli\volt}) measured value of the standard deviation of the source potential is much larger than the line width, such that uncertainties of the width have a minor effect on the determination of the standard deviation. For all other lines of the krypton spectrum this ratio is reversed, which means that they are not optimally suited for determining the source potential. However, KATRIN has the capability to determine their line widths even given uncertainties in the modeling of the source potential.

The determination of the absolute line positions requires knowledge of the absolute electric potential of the source, which in turn is calibrated using \textsuperscript{83m}Kr measurements. Conversely, the determination of the absolute electric potential is limited by the literature uncertainties of the line positions. These uncertainties arise from uncertainties in the binding energies of krypton and the uncertainty on the $\upgamma$ transition energy (Eq.~\ref{Eq:KrConversionElectronEnergy}), with the latter strongly dominating. KATRIN plans to decrease this uncertainty using the excellent linearity of the KATRIN spectrometer~\cite{Rest2019}: by determining the relative positions of lines of the conversion electrons emitted from the same atomic subshells of all three $\upgamma$ transitions (\num{9.4}, \num{32.2}, \SI{41.6}{\kilo\electronvolt}) KATRIN can measure the energy with high precision~\cite{RodenbeckKr83m:2022}.

The determination of the absolute activity or half-life requires precise modeling of both the gas dynamics and emanation of the \textsuperscript{83m}Kr from the rubidium generator. Due to the continuous supply of \textsuperscript{83m}Kr from the generator, it is in quasi-equilibrium and only the half-life of rubidium is actually observed. The expected and measured krypton activity agree within uncertainties~\cite{Marsteller2022}, but the modeling is not tailored for precise determination of the absolute activity.

Using a $^{83\rm{m}}$Kr source in the \si{GBq} range, KATRIN is capable of achieving a statistical sensitivity to line positions of a few \si{\milli\electronvolt}, to line width in the \SI{10}{\milli\electronvolt} range and to intensities in the per mil range in a few hours, depending on the chosen lines and mode of \textsuperscript{83m}Kr operation. In the literature, the uncertainty of the absolute line positions is in the \SI{500}{\milli\electronvolt} range (dominated by the uncertainty of the $\upgamma$ energy), and that of the relative line positions within one $\upgamma$ transition in the \SI{40}{\milli\electronvolt} range (dominated by the uncertainties of the binding energies), while the uncertainties of the line widths are in the \SI{100}{\milli\electronvolt} range~\cite{Venos2018}. Thus, KATRIN could in principle significantly improve these values.

In summary, using KATRIN for spectroscopy of \textsuperscript{83m}Kr (or other electron spectra beyond tritium) is possible but it requires an assessment of systematic effects which is specific for each chosen line. For the N$_{2,3}$-32 doublet, measurements are performed regularly and the relative intensity and position of N$_2$ and N$_3$ can be obtained by KATRIN with unprecedented sensitivity. In addition, a measurement of the N$_1$-32 line region was performed in 2021 using helium as carrier gas. The latter requires at least \SI{21.2}{\electronvolt} energy for incident electrons to scatter inelastically, so that scattered electrons from the N$_{2,3}$-32 doublet appear only below the N$_1$-32 line. Thus, while the current application of \textsuperscript{83m}Kr measurements is focused on reaching KATRIN's physics goal, some spectroscopic byproducts have already been obtained and will be part of a future publication.

% ############################################################
% R&D for future of KATRIN
%# ##########################################################
\section{R\&D for the future of KATRIN}
\label{Sec:R&DForTheFutureOfKatrin}

\begin{figure}[tb]
    \centering
    \includegraphics[width=1.0\textwidth]{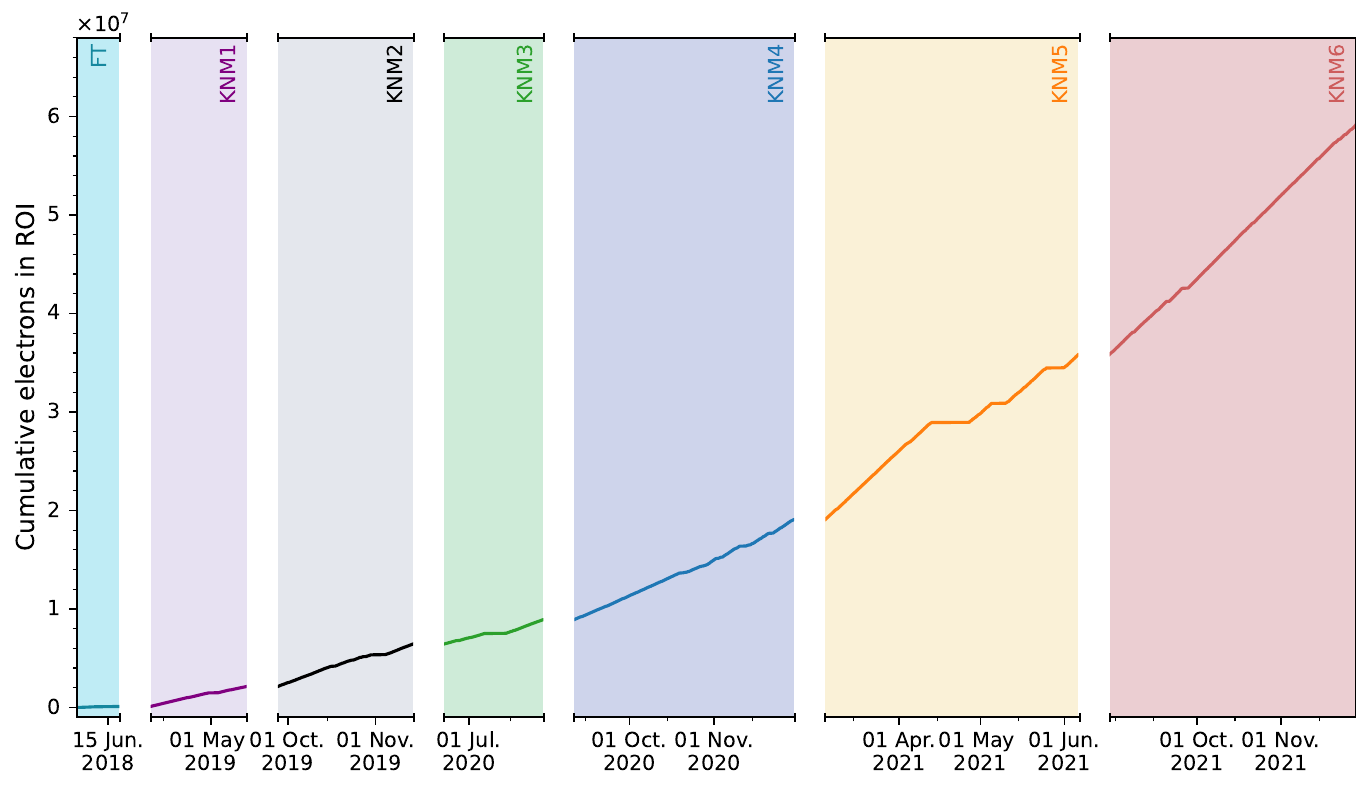}
    \caption{Integrated luminosity in the analysis interval since the beginning of tritium data-taking with KATRIN (Sec.~\ref{Subsec:OperationalHistory}). FT denotes the First Tritium campaign; each neutrino-mass measurement campaign is numbered sequentially in the format KNMx . White gaps correspond to scheduled maintenance periods.}
    \label{fig:integratedlumi}
\end{figure}

In 2005, the KATRIN design report set a sensitivity goal of \SI{0.2}{\electronvolt} at \SI{90}{\percent} confidence~\cite{KATRIN2005}. This goal was based on an assumption of three full years of running, and a set of projections about achievable backgrounds and systematic uncertainties. KATRIN has now successfully acquired neutrino-mass data during three calendar years (Fig.~\ref{fig:integratedlumi}). Several systematics once regarded as especially difficult are now exceptionally well controlled, e.g. the description of $\upbeta$ energy loss in the source; others, not originally anticipated, are now recognized as particularly important (Sec.~\ref{Subsec:SourcesOfSystematicUncertainty}). Meanwhile, KATRIN's designed countermeasures against the dominant backgrounds of prior generations, including cosmic-ray muons~\cite{Altenmueller2018} and environmental gammas~\cite{Altenmuller2019}, have been largely successful -- but previously unforeseen backgrounds (Sec.~\ref{Subsec:Backgrounds}) far exceed the original design budget. 

In this section, we discuss R\&D efforts -- some quite early, and some advanced -- with the goal of improving KATRIN's sensitivity to $m_\nu^2$ and other physics observables. One of the foremost efforts is toward background reduction (Sec.~\ref{Subsec:BackgroundMitigation}). Several countermeasures have already been successfully applied (Sec.~\ref{Subsec:Backgrounds}), including more frequent regeneration of liquid-nitrogen-cooled baffles on the main-spectrometer pump ports; reduction of the pre-spectrometer operating voltage; and operation of the main spectrometer with a shifted analyzing plane~\cite{lokhov2022background} that more than  halves the spectrometer volume imaged by the detector (Sec.~\ref{Subsubsec:MACEfilter}). In addition, KATRIN is investigating both active (Sec.~\ref{Subsubsec:ActiveTransverseEnergyFilter}) and passive (Sec.~\ref{Subsubsec:PassiveTransverseEnergyFilter}) transverse filters, to distinguish background Rydberg electrons from signal $\upbeta$s. \si{\tera\hertz} radiation is a possible means of actively de-exciting Rydberg atoms, reducing the number of background electrons produced (Sec.~\ref{Subsubsec:DeexcitationOfRydbergAtomsWithTHzRadiation}). Electron taggers (Sec.~\ref{Subsubsec:ElectronTaggers}) could flag electrons entering the main spectrometer, allowing straightforward rejection of backgrounds originating inside the main spectrometer.

An increase in the KATRIN acceptance angle would improve statistics for any given amount of running time; investigations are underway to establish that this would not result in an unacceptable worsening of systematics (Sec.~\ref{Subsubsec:AcceptanceImprovements}). A further statistics gain could be realized by increasing the size of the analysis interval by reducing the lower limit of energies considered to, for example, $E_0 -$\SI{60}{\electronvolt} from the present $E_0 -$\SI{40}{\electronvolt}; this requires establishing high confidence in the models for electron energy loss and the molecular final-state distribution (Sec.~\ref{Subsec:SourcesOfSystematicUncertainty}). 

Dedicated searches for \si{\kilo\electronvolt}-scale sterile neutrinos (Sec.~\ref{sec:keV}) require upgrades to the KATRIN apparatus. To scan deeper into the spectrum, the rear wall must be redesigned to reduce backscattering (Sec.~\ref{Subsubsec:ModifiedRearWall}). Significantly higher rates will require a new detector system, which we call TRISTAN (Sec.~\ref{Subsubsec:TristanDetector}).

\subsection{Background mitigation}
\label{Subsec:BackgroundMitigation}

After the successful implementation of initial measures to reduce the Rydberg background component~\cite{lokhov2022background} (Sec.~\ref{Subsubsec:MACEfilter}), KATRIN's background is still an order of magnitude higher than anticipated. Therefore we are investigating several methods to reduce the background further.

\subsubsection{Active transverse-energy filter}
\label{Subsubsec:ActiveTransverseEnergyFilter}

\begin{figure}[htb]
    \centering 
    \includegraphics[width=0.6\textwidth]{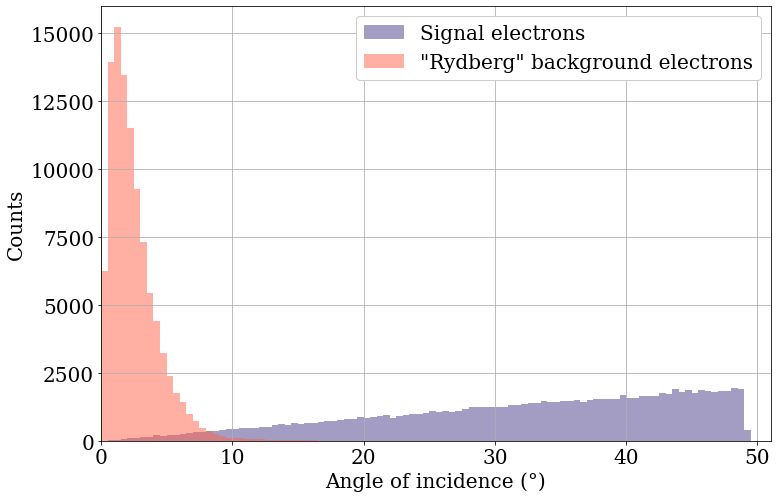}
    \caption{Simulated angular distributions of signal (purple) and ``Rydberg'' background electrons (orange) at the detector of the KATRIN experiment simulated for \SI{1e5} electrons using input from~\cite{PhDTrost2019}. The simulation considers the energy spectrum of background electrons that are liberated by room-temperature blackbody radiation from hydrogen (\SI{36}{\percent}) and oxygen atoms (\SI{64}{\percent}) in highly excited Rydberg states. These Rydberg atoms are sputtered from the spectrometer walls % with a $\cos(\theta)$-distribution 
    due to $\upalpha$-decays of implanted $^{210}$Pb progenies from the $^{222}$Rn decay chain~\cite{Fraenkle2022} (Sec.~\ref{Subsec:Backgrounds}). Adapted with permission from Ref.~\cite{atef_paper}.}
    \label{fig:background_at_KATRIN} 
\end{figure}

As described in Sec.~\ref{Subsec:Backgrounds}, our evidence suggests that a significant share of the background electrons in the KATRIN spectrometer originate from highly excited Rydberg atoms, which are ionized by the blackbody radiation emitted from the spectrometer walls at room temperature~\cite{Fraenkle2022}. These secondary electrons possess very low starting energies~\cite{PhDTrost2019}, but are accelerated by the electric-potential gradient of the spectrometer and arrive with essentially the same kinetic energy at the detector as the signal electrons. However, since the initial transverse energy of the electrons from this process is only ${\cal O}(k_B T \approx$ \SI{25}{\milli\electronvolt}) and they do not acquire transverse energy in the electric-field gradient towards the exit of the spectrometer due to the strong magnetic field there, they reach the detector with incidence angles of typically less than 10° in contrast to the signal electrons which fill the phase space up to the maximum angle of 51° allowed by the pinch magnet (Fig.~\ref{fig:background_at_KATRIN}).

The idea of the active Transverse Energy Filter (aTEF) is to utilize the different angular distributions of signal and background electrons at the detector to further reduce the background rate. Standard methods to differentiate different angles of incidence, like low-density tracking detectors (e.g. gas-filled drift chambers) or a $\Delta E$-$E$ arrangement, are not applicable because the electrons in the strong magnetic field at the KATRIN detector possess sub-\si{\milli\metre} cyclotron radii and ranges of a few \si{\micro\metre}. In addition, a gas-filled detector with a thin entrance window is not ultra-high-vacuum-compatible. Therefore, we propose to geometrically distinguish the impact angles %or the transverse energy of the particles 
via the cyclotron radius of the spiraling motion around the magnetic field lines. At first glance a suitable device for this purpose could be a micro-channel plate (MCP) which we will thus use to explain the idea~\cite{atef_paper}. 

\begin{figure}[htb]
    \centering 
    \includegraphics[width=0.6\textwidth]{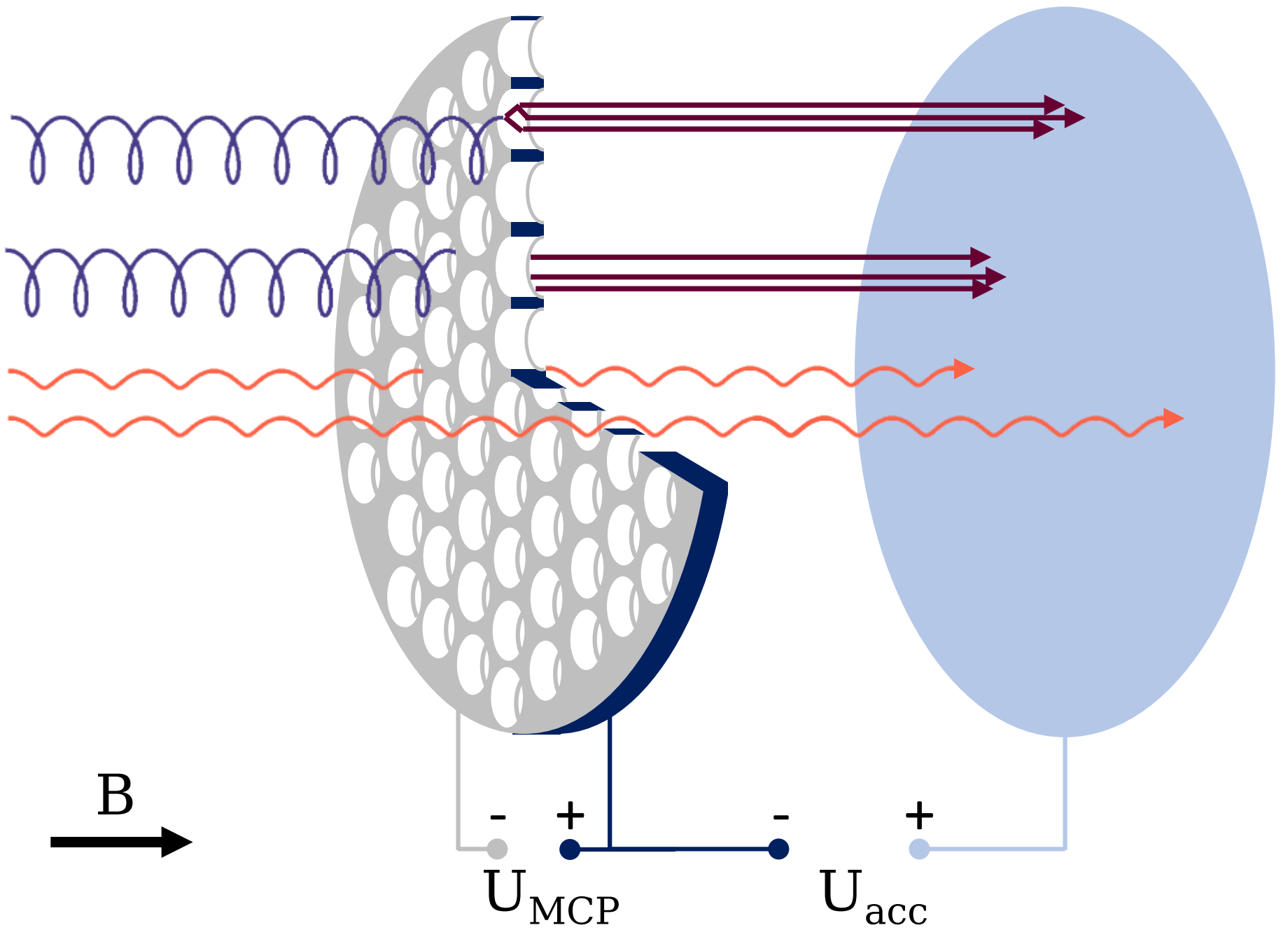}
    \caption{Principle of an MCP-based active transverse energy filter in front of the KATRIN detector (light blue) in a horizontal magnetic field $B$. Electrons with a small cyclotron radius (orange) pass the MCP without interaction and hit the detector.  Electrons with a large cyclotron radius (purple) hit the MCP walls, giving rise to a cascade of secondary electrons (dark red) that is accelerated via $U_\mathrm{acc}$ towards the detector. Since the secondary electrons are produced with practically no transverse momentum, they follow the magnetic-field lines without a sizable cyclotron radius. Adapted with permission from Ref.~\cite{atef_paper}.}
 \label{fig:mcp_atef_principle} 
\end{figure}

Typically, when a charged particle such as an electron or ion 
%(or even neutral particles such as atoms or photons) 
hits the side walls of a channel within the MCP, it triggers the emission of several secondary electrons from the wall, which is coated with a material with a particularly high secondary-electron yield. These secondary electrons are accelerated by the electric field applied across the channel, hit the wall again and, thus, trigger a secondary-electron cascade~\cite{Wiza1979}. In commercial MCPs, the channels with diameter $d$ are tilted by \SIrange{8}{15}{\degree}~\cite{Wiza1979} with respect to the surface normal. We deliberately neglect this tilt angle of the channels in Fig.~\ref{fig:mcp_atef_principle} to illustrate the principle of our idea. 

If an electron with a small cyclotron radius $r_\mathrm{c} \ll d/2$ (orange spiral track) enters a channel of the MCP, it usually passes through unaffected and causes a small signal in the detector (here in light blue). If, on the other hand, an electron with a large cyclotron radius $r_\mathrm{c} > d/2$ (purple spiral track) enters a channel, the particle will most likely hit the channel walls and trigger a secondary-electron avalanche (dark red arrows). An applied voltage $U_\mathrm{acc}$ will accelerate the electron avalanche towards the detector where it will cause a comparably larger signal. When we discriminate on the measured signal height, the channels thus act as a ``transverse energy filter''. Given that the filter is not a passive element but aids in the detection of the signal electrons, we name this principle ``active Transverse Energy Filter'' (aTEF)~\cite{atef_paper}.

Of course a standard MCP does not meet the KATRIN requirements. First, we have to adapt the diameter-to-length ratio of the channels to allow background electrons with small transverse energy to pass the filter unhindered while most of the signal electrons are detected. A channel bias angle is disfavored, unlike in typical commercial MCPs where it is mainly used to reduce ion feedback~\cite{vavra2004}. In contrast, the channels for the proposed aTEF have to be parallel to the magnetic-field lines. Secondly, the open-area  ratio (OAR) needs to be maximized to reduce losses in non-back\-ground events due to signal electrons being blocked by the material in between the channels. Using a hexagonal channel shape instead of the circular shape of regular MCPs allows much larger open-area ratios. To overcome the inherent dark-count rate of the glass materials usually used for MCP production, a radiopure material like silicon with an intrinsic low background rate could alternatively be used. The targeted geometry can be manufactured in silicon via a highly anisotropic cryo-etching procedure for deep silicon structures. 

There are obvious disadvantages of an MCP-based aTEF for our application, like the low secondary-electron yield of most materials for \SI{18.6}{\kilo\electronvolt} electrons or the sensitivity to the very strong magnetic field of \SI{2.4}{\tesla}. We do not discuss these here, since the aTEF idea is more general. This principle can be realized on any kind of geometric aperture that is instrumented with a suitable method of particle detection. A generally suitable aTEF-aperture geometry is a honeycomb structure due to its ideal open-area ratio. The instrumentation of such a honeycomb structure could, for instance, also be realized using scintillator or semiconductor materials to generate photons or electron-hole pairs and thereby detectable signals. The advantage of the latter is that the energy of the signal electrons could still be determined, an interesting feature for suppression of other kinds of backgrounds. 

\begin{figure}[tb]
    \centering 
    \includegraphics[width=0.6\textwidth]{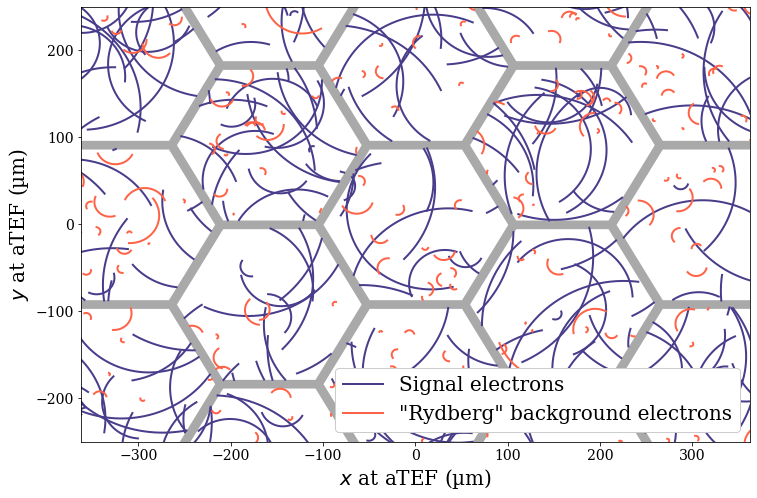}
    \caption{Simulation of signal (purple) and background (orange) electron tracks in a \SI{2.4}{\tesla} axial magnetic field with the angular distributions shown in Fig.~\ref{fig:background_at_KATRIN} for a honeycomb-like aTEF structure with a side length of \SI{100}{\micro\metre}, a wall thickness of \SI{10}{\micro\metre}, and a depth of \SI{400}{\micro\metre}. The tracking of an electron is stopped when it either hits the aTEF surface or when it reaches the full depth without any hit. Adapted with permission from Ref.~\cite{atef_paper}.}
    \label{fig:aTEF_simulation} 
\end{figure}

A Monte Carlo simulation of electron tracks through a filter design with hexagonal channels, a side length of \SI{100}{\micro\metre}, a channel wall thickness of \SI{10}{\micro\metre} and a depth of \SI{400}{\micro\metre}  is illustrated in Fig.~\ref{fig:aTEF_simulation} (see Fig.~\ref{fig:pTEFGeo} for definitions of these parameters). Electrons are generated with angles drawn from the two angular distributions shown in Fig.~\ref{fig:background_at_KATRIN}, relative to the \SI{2.4}{\tesla} magnetic field  at the entrance of the three-dimensional hexagonal structure. Corresponding to the angular distribution used, the simulated electrons are either of Rydberg-background type (orange tracks) or of signal type (purple tracks). The magnetic-field lines and, therefore, the guiding centers of the electron's motion are directed perpendicular to the viewing plane. The tracking stops when the electron either hits the wall or reaches the full depth of the channel of \SI{400}{\micro\meter}. Electrons are counted as ``detected'' in the former and as ``not detected'' in the latter case. The percentage of detected electrons differs noticeably between the two types of electrons: \SI{90}{\percent} of the signal electrons and \SI{11}{\percent} of Rydberg background electrons interact with the wall. Taking into account an open-area-ratio of 90\%, we therefore retain \SI{81}{\percent} of the signal events and \SI{10}{\percent} of the background events. Research and development on this concept is continuing.

\subsubsection{Passive transverse-energy filter}
\label{Subsubsec:PassiveTransverseEnergyFilter}
%responsible Dominic Hinz

\begin{figure}[tb]
    \centering
    \includegraphics[width=0.7\textwidth]{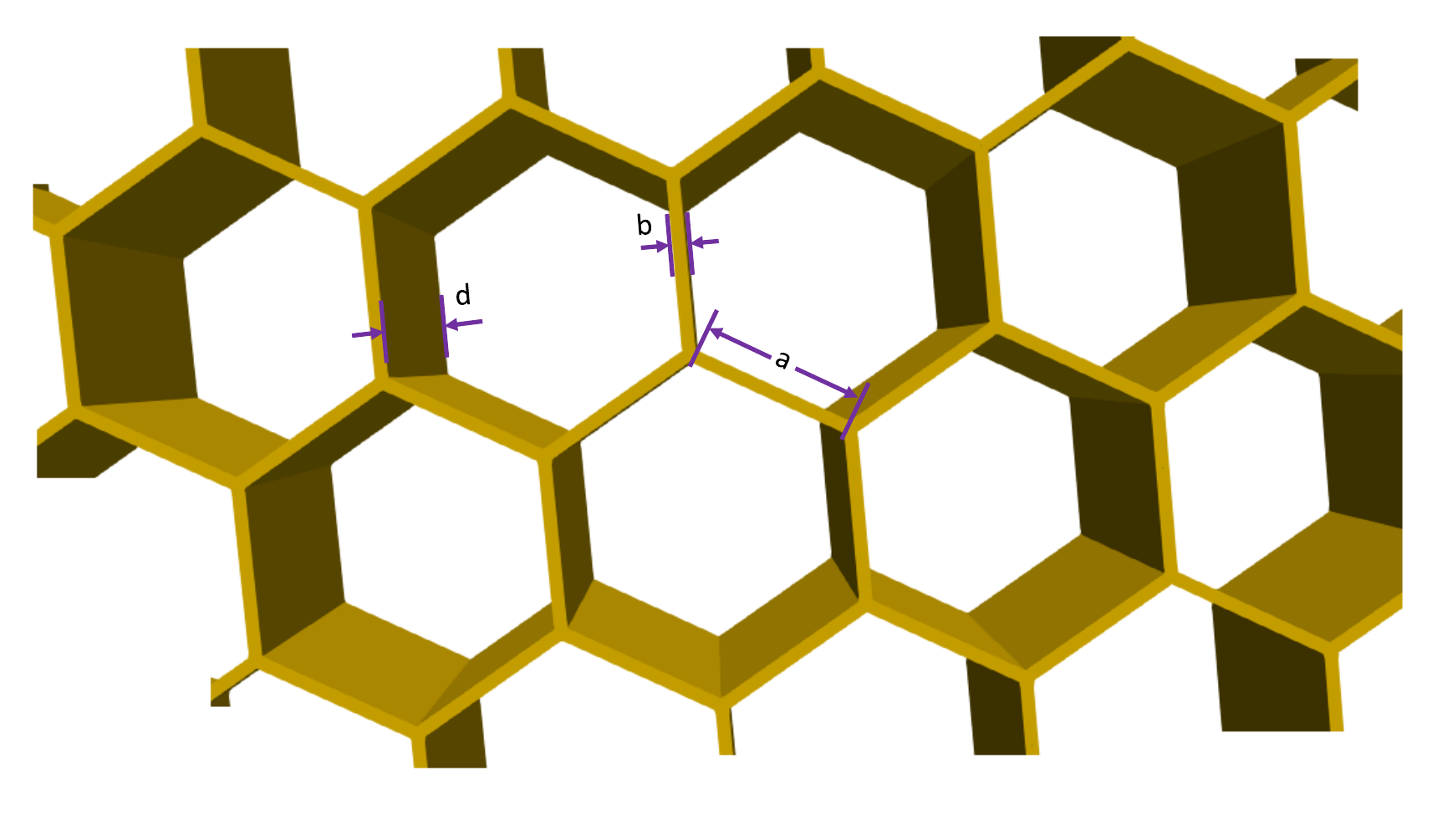}
    \caption{Cutout of the geometric pTEF model as simulated in Kassiopeia. The hexagonal channels are described with the parameters side length $a$, wall thickness $b$, and depth $d$.}
    \label{fig:pTEFGeo}
\end{figure}

\begin{figure}[tb]
    \centering
    \includegraphics[width=.9\textwidth]{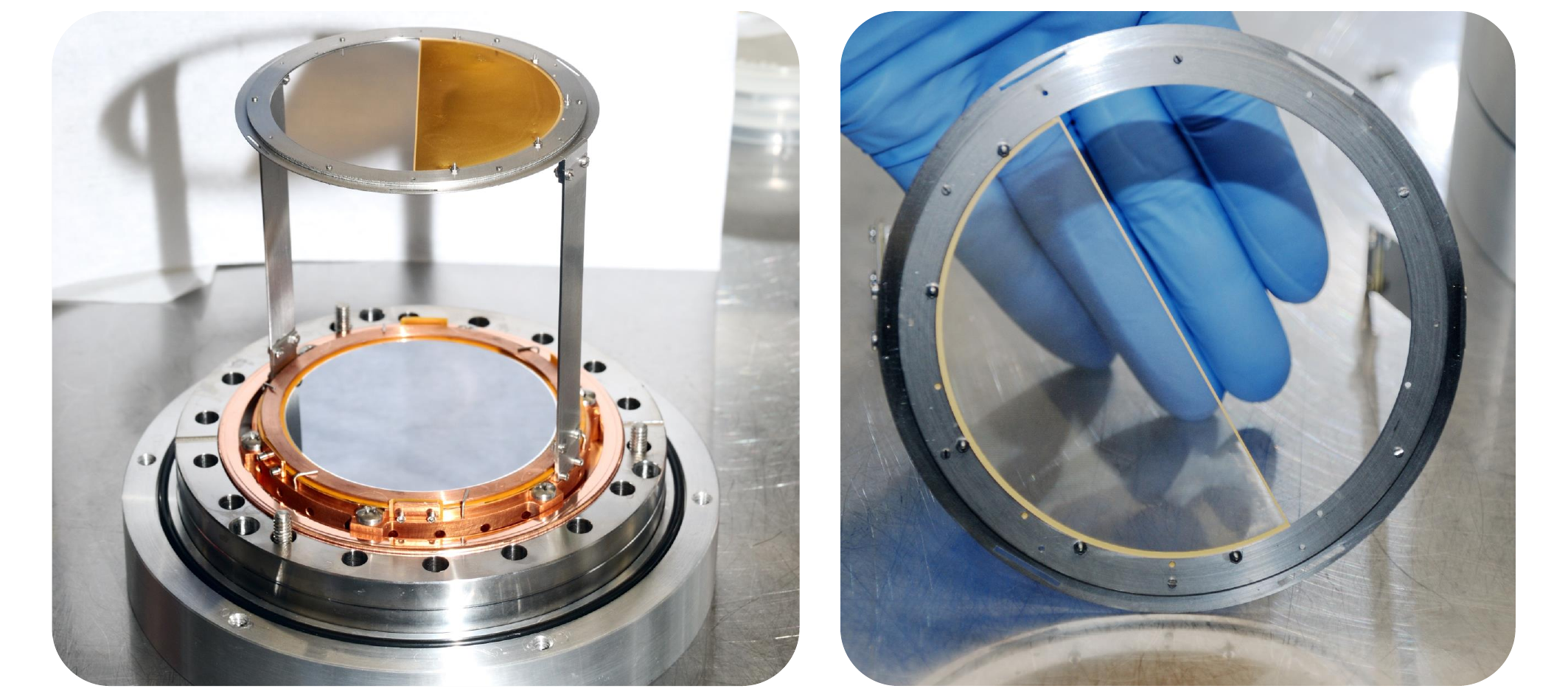}
    \caption{\textbf{Left:} pTEF (gold plate) mounted with a dedicated holding structure on the focal-plane-detector wafer flange before installation into the KATRIN beamline. \textbf{Right:} pTEF installed in holding structure.}
    \label{fig:MountefpTEF}
\end{figure}

To inform the development of the aTEF (Sec.~\ref{Subsubsec:ActiveTransverseEnergyFilter}), we have built a passive transverse-energy filter (pTEF) to measure the angular distribution of the background electrons, testing the prediction in Fig.~\ref{fig:background_at_KATRIN}. The pTEF is a gold semicircle with a lithographed, hexagonal honeycomb structure, designed to block electrons within a certain range of transverse kinetic energy. This semicircle is installed to shadow one half of the detector while the other half is left clear (Fig.~\ref{fig:MountefpTEF}). In this way, different pitch-angle distributions are distinguishable based on the rate on the respective detector sections. The geometry (Fig.~\ref{fig:pTEFGeo}) is described by three parameters: the side length $a$, wall thickness $b$ and depth $d$. These parameters were optimized in terms of signal-to-background ratio, based on simulations of the KATRIN beamline performed with Kassiopeia~\cite{Furse2017} code for tracking charged particles in varying electromagnetic fields. Additional parameters, such as diameter or open-area ratio (OAR), are derived from these. The values of the pTEF design are: 
\begin{equation}
    a = \SI{100}{\micro\metre},~~
    b = \SI{8}{\micro\metre},~~ 
    d = \SI{250}{\micro\metre},~and~~ 
    \mathrm{OAR} = \SI{91.4}{\percent}. \label{OAR}
\end{equation}

\begin{figure}[tb]
    \centering
    \includegraphics[width=0.9\textwidth]{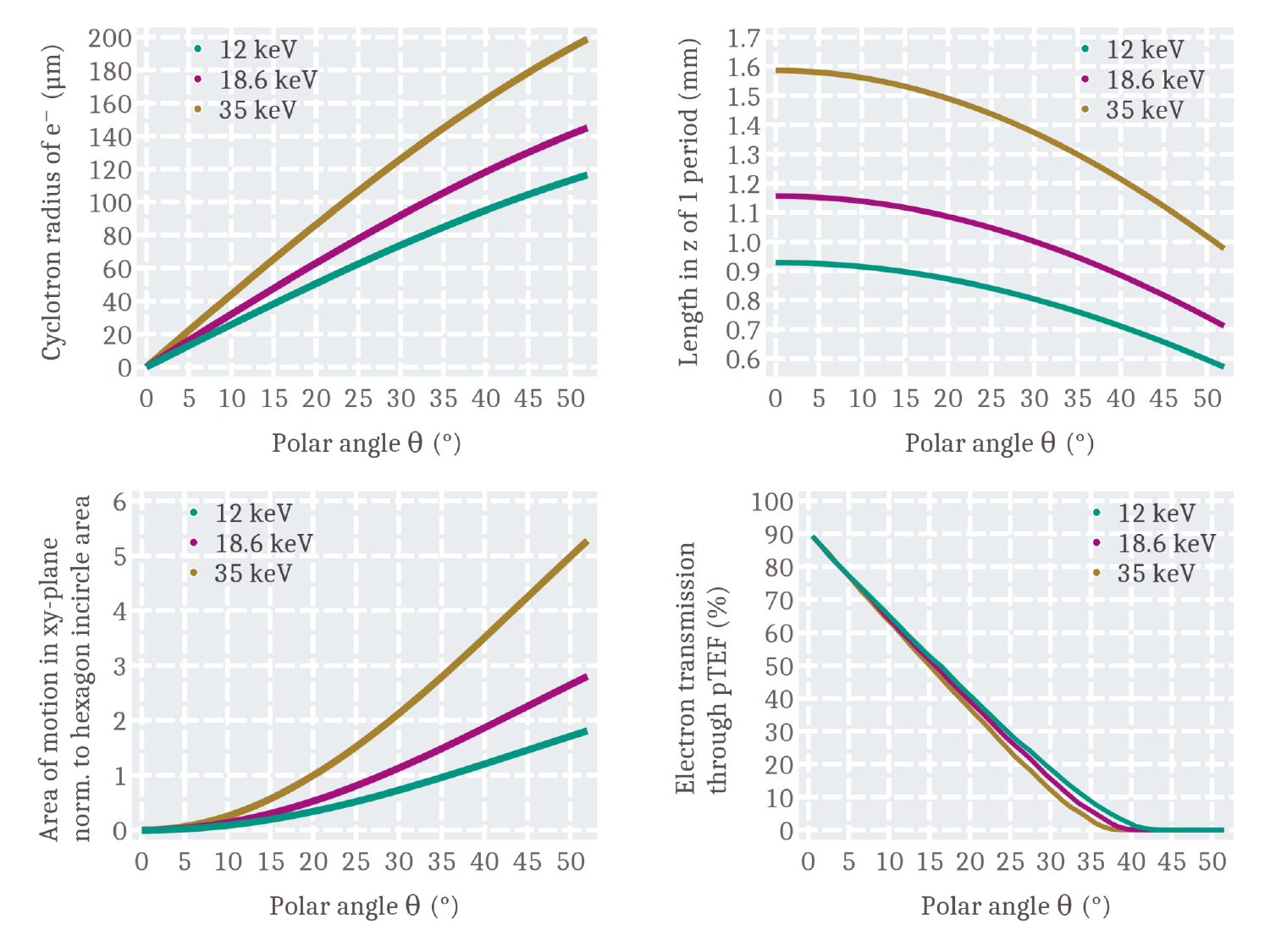}
    \caption{Dependence of relevant quantities on pitch angle $\theta$ for electron energies of \num{12}, \num{18.6}, and \SI{35}{\kilo\electronvolt} at \SI{2.5}{\tesla} magnetic field. The minimum and maximum energies give the largest spread that can be measured in a practical setup. \textbf{Top left:} Cyclotron radius. \textbf{Top right:} Axial path length of one full circular motion. \textbf{Bottom left:} Area of circular motion in $xy$ plane, normalized to the area of a circle inscribed in a hexagonal pTEF channel. \textbf{Bottom right:} Simulation of electron transmission through a hexagonal micro-structured plate. Refer to Fig.~\ref{fig:background_at_KATRIN} for the predicted angular distributions of signal and background electrons.}
    \label{fig:pTEF-electrons}
\end{figure}

\vspace{1pt}\hspace{1pt}\\
The electron motion can be simplified to a two-dimensional case. Electrons with non-zero pitch angles circulate around magnetic-field lines due to the Lorentz force. If we let the motion and magnetic-field lines lie along the beam axis ($z$), the circular motion leads to a circle in the $xy$ plane. Therefore, the electron direction has components parallel and perpendicular to $\vec{B}$. The size of the circular motion is described by the cyclotron radius 
\begin{equation}
    r_{c} = \frac{\gamma\cdot m\cdot v_\perp}{\lvert q \rvert \cdot B}, 
\end{equation}
and its frequency $\omega = q\cdot B / m$, where $B$ is the magnetic field strength, $q$ the charge and $m$ the mass of the particle, $\gamma$ the Lorentz factor, and $v_\perp$ the velocity perpendicular to the magnetic field. The pitch angle $\theta$, between the electron velocity and the magnetic-field line, gives the relative size of the $v_\perp$ component. The energy due to transverse motion, along with the cyclotron radius, is higher for larger $\theta$. Figure~\ref{fig:pTEF-electrons} shows the dependence of various parameters on $\theta$ for selected electron energies. The cyclotron radius as a function of the polar angle $\theta$ is proportional to $\sin\theta$ (top left). The top right shows the $\cos\theta$ dependence of the projection of the electron path along the $z$-axis, during one period of circular motion. For energies relevant to KATRIN, these path lengths always exceed the \SI{250}{\micro\meter} pTEF depth. By comparing the electrons' area of motion in the $xy$-plane to the area of the circle inscribed in a hexagonal cell (bottom left), we find that for \SI{18.6}{\kilo\electronvolt} the area of motion already exceeds the open area for $\theta=$~\SI{27}{\degree}. However, the depth of the filter is insufficient for a full period of circular motion, so electrons with even higher pitch angles can be transmitted. Figure~\ref{fig:pTEF-electrons} (bottom right) shows a simulation of electron transmission through a hexagonal filter. For each \SI{1}{\degree} angular bin, \num{50000} electrons were simulated and counted as transmitted if they passed the filter without striking its surface. Electrons with $0^\circ$ pitch are not \SI{100}{\percent} transmitted since the boundaries between the hexagons fill about \SI{8.6}{\percent} of the area. Small variations of the transmission occur between different electron kinetic energies, resulting from a combination of the different cyclotron radii and paths along $z$. 

A multifaceted measurement campaign was performed in winter 2021, including detailed investigations into transmission through the pTEF and therefore into the $\theta$ distribution of the background electrons generated in the spectrometers. The results are under analysis.

\subsubsection{De-excitation of Rydberg atoms with \si{THz} radiation}
\label{Subsubsec:DeexcitationOfRydbergAtomsWithTHzRadiation}
% Responsible: Enrico Ellinger (BUW)

An alternate approach to mitigate the Rydberg background is to manipulate the atomic states of the Rydberg atoms so as to reduce the ionization process by thermal radiation (Sec.~\ref{Subsec:Backgrounds}). This method was pioneered by the CERN antihydrogen group~\cite{Wolz:2020, VielleGrosjean:2021}. KATRIN is currently investigating this idea via numerical simulations. 

\begin{figure}[tb]
    \centering
     \includegraphics[width=0.55\textwidth]{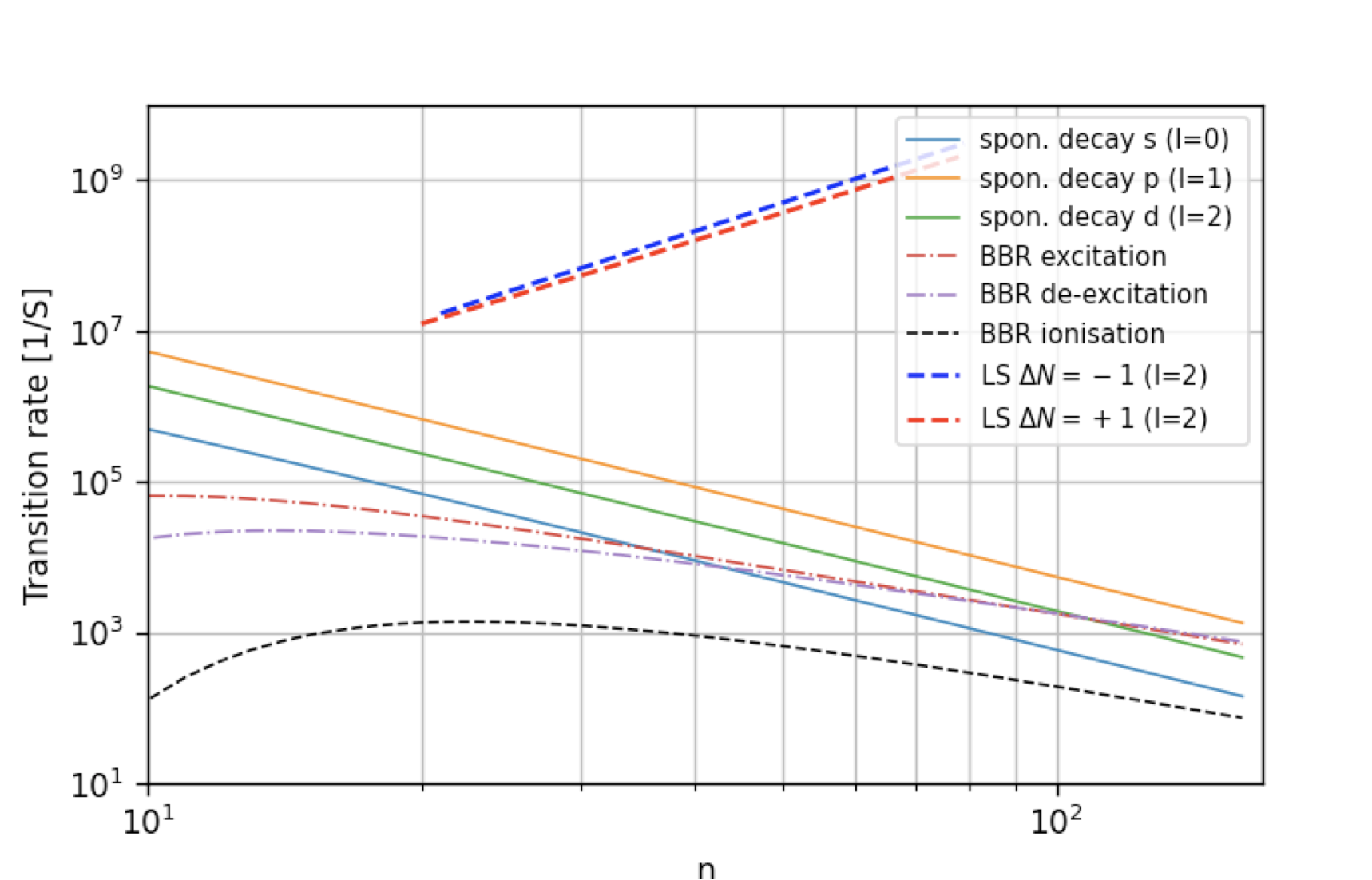}
    \includegraphics[width=0.43\textwidth]{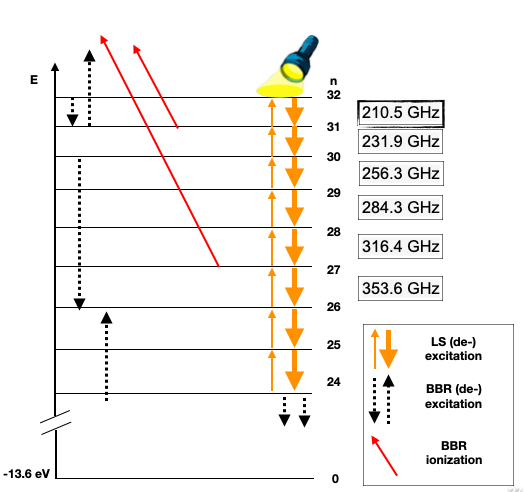}
    \caption{\textbf{Left:} Transition rates for the different processes the Rydberg atoms undergo in the main spectrometer for low angular momenta $l$. The blackbody-radiation-induced ionization peaks at about $n=25$. Even at a radiance of only \SI{0.005}{W/m^2}, the transition rate for light-source-induced de-excitation is about \num{5} orders of magnitude higher than for ionization induced by blackbody radiation. \textbf{Right:} Simplified transitions and binding-energy diagram of hydrogen. The thicker arrow for light-source-induced de-excitation indicates the higher ionization rates for this process compared to the excitation channel (compare with left panel).}
    \label{fig:rydberg_rates}
\end{figure}

Here, we assume that the Rydberg atoms are produced by charge-exchange mechanisms during the sputtering processes in the main-spectrometer walls~\cite{PhDTrost2019} (see Fig.~\ref{fig:RydbergProcess} above). Such mechanisms tend to produce a Rydberg-state distribution favouring low principal quantum numbers $n$ and high angular quantum numbers $l$, with probabilities $P$ according to $P(n,l)=(2l+1)\cdot n^{-5} $. During their transits though the main spectrometer, the Rydberg atoms undergo one or more of the following processes: spontaneous decay with transition rate $\Gamma_{\mathrm{sp}}$, blackbody-radiation-induced excitation or de-excitation ($\Gamma_{\mathrm{trans}}^{\mathrm{BBR}}$) and blackbody-radiation-induced ionization ($\Gamma_{\mathrm{ion}}^{\mathrm{BBR}}$). $\Gamma_{\mathrm{ion}}^{\mathrm{BBR}}$ peaks for all $l$ at about $n=25$ (Fig.~\ref{fig:rydberg_rates}, left). 

Therefore, we aim to mitigate the background with stimulated de-excitation after creation of the Rydberg atoms, quickly depopulating these levels. The goal is to drive multiple $\Delta n=-1$ transitions that de-excite the Rydberg atoms to states from which spontaneous decay is fast ($\tau\approx n^3l(l+1)\cdot$\SI{e-10}{\second}). The frequencies of these transitions range from \SI{2.2}{\tera\hertz} for $n=15\rightarrow14$ to \SI{0.26}{\tera\hertz} for $n=30\rightarrow29$. However, a light source (LS) will not only cause LS-induced excitation/ de-excitation ($\Gamma_{\mathrm{trans}}^{\mathrm{LS}}$) but also LS-induced ionization ($\Gamma_{\mathrm{ion}}^{\mathrm{LS}}$). To reduce additional ionization, narrow-band light sources can be chosen to target specific transitions. The LS is simulated with a Lorentzian spectral distribution described by the center frequency, the radiance and the spectral width.

The Monte Carlo code calculates the Rydberg-state development according to the full transition rate
\begin{equation}
    \Gamma_{\mathrm{full}}=\Gamma_{\mathrm{sp}}+\Gamma_{\mathrm{trans}}^{\mathrm{BBR}}+\Gamma_{\mathrm{ion}}^{\mathrm{BBR}}+\Gamma_{\mathrm{trans}}^{\mathrm{LS}}+\Gamma_{\mathrm{ion}}^{\mathrm{LS}}.
\end{equation}
Fig.~\ref{fig:rydberg_rates} (left) shows the transition rates for low angular momenta.

Combined with a dedicated tracking code, which includes the initial kinetic energies and angular distribution of the different species of Rydberg atoms from SRIM~\cite{srim}, the ionization location and hence the background contribution can be calculated. We investigated a large parameter space, varying the number of targeted transitions and the intensities of the corresponding light sources. As a first, promising result, we found that a $>$\SI{50}{\percent} reduction of the Rydberg background is possible with a set of \num{8} equal-radiance, \SI{0.005}{W/m^2} light sources driving $\Delta n=$ \num{-1} transitions from $n=32$ to $n=24$ (compare with Fig.~\ref{fig:rydberg_rates} right).

So far the Stark and Zeeman effects are not considered in our model. Full inter-$l$-mixing with microwave radiation might be beneficial to our de-excitation goal, but will also induce potential losses through new excitation and ionization channels. However, these effects are expected to be very small as both the magnetic and the electric fields are weak in the bulk of the main spectrometer.

The practical challenge of this approach is that powerful sources with the necessary frequencies for the targeted transition are not easily available. Although \si{THz} technology is a fast-moving field, \si{mW} power is roughly the maximum achievable. 

\subsubsection{Electron taggers}
\label{Subsubsec:ElectronTaggers}

As described in Sec.~\ref{Subsec:Backgrounds}, most of the KATRIN background electrons are produced within the volume of the main spectrometer. These backgrounds could, in principle, be vetoed based on the fact that they do not pass through any of the beamline components upstream of the main spectrometer. Detecting the passage of $\upbeta$ electrons upstream of the main spectrometer allows electron times of flight to be measured for electrons incident on the detector. The majority of tritium $\upbeta$ electrons have times of flight of less than \SI{100}{\micro\second} through the main spectrometer. Detector events from background electrons produced in the main spectrometer could then be vetoed if an electron passage has not been detected within a suitable time window. 

An electron tagger to perform this coincidence measurement must provide timing resolution better than $\sim$\SI{10}{\micro\second}. The rate of electron tagging must be less than the inverse of the required timing resolution in order to avoid random coincidence rates dominating the time-of-flight measurement. This requires an electron tagger to be implemented in the limited space between an active pre-spectrometer and the main spectrometer. Therefore, the tagger would be sensitive to electrons in the pre-spectrometer Penning trap, but this would produce a distinctive oscillating signal in the tagger that could be vetoed. The electron energy loss to the tagger must also be known to better than \SI{1}{\electronvolt} so as not to degrade the MAC-E filter width excessively. Previous studies have shown that there are no fundamental physical impediments to these requirements~\cite{Steinbrink2013}. Further, with a timing resolution of \SI{<100}{\nano\second}, a tagger would enable a differential, time-of-flight spectral measurement that would improve the KATRIN statistical sensitivity at a given run time~\cite{PhDMartin2017}.  

The signal-to-noise ratio and required timing resolution are challenges for measuring the induced signal of the electron by an electron tagger. The development of an electron tagger involves two distinct sections: the pickup structure and the amplifier system. Loop pickups, plate capacitors, and resonators are being investigated as possible pickup structures used to couple with the passing electrons. Fundamental limits of a number of different amplifier systems, including SQUIDs, amplifiers operating at and below the Standard Quantum Limit, and qubit state measurements, are also being investigated. The final goal is to find a combination of pickup and amplifier that provides a sufficient signal-to-noise ratio.

\subsection{Acceptance improvements}
\label{Subsubsec:AcceptanceImprovements}
% Responsible: Jan Behrens (KIT)

KATRIN's design limits the acceptance angle of tritium electrons to $\theta_\mathrm{max} = \arcsin \sqrt{ B_\mathrm{src} / B_\mathrm{max} }$, since the source magnetic field $B_\mathrm{src}$ is smaller than the pinch magnetic field $B_\mathrm{max}$. The magnetic-field ratio is chosen to reject electrons with large starting angles and therefore larger systematic uncertainties. The goal is to balance out the statistical and systematic contributions to the neutrino-mass uncertainty.

One approach to improve the statistical uncertainty of the experiment is to increase the acceptance angle, so that more electrons reach the detector during a neutrino-mass scan. This can be achieved by changing the ratio of the magnetic fields, $B_\mathrm{src} / B_\mathrm{max}$. In our particular case, it is desirable to keep the source at its nominal field strength of $\approx \SI{2.5}{T}$, while reducing the pinch magnetic field together with all other beamline magnetic fields between source and detector. Hence, the entire beamline downstream of the source is operated at lower magnetic field compared to the nominal configuration. This has several advantages. First, by using the same ratio of analyzing and pinch magnetic fields, $B_\mathrm{min} / B_\mathrm{max}$, the active spectrometer volume that is viewed by the detector remains unchanged. Thus, the volume-dependent background from the spectrometer is the same as under nominal conditions. The main-spectrometer filter width is also unchanged. Furthermore, the systematic uncertainty on $B_\mathrm{min}$ depends partially on the total magnetic field strength in the analyzing plane. Lowering the analyzing magnetic field reduces this uncertainty accordingly. Finally, reducing the magnetic field in the transport section between source and spectrometers reduces the amount of synchrotron energy loss and its impact on the total  systematic uncertainty.

On the other hand, the magnetic field in the source and in the rear section is at nominal strength. Contributions to the total systematic uncertainty from these sections have been thoroughly investigated over past measurement campaigns. If these magnetic fields remain unchanged, these time-consuming commissioning measurements remain applicable. 

% Note: there are some discrepancies with the "nominal" acceptance angle. In arXiv:2101.05253 we write 50.5°, in arXiv:2103.04755 we have 51°, but in arXiv:2105.08533 we say 50.4°.
By reducing the pinch magnetic field from its nominal strength of \SI{4.2}{T} to \SI{2.8}{T}, the acceptance angle increases from \ang{50.4} to \ang{70.9} at a source magnetic field of \SI{2.5}{T}. The accepted forward solid angle is then $\Delta\Omega / 2\pi = 1 - \cos \theta_\mathrm{max} =$ \num{0.67}, about \num{1.9}~times larger than in the nominal configuration~\cite{Aker2021Knm1}.

\begin{figure}[tb]
    \centering
    \includegraphics[width=0.7\textwidth]{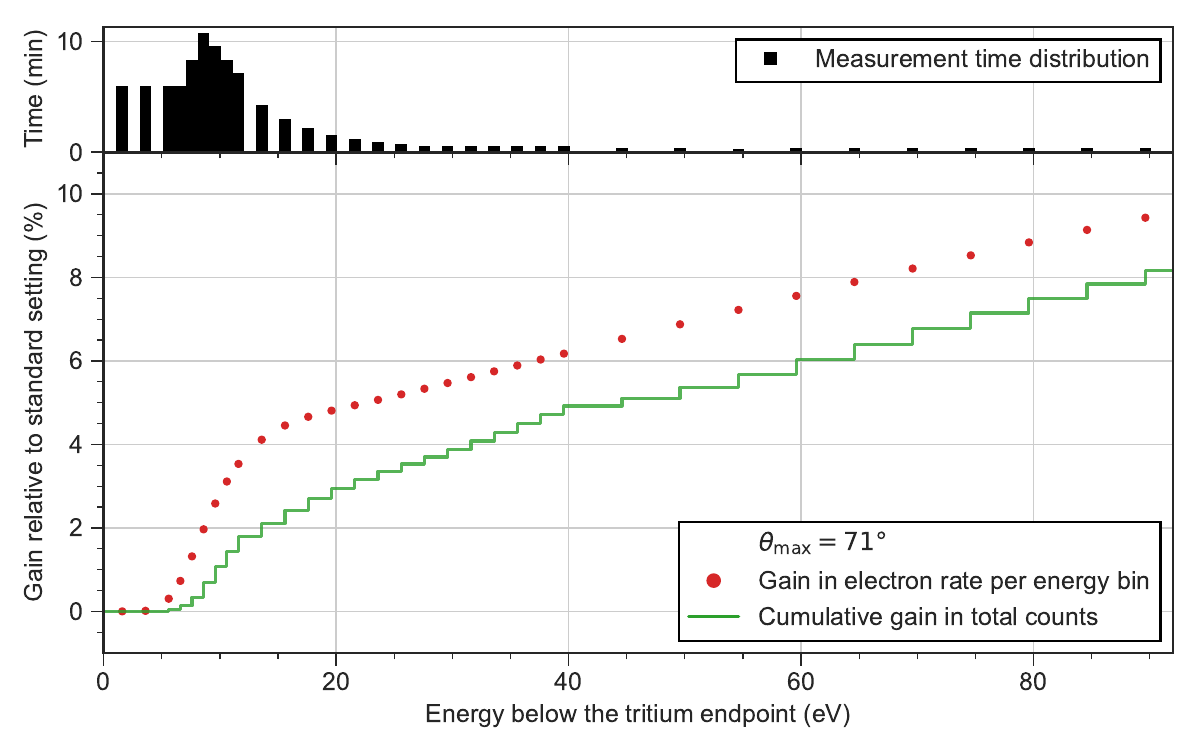}
    \caption{Expected statistics gain from an increased acceptance angle over an energy range of \SI{90}{eV} below the tritium endpoint, at $\theta_\mathrm{max} = \ang{71}$. The red data points show the rate gain at each filter energy, compared to nominal conditions from KATRIN's third measurement campaign. The green stepped line shows the cumulative gain in total statistics (counts) for the corresponding energy window $[E; 0]$, with the distribution of measurement time shown in the top panel.}
    \label{fig:AcceptanceAngle}
\end{figure}

In addition to the accepted forward solid angle, the maximum electron rate at the detector is defined by the magnetic flux $\Phi$. According to the equation $\Phi = \int B(r) dr \approx B \cdot \pi r^2$ the flux is conserved in the beam line. The flux visible by the detector is given by the detector magnetic field, $B_\mathrm{det}$, and the detector radius of \SI{4.5}{cm}. %When changing the magnetic field configuration to increase the acceptance angle one can keep $B_\mathrm{det}$ at its nominal strength of \SI{2.4}{T}. The flux transmitted from source to detector is then the same as under nominal conditions: $\Phi =$ \SI{153}{T.cm^2}.
Scaling down the detector magnetic field by the same ratio as the pinch magnetic field ($B_\mathrm{det}' =$ \SI{1.6}{T}) reduces the transmitted flux to \SI{102}{T.cm^2}. This setting has the advantage of avoiding collisions of the flux tube with the beam-tube walls in the source due to the lower magnetic field in the beamline.

Figure~\ref{fig:AcceptanceAngle} shows the expected gain in statistics relative to the nominal magnetic-field configuration, at different energies below the tritium endpoint $E_0$. To estimate the statistics gain, we determine the total electron counts (including background electrons) at the detector for each retarding potential, using the scenario from KATRIN's third measurement campaign with \num{125} active pixels at $B_\mathrm{max} = \SI{2.8}{T}$ and $B_\mathrm{min} = \SI{0.4}{mT}$.
The rate gain increases steeply over the first \SI{15}{eV} due to the increasing contribution of single-scattered electrons. The cumulative gain over the full energy window depends on the measurement-time distribution and reaches an improvement in statistics by \SI{8}{\%} over the investigated \SI{90}{eV} energy window.

A higher acceptance angle allows electrons that have a high probability of scattering to reach the detector. Electrons lose a fraction of their kinetic energy in inelastic scattering processes on source-gas molecules. Hence, electrons with higher angles $\theta$ contribute mainly to the energy region that is several tens of \si{eV} below the tritium endpoint. A maximum gain in statistics therefore requires increasing the energy window of the neutrino-mass analysis to $\gtrsim$\SI{100}{eV} as shown in Fig.~\ref{fig:AcceptanceAngle}. This in turn requires an excellent understanding of the systematic uncertainties associated with source scattering processes and the tritium molecular final-state distribution (Sec.~\ref{Subsec:SourcesOfSystematicUncertainty}). Fortunately, recent developments in these areas indicate a much smaller contribution to the systematic uncertainty budget than originally anticipated. Hence, the gain in statistics from an increased acceptance angle becomes favorable in terms of balancing the statistical and systematic uncertainties.

\subsection{R\&D for a keV-scale sterile-neutrino search}
\label{Sec:keVRandD}

This section describes research and development efforts in support of a high-statistics search for keV-scale sterile neutrinos (Sec.~\ref{sec:keV}), anticipated after the conclusion of neutrino-mass running.

\subsubsection{Modified rear wall}
\label{Subsubsec:ModifiedRearWall}

A \si{\kilo\electronvolt}-scale sterile-neutrino search requires scanning over a much wider energy range than for the neutrino-mass determination~\cite{Mertens:2014nha}. Simulation studies show that backscattering of signal electrons on the rear wall of the source leads to a significant systematic uncertainty. By replacing the gold-plated rear wall with a beryllium disk, the backscattering coefficient can be suppressed, which leads to a reduction of the absolute size of this systematic effect by up to one order of magnitude~\cite{PhDHuber2020}.

\subsubsection{TRISTAN detector}\label{Subsubsec:TristanDetector}
Since the mass associated with a possible sterile neutrino is unconstrained, the kink-like distortion of the tritium $\upbeta$-decay spectrum, caused by the emission of a sterile neutrino, could be located several \si{\kilo\electronvolt} away from the endpoint, see Fig.~\ref{fig:BetaSpectrum}. As a consequence, the electron count rate is increased up to levels of~$\mathcal{O}$(\SI{e8}{cps}), yet the current focal-plane detector is limited to a total rate of~\SI{e5}{cps} integrated over all~\num{148} pixels. To handle the exceedingly high rate and maintain excellent spectroscopic properties at the same time, i.e.~an energy resolution of~\SI{300}{\electronvolt}~FWHM at~\SI{20}{\kilo\electronvolt}, a multi-pixel silicon drift detector~(SDD) and readout system is currently being developed. The ultimate goal of the TRISTAN detector is to enable KATRIN to reach a sensitivity to the active-to-sterile mixing amplitude at the ppm level. With this, current laboratory limits could be improved by three orders of magnitude and the parameter space of cosmological interest could possibly be reached. 

\begin{figure}[tbp]
    \centering
     \includegraphics[width=0.6\textwidth]{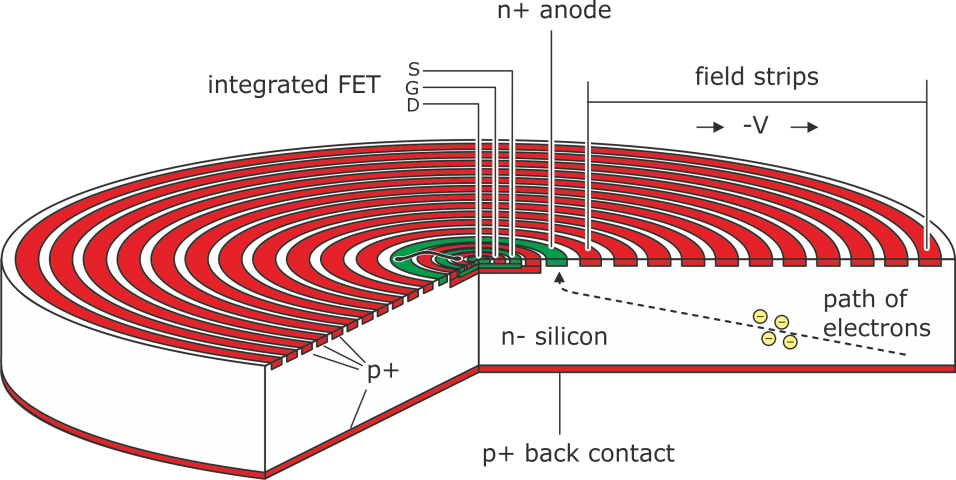}
    \caption{Schematic of a single pixel of the TRISTAN silicon drift detector, reproduced with permission from Ref.~\cite{Lechner:1996sdd}. Electrons are guided by an electric field to the collecting n+~anode in the center of the detector, which is surrounded by several drift rings. The entrance window for radiation is on the opposite side~(p+ back contact). The detector features an integrated field-effect transistor~(FET) close to the anode which forms the first stage of the readout electronics.}
    \label{fig:tristan_sdd}
\end{figure}
%\FloatBarrier

\begin{figure}[tbp]
    \centering
     \includegraphics[width=0.65\textwidth]{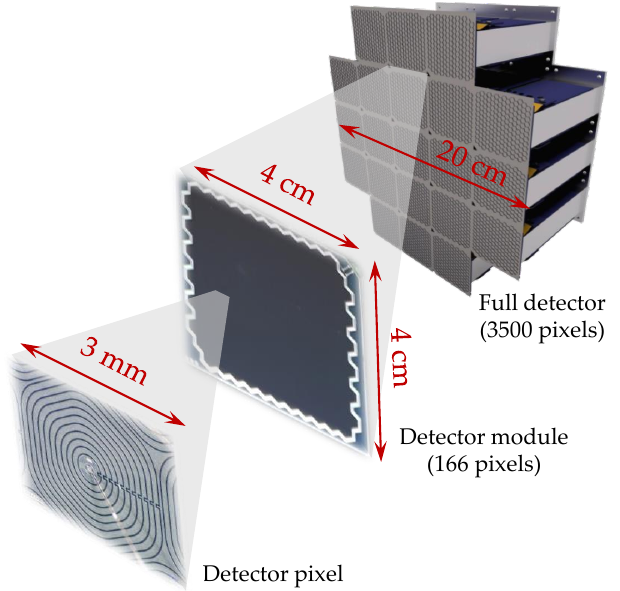}
    \caption{Design of the TRISTAN detector. The full detector array consists of 21 identical modules each consisting of 166~pixels. One module has a size of $4\times4\,$cm$^2$ with the individual pixels having a diameter of $3\,$mm. The pixel anodes are connected to the charge-sensitive amplifier via wire bonds.}
    \label{fig:tristan_module}
\end{figure}
%\FloatBarrier

The TRISTAN detector follows the general design idea of SDDs used for X-ray spectroscopy~\cite{lechner_01}. The detectors are being produced at the Semiconductor Laboratory of the Max Planck Society (HLL). SDDs use the basic principle of sideways depletion, see Fig.~\ref{fig:tristan_sdd}. A volume of a high-resistivity semiconductor material, $n$-type silicon, is covered by rectifying p-doped junctions on both surfaces. A small substrate contact in reverse bias to the p‑regions depletes the silicon bulk. The $p$-junctions are segmented strip-like and biased such that they generate an electric field with a strong component parallel to the surface. Signal electrons released within the depleted volume by the absorption of ionizing radiation drift towards the readout contact, i.e.~the collecting anode. Due to the small physical dimensions of the anode, the detector has a small capacitance which is almost independent of the detector area. Compared to a conventional diode of equal area, this feature translates into larger amplitudes of the output signals. The anode of every pixel is read out by a charge-sensitive amplifier~(CSA) with a JFET integrated into the anode structure of the chip, followed by a low-noise application-specific integrated circuit~(ASIC) specifically developed for the TRISTAN detector~\cite{trigilio_18}. The integrated JFET allows the ASIC chip to be placed at several~\si{cm} distance to the detector chip, while keeping the total anode capacitance at only~\SI{180}{fF}. This provides an excellent signal-to-noise ratio. The read-out components are developed by the company XGLab, Politechnico di Milano \cite{King:2021hcx}, the Institute of Data Processing and Electronics at KIT, and the Max Planck Institute of Physics. 

The SDD concept is very flexible in shape and size. However, a large single-pixel detector would have inherent limitations in terms of drift time and count rate. Therefore, the TRISTAN detector system will be based on multi-cell SDDs combining a large sensitive area with the energy resolution and the count rate capability of a single SDD~(Fig.~\ref{fig:tristan_module}). A multi-cell SDD is a continuous, gapless arrangement of a number of SDDs with individual readout, but with common voltage supply, entrance window, and guard ring structure. The novel detector system is optimized to minimize effects which can alter the shape of the detector response. The pixel size is chosen to be~\SI{3}{\milli\metre} in order to minimize charge-sharing and pixel changes after backscattering and back-reflection~\cite{PhDKorzeczek2020}. The entrance window has a minimal thickness of about~\SI{50}{\nano\metre} in order to minimize energy loss. Detector and readout electronics are optimized for low noise, and thus for good energy resolution. Finally, a full waveform digitization is chosen to minimize the effect of ADC non-linearities.

The development of the TRISTAN detector system follows a staged approach. Starting with 7-pixel detector prototypes with a simple mechanical design~(first without an integrated JFET into the anode structure of the FET), the detector chip was scaled to a more complex module consisting of \SI{47}{pixels}, which has already been tested successfully~\cite{Gugiatti2022}. The final focal-plane array will consist of~\num{9}~(phase 1) and \num{21}~(phase 2) detector modules of \SI{166}{pixels} each.

\paragraph{Detector characterization}
To test the general performance of the novel detector system, several measurements were performed with 1-, 7- and 47-pixel prototype detectors in dedicated setups \cite{Gugiatti:2020wad, Mertens_2019, Mertens_2020, Biassoni:2020oaj}. In particular, we investigated the energy resolution, linearity, and electronic noise of the system. For all measurements, suitable electron, X-ray, and gamma calibration sources were used. For the 7-pixel detectors, we used devices without a JFET integrated into the anode structure, as this feature was only added at a later development stage. Instead, the anode of each pixel was directly connected to the input of an external ASIC via a wire bond.

\begin{figure}[tbp]
    \centering
     \includegraphics[width=0.7\textwidth]{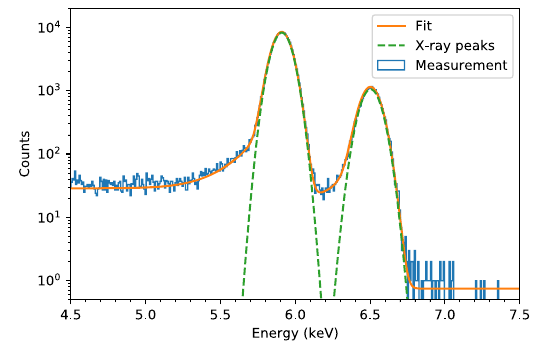}
    \caption{Energy spectrum of $^{55}$Fe recorded with a \SI{2}{\milli\metre} pixel-diameter SDD array at a temperature of \SI{-30}{\celsius}. The MnK$\upalpha$ line at \SI{5.9}{\kilo\electronvolt} has an energy resolution of \SI{139}{\electronvolt}~FWHM and is clearly separated from the MnK$\upbeta$ line at \SI{6.5}{\kilo\electronvolt}. Two phenomenological functions similar to a hypermet~\cite{PhDEggert2004} were used to fit the spectrum. Reproduced from Ref.~\cite{Mertens_2019}.}
    \label{fig:tristan_energy_spec}
\end{figure}
%\FloatBarrier

\begin{figure}[tbp]
    \centering
     \includegraphics[width=0.7\textwidth]{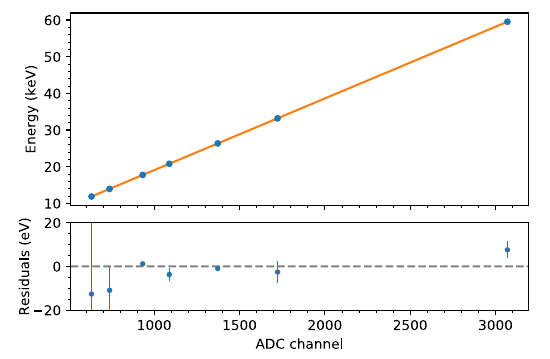}
    \caption{Linear calibration curve based on a measurement with an $^{241}$Am source. The blue data points correspond to several photon lines. The error bars include the uncertainties of the line positions and of the fit. The maximal deviation from linearity is \SI{0.1}{\percent} at the lowest-energy peak of \SI{11.9}{\kilo\electronvolt}. Reproduced from Ref.~\cite{Mertens_2019}.}
    \label{fig:tristan_linearity}
\end{figure}
%\FloatBarrier

\begin{figure}[tbp]
    \centering
     \includegraphics[width=0.7\textwidth]{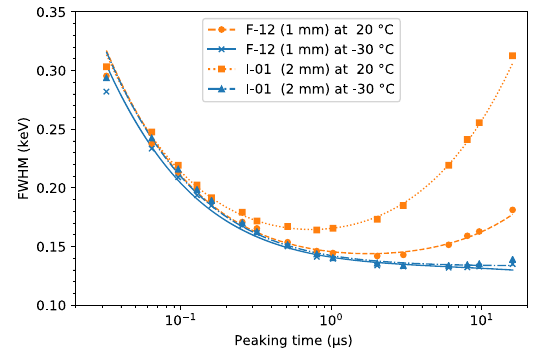}
    \caption{Energy resolution (FWHM) of the \SI{5.9}{\kilo\electronvolt} line of $^{55}$Fe as a function of the trapezoidal-filter peaking time, comparing the performance of the \SI{1}{\milli\metre} and \SI{2}{\milli\metre} pixel diameter detectors at room temperature~(orange) and \SI{-30}{\celsius} (blue). Reproduced from Ref.~\cite{Mertens_2019}.}
    \label{fig:tristan_noise}
\end{figure}
%\FloatBarrier

A typical $^{55}$Fe energy spectrum recorded with a \SI{2}{\milli\metre} pixel-diameter SDD array at \SI{-30}{\celsius} is shown in Fig.~\ref{fig:tristan_energy_spec}. The MnK$\upalpha$ and MnK$\upbeta$ lines at \SI{5.9}{\kilo\electronvolt} and \SI{6.5}{\kilo\electronvolt}, respectively, can be well approximated by Gaussian functions, with the entire structure best described by hypermet-type functions~\cite{PhDEggert2004}. At the \SI{5.9}{\kilo\electronvolt} X-ray line, an excellent energy resolution of \SI{139}{\electronvolt}~FWHM was obtained. To test the linearity of the system, an $^{241}$Am source was used. As can be seen in Fig.~\ref{fig:tristan_linearity}, the system shows  excellent linearity over the entire energy range from \SIrange{10}{60}{\kilo\electronvolt}. All measured line positions deviate from linearity by~\SI{<0.1}{\percent}. The noise performance of the detector system was investigated by measuring the energy resolution of the~\SI{5.9}{\kilo\electronvolt} MnK$\upalpha$ line of~$^{55}$Fe as a function of the peaking time, which is a parameter of the trapezoidal filter applied to determine the energy of an event from the waveform. Figure~\ref{fig:tristan_noise} shows the typical dependence of the resolution on the peaking time for \SI{1}{\milli\metre}- and \SI{2}{\milli\metre}-pixel-diameter detectors, at room temperature and at \SI{-30}{\celsius}. The increased resolution at higher peaking times at room temperature is due to leakage current. As expected, it is more pronounced for the~\SI{2}{\milli\metre} detector since the leakage current grows with the area of the pixel. In contrast, at~\SI{-30}{\celsius}, the leakage current is suppressed to a negligible level for both detectors. This shows that the performance is not degraded for larger pixel areas. An energy resolution of~$\leq$\SI{140}{\electronvolt}~FWHM is reached for peaking times of about~\SI{1}{\micro\second}. More details on the X-ray characterization measurements can be found in Ref.~\cite{Mertens_2019}.

The application of SDDs to high-precision electron spectroscopy is novel to the TRISTAN detector. The energy-deposition profile is different for massive charged particles and for photons. This fact has several consequences: first, electrons deposit a significant fraction of their energy close to the entrance-window surface, where the electric fields are too weak to transport charge carriers to the readout electrode. Therefore, the energy of the electron is not fully detected. Second, low-energy electrons have a certain probability to scatter back from the detector surface, again resulting in a partial energy measurement. As a consequence, a detailed characterization of the detection system with electrons is of major importance.

\begin{figure}[tbp]
    \centering
     \includegraphics[width=0.7\textwidth]{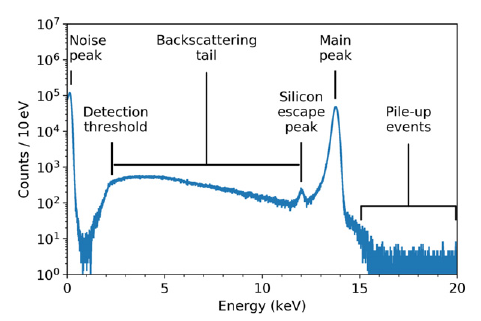}
    \caption{Spectrum of a \SI{14}{\kilo\electronvolt} monoenergetic electron beam from an electron microscope measured with a TRISTAN prototype detector. Taken from Ref.~\cite{Mertens_2020}.}
    \label{fig:tristan_electron_spectrum}
\end{figure}

\begin{figure}[tbp]
    \centering
     \includegraphics[width=0.7\textwidth]{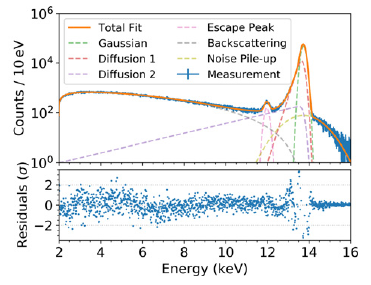}
    \caption{Fit of the electron response model to the measured energy spectrum. The energy of the electron microscope was set to \SI{14}{\kilo\electronvolt}. The silicon escape peak and noise pile-up are approximated by two additional functions. Taken from Ref.~\cite{Mertens_2020}.}
    \label{fig:tristan_electron_spectrum_fit}
\end{figure}

Here, we present first characterization measurements of a~\SI{7}{-pixel} prototype SDD with monoenergetic electrons at room temperature~\cite{Mertens_2020, Gugiatti:2020wad, Biassoni:2020oaj}. A scanning electron microscope was used as a calibration source. A typical electron energy spectrum is shown in Fig.~\ref{fig:tristan_electron_spectrum}. An empirical model was developed to describe the detector's response to electrons~\cite{Mertens_2020}. Each physical effect is modeled by a separate term including a Gaussian function, a low-energy tail, a silicon escape peak, and a backscattering tail. An example of the fit to the obtained energy spectrum is shown in Fig.~\ref{fig:tristan_electron_spectrum_fit}, indicating very good agreement of the model with the data. More details on the electron-characterization measurements can be found in Ref.~\cite{Mertens_2020, Gugiatti:2020wad, Biassoni:2020oaj}, including an estimation of the entrance-window thickness as well as the impact of the detector response on the sterile-neutrino sensitivity.

%\FloatBarrier

%\FloatBarrier

\begin{figure}[tb]
    \centering
     \includegraphics[width=0.7\textwidth]{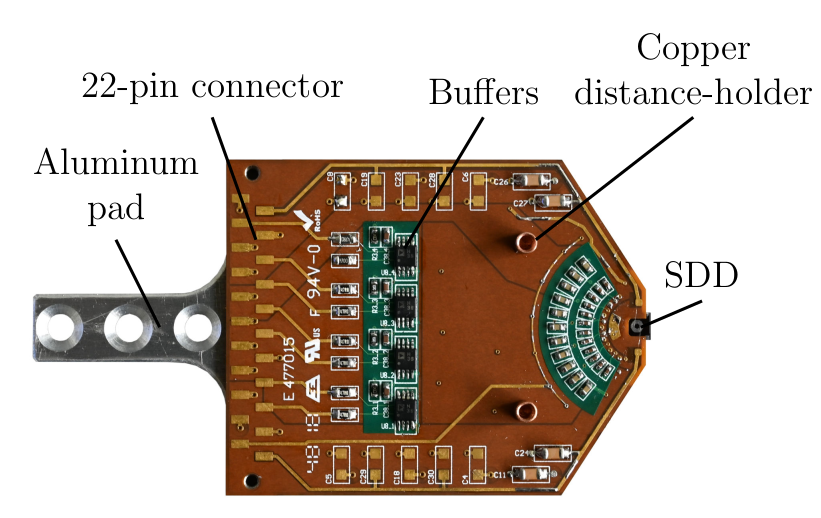}
    \caption{Photograph of the board hosting the TRISTAN FBM detector. Reproduced with permission from Ref.~\cite{MasterUrban2019}.}
    \label{fig:tristan_fbm}
\end{figure}

\paragraph{TRISTAN detector in the KATRIN beamline}
A \num{7}-pixel TRISTAN prototype detector with a pixel diameter of~\SI{250}{\micro\metre} has been integrated as the Forward Beam Monitor~(FBM) in the KATRIN beam line. A photograph of the printed circuit board hosting the SDD is shown in Fig.~\ref{fig:tristan_fbm}. Located upstream of the main spectrometer, the FBM continuously monitors the flux of $\upbeta$-electrons. In addition, the excellent energy resolution of the SDD permits measurement of the spectral shape of the $\upbeta$-decay spectrum. In Ref.~\cite{MasterUrban2019} it was demonstrated that the activity of the tritium source can be monitored on the~\SI{<0.1}{\percent} level on time scales of minutes. Moreover, measurements with the FBM can be used for initial searches for keV-scale sterile neutrinos with limited statistics and systematic uncertainties. These measurements can help to identify challenges and prepare for operations with the full TRISTAN detector.

%\FloatBarrier

\begin{figure}[tbp]
    \subfloat[47ch planar TRISTAN detector.] {\label{fig:tristan_47ch_planar}
    \includegraphics[width=.50\linewidth]{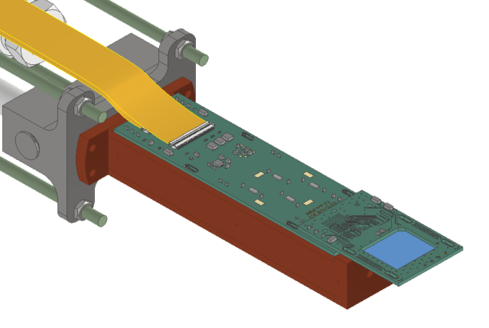}} %\quad
    \subfloat[47ch 3D TRISTAN detector.] {\label{fig:tristan_47ch_3d}
    \includegraphics[width=.50\linewidth]{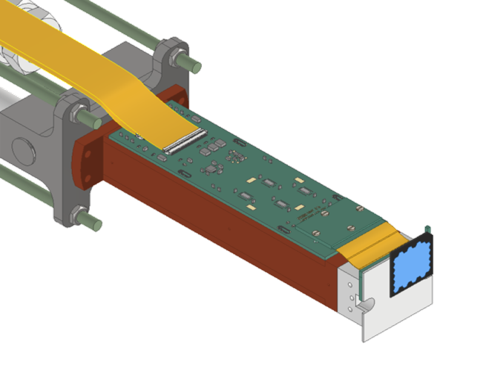}}
    \caption{CAD drawings of the \num{47}-pixel TRISTAN detector. The left figure shows the planar detector design, whereas the right one shows the three-dimensional design. The green printed circuit board hosts the signal readout electronics and is mounted on a copper cooling block.}
    \label{fig:tristan_modules}
\end{figure}
%\FloatBarrier

\begin{figure}[tbp]
    \subfloat[47ch planar TRISTAN detector.] {\label{fig:tristan_47ch_planar_resolution}
    \includegraphics[width=.50\linewidth]{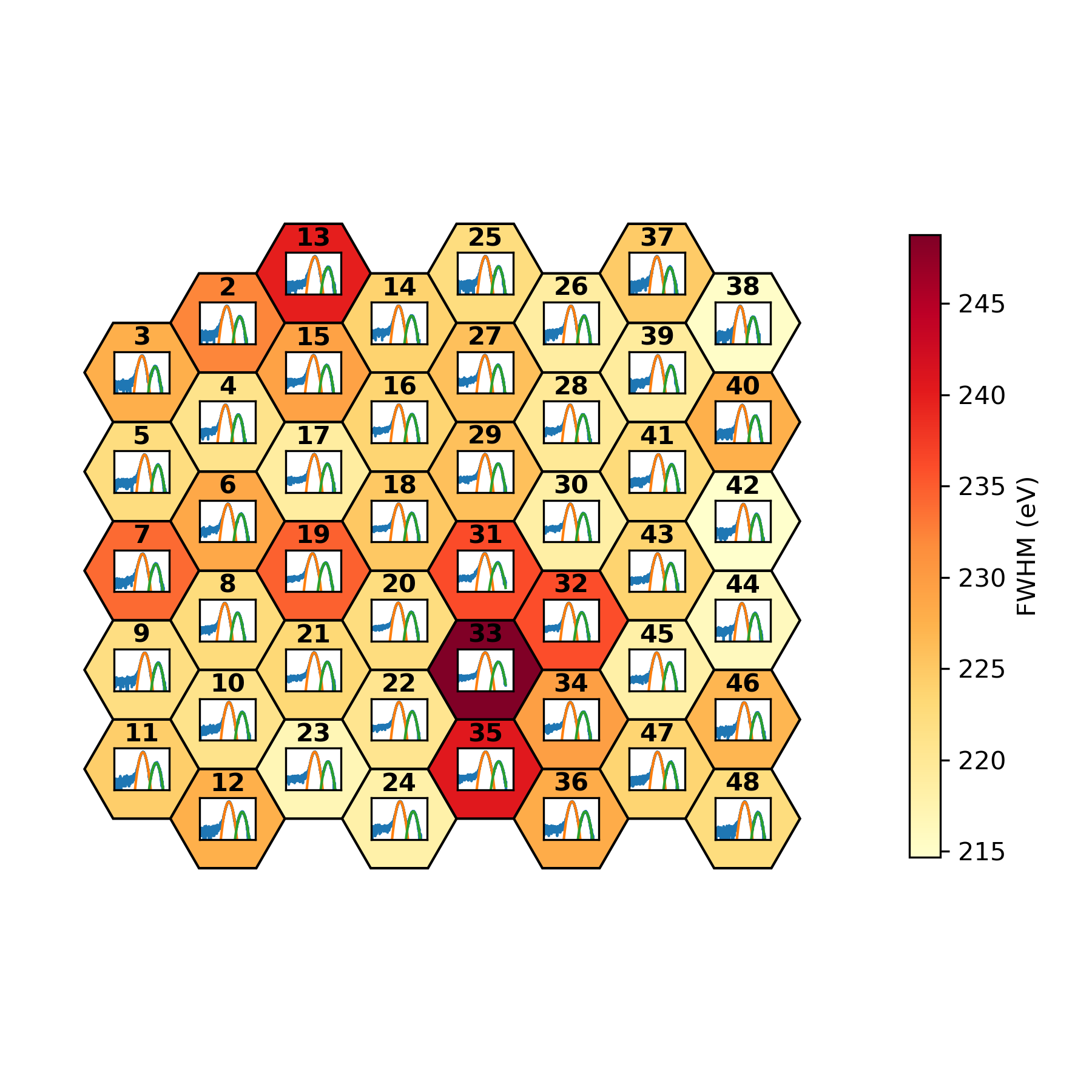}} %\quad
    \subfloat[47ch 3D TRISTAN detector.] {\label{fig:tristan_47ch_3d_resolution}
    \includegraphics[width=.50\linewidth]{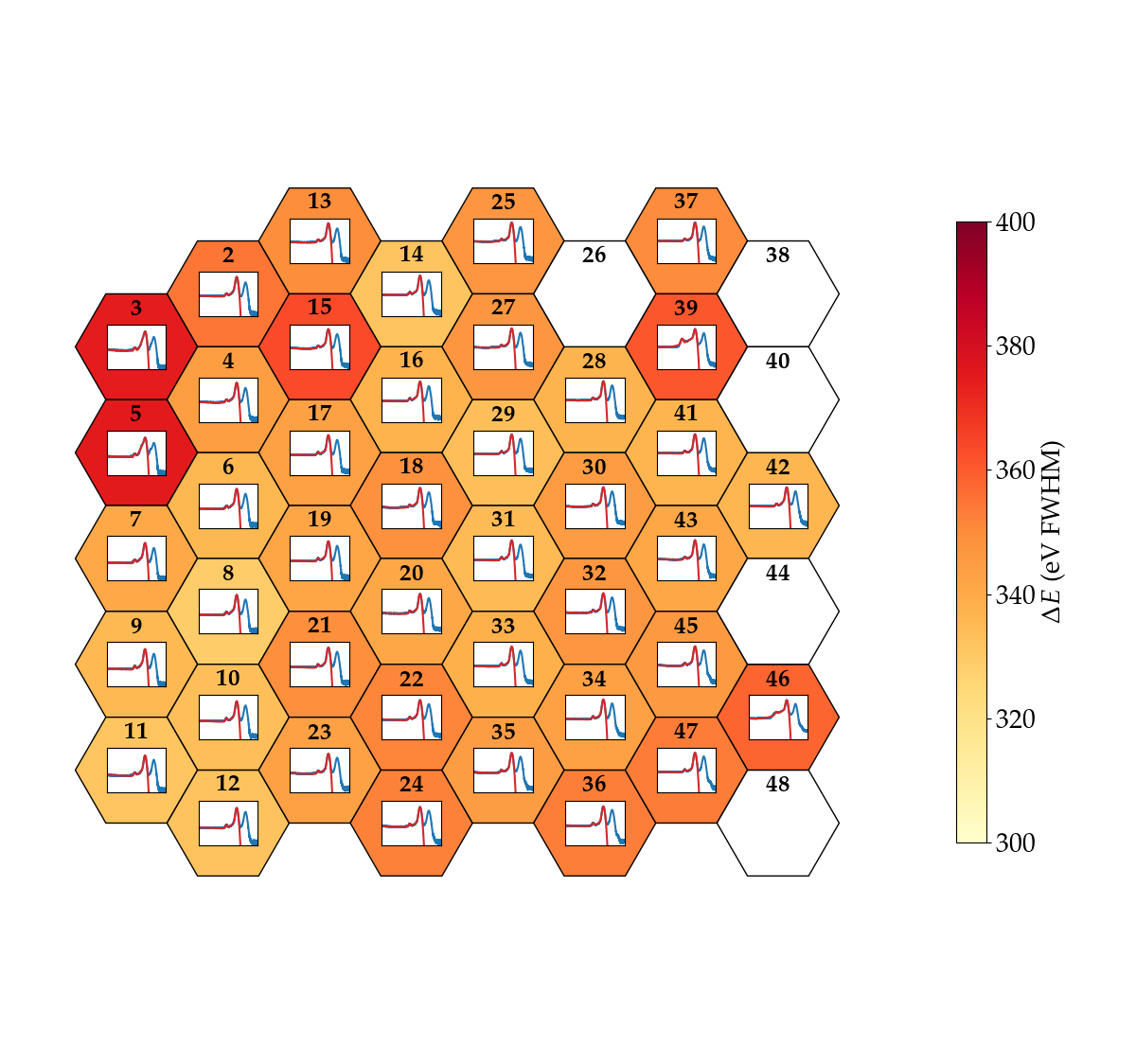}}
    \caption{Pixel maps of the \num{47}-pixel TRISTAN detector~(planar and 3D design) in the KATRIN monitor spectrometer. The color bar indicates the energy resolution of the individual pixels. The plot insets show typical energy spectra ($^{55}$Fe calibration spectra for the planar detector and $^{83\mathrm{m}}$Kr spectra for the 3D detector). The spectra were fit with suitable functions. For the 3D design, five pixels did not work properly and were excluded from the analysis; they are marked here in white.}
    \label{fig:tristan_modules_resolution}
\end{figure}
%\FloatBarrier

\begin{figure}[tpb]
    \centering
     \includegraphics[width=0.9\textwidth]{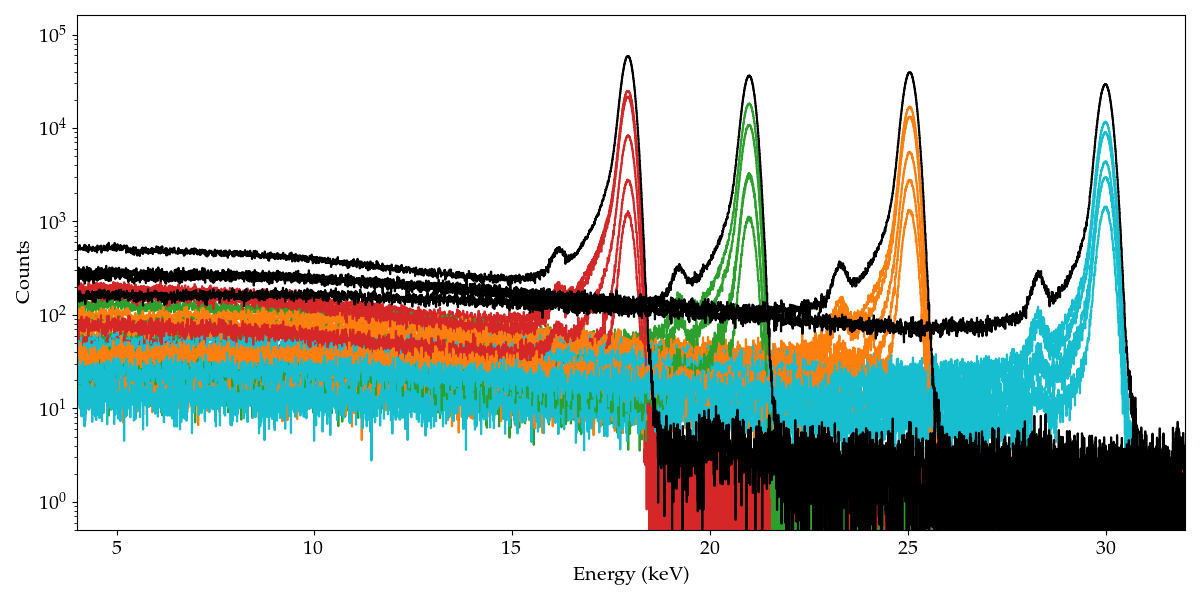}
    \caption{Energy spectra (wall electrons) measured with a \num{47}-pixel TRISTAN detector in the KATRIN monitor spectrometer at different high voltages. The black spectra represent stacked spectra of several pixels.}
    \label{fig:tristan_wall_electrons}
\end{figure}
%\FloatBarrier

A \num{47}-channel TRISTAN prototype detector was successfully integrated into the KATRIN monitor spectrometer~(MoS) in November~2020. The main objective was to gain a better understanding of the response of the SDD in a realistic MAC-E filter environment, i.e.~operation in a strong magnetic field~(up to \SI{0.4}{\tesla}), at high temperatures~(\SI{40}{\celsius}), and in vacuum~(\SI{e-9}{\milli\bar}). In a first step for initial tests, a planar detector was installed (Fig.~\ref{fig:tristan_47ch_planar}). It was demonstrated that all pixels are functional with an energy resolution of about~$\lesssim$\SI{300}{\electronvolt}~FWHM at the \SI{5.9}{\kilo\electronvolt} line of $^{55}$Fe, (Fig.~\ref{fig:tristan_47ch_planar_resolution}). In the next step, a three-dimensional version~(SDD and readout electronics mounted orthogonally to each other) was installed in the MoS (Fig.~\ref{fig:tristan_47ch_3d}). Eventually, this device was used to detect electrons from the spectrometer walls, from an electron gun, and from a \textsuperscript{83m}Kr source. Fig.~\ref{fig:tristan_wall_electrons} shows the recorded energy spectra of the wall electrons with the spectrometer set to different retarding potentials. For \SI{30}{\kilo\electronvolt} wall electrons, an energy resolution of about~$\lesssim$\SI{440}{\electronvolt}~FWHM was obtained (Fig.~\ref{fig:tristan_47ch_3d_resolution}). In 2022, a TRISTAN 3D~detector with \num{166}~pixels will be deployed for the first time in the MoS.

\section{Discussion and conclusions}
\label{Sec:DiscussionAndConclusions}

First envisioned in 2001, with its conceptual design formalized in 2005~\cite{KATRIN2005}, the KATRIN experiment is the world-leading effort to measure the absolute mass scale of the neutrino. Via an exquisite measurement of the endpoint region of the tritium $\upbeta$-decay spectrum, KATRIN probes for the telltale shape distortion induced by the tiny neutrino rest mass. After the first two neutrino-mass data-taking campaigns, both in 2019, KATRIN has set an upper limit of $m_\nu <$ \SI{0.8}{\electronvolt} (\SI{90}{\percent} C.L.) (Sec.~\ref{subsec:RecentResults}). The experiment continues to acquire data and is expected to run for at least two more years; the collaboration is continuing to develop increasingly sophisticated analysis methods (Secs.~\ref{Sec:KatrinAnalysisToolsAndStrategies},~\ref{Subsec:FittingStrategies} and~\ref{sec:combinecampaigns}). Several promising upgrades, including active background-mitigation measures (Sec.~\ref{Subsec:BackgroundMitigation}) and improvements to the acceptance (Sec.~\ref{Subsubsec:AcceptanceImprovements}), are currently in development with the aim of reducing or compensating for the observed KATRIN background, and thereby further improving the neutrino-mass sensitivity towards the 0.2~eV sensitivity goal.

In addition to the effective neutrino mass, KATRIN's measurement of the tritium endpoint $\upbeta$ spectrum provides sensitivity to a variety of hypothesized beyond-Standard-Model phenomena, including sterile neutrinos at \si{\electronvolt} (Sec.~\ref{sec:eV}) and \si{\kilo\electronvolt} scales (Sec.~\ref{sec:keV}); generalized neutrino interactions and exotic weak currents (Sec.~\ref{sec:CCBSM}); new light bosons (Sec.~\ref{Subsec:NewLightBosons}); Lorentz-invariance violation (Sec.~\ref{Subsec:LorentzInvarianceViolation}); local relic-neutrino overdensities (Sec.~\ref{Subsec:RelicNeutrinos}); and other new physics ideas (Sec.~\ref{Subsec:OtherNewPhysics}). In several cases, initial searches have already been performed with the first data sets. The KATRIN data sets also hold promise for spectroscopy of tritium molecular final states (Sec.~\ref{Subsec:MolecularFinalStatesFromT2Decay}) and $^{83m}$Kr conversion electrons (Sec.~\ref{Subsec:83mKrSpectroscopy}), used for calibration.

Once KATRIN neutrino-mass running is complete, the next phase of the experiment will be a dedicated search for keV-scale sterile neutrinos  (Sec.~\ref{sec:keV}). Optimizing KATRIN's sensitivity in this mass range requires running with a lower-luminosity source, a modified rear wall, and a new detector with improved energy resolution at high electron rates. The effort to realize such a detector is well underway (Sec.~\ref{Subsubsec:TristanDetector}). With these improvements, KATRIN will achieve a world-leading laboratory sensitivity to the presence of a sterile neutrino with $m_4$ in the \si{\kilo\electronvolt} mass range, setting complementary limits to those from astrophysics measurements. The future of KATRIN is bright.

\ack
We acknowledge the support of Helmholtz Association (HGF), Ministry for Education and Research BMBF (05A20PMA, 05A20PX3, 05A20VK3), Helmholtz Alliance for Astroparticle Physics (HAP), the doctoral school KSETA at KIT, and Helmholtz Young Investigator Group (VH-NG-1055), Max Planck Research Group (MaxPlanck@TUM), and Deutsche Forschungsgemeinschaft DFG (Research Training Groups Grants No., GRK 1694 and GRK 2149, Graduate School Grant No. GSC 1085-KSETA, and SFB-1258) in Germany; Ministry of Education, Youth and Sport (CANAM-LM2015056, LTT19005) in the Czech Republic; and the Department of Energy through grants DE-FG02-97ER41020, DE-FG02-94ER40818, DE-SC0004036, DE-FG02-97ER41033, DE-FG02-97ER41041,  {DE-SC0011091 and DE-SC0019304 and the Federal Prime Agreement DE-AC02-05CH11231} in the United States. This project has received funding from the European Research Council (ERC) under the European Union Horizon 2020 research and innovation programme (grant agreement No. 852845). The development of the TRISTAN detector has received funding from Istituto Nazionale di Fisica Nucleare (Italy) CSN2. We thank the computing cluster support at the Institute for Astroparticle Physics at Karlsruhe Institute of Technology, Max Planck Computing and Data Facility (MPCDF), and National Energy Research Scientific Computing Center (NERSC) at Lawrence Berkeley National Laboratory.

\end{document}

%% file: authorlist-IOP.tex
% autogenerated by authorlist.py from input file 'authorlist-APS.tex' using iopart format

% Authors:
\author{%
M~Aker$^{1}$, 
M~Balzer$^{2}$, 
D~Batzler$^{1}$, 
A~Beglarian$^{2}$, 
J~Behrens$^{1}$, 
A~Berlev$^{3}$, 
U~Besserer$^{1}$, 
M~Biassoni$^{4}$, 
B~Bieringer$^{5}$, 
F~Block$^{6}$
% ...
}

% Affiliations:
\address{$^{1}$ Institute for Astroparticle Physics~(IAP), Karlsruhe Institute of Technology~(KIT), Hermann-von-Helmholtz-Platz 1, 76344 Eggenstein-Leopoldshafen, Germany} 
\address{$^{2}$ Institute for Data Processing and Electronics~(IPE), Karlsruhe Institute of Technology~(KIT), Hermann-von-Helmholtz-Platz 1, 76344 Eggenstein-Leopoldshafen, Germany}
\address{$^{3}$ Institute for Nuclear Research of Russian Academy of Sciences, 60th October Anniversary Prospect 7a, 117312 Moscow, Russia}
\address{$^{4}$ INFN - Sezione di Milano - Bicocca, 20126 Milano, Italy}
\address{$^{5}$ Institute for Nuclear Physics, University of M\"{u}nster, Wilhelm-Klemm-Str.~9, 48149 M\"{u}nster, Germany}
\address{$^{6}$ Institute of Experimental Particle Physics~(ETP), Karlsruhe Institute of Technology~(KIT), Wolfgang-Gaede-Str.~1, 76131 Karlsruhe, Germany}
% ...
\eads{\mailto{dparno@cmu.edu}, \mailto{magnus.schloesser@kit.edu}}